\documentclass[structabstract]{aa}
\usepackage{graphicx}
\usepackage{url}            % Fuer Internetadressen
\usepackage{txfonts}
\usepackage{natbib}
\bibpunct{(}{)}{;}{a}{}{,}  % A&A style
\usepackage{color}
\usepackage[percent]{overpic} % RF: for plot labelling

\begin{document}
   \title{A pan-chromatic view of the galaxy cluster \\
   \object{XMMU\,J1230.3+1339}
   at $z\!=\!0.975$\thanks{Based on observations obtained with ESO Telescopes at the
   Paranal Observatory under program ID 078.A-0265 and 081.A-0312, and observations with the
   Large Binocular Telescope (LBT), %APEX-SZ, 
   and the X-ray observatories XMM-{\it Newton} and {\it Chandra}.}}

   \subtitle{Observing the assembly of a massive %Coma-like 
   system}

   \author{R. Fassbender
          \inst{1}
          \and
          H. B\"ohringer
          \inst{1}
          \and
          J.S. Santos
          \inst{2}
          \and
          G.W. Pratt
          \inst{3}
          \and
          R. \v{S}uhada
          \inst{1}
          \and
          J. Kohnert
          \inst{4}
%          \and
%         M.W. Sommer
%          \inst{4,10}
%          \and
%          J. Kennedy
%          \inst{12}
          \and
          M. Lerchster
          \inst{1,5}
          \and
          E. Rovilos
          \inst{1}
          \and
          D. Pierini
          \inst{1}
          \and
          G. Chon
          \inst{1}
          \and
          A.D. Schwope
          \inst{4}
          \and
          G. Lamer
          \inst{4}
          \and
          M. M\"uhlegger
          \inst{1}
          \and
          P. Rosati
          \inst{6}
          \and
          H. Quintana
          \inst{7}
          \and
          A. Nastasi
          \inst{1}
          \and
          A. de Hoon
          \inst{4}
         \and
          S. Seitz
          \inst{5}
           \and
          J.J. Mohr
          \inst{1,5,8}          
%          \and
%          K. Basu
%          \inst{4,10}
%          \and
%          A.N. Bender
%          \inst{11}
%          \and
%          F. Bertoldi
%          \inst{10}
%          \and
%          M. Dobbs
%          \inst{12}
%          \and
%          N.W. Halverson
%          \inst{11}
%          \and
%          W.L. Holzapfel
%          \inst{13}
%          \and
%          C. Horellou
%          \inst{14}
%          \and
%          D. Johannson
%          \inst{14}
%          \and
%          B.R. Johnson
%          \inst{13}
%          \and
%          R. Kneissl
%          \inst{4}
%          \and
%          A.T. Lee
%          \inst{13,15}
%          \and
%          K.M. Menten
%          \inst{4}
%          \and
%          F. Pacaud
%          \inst{10}
%          \and
%          C.L. Reichert
%          \inst{13}
%          \and
%          B. Westbrook
%         \inst{13}
          }
% possibly: + APEX people

  \institute{Max-Planck-Institut f\"ur extraterrestrische Physik (MPE),
              Giessenbachstrasse~1, 85748 Garching, Germany \\
              \email{rfassben@mpe.mpg.de}
            \and
         INAF-Osservatorio Astronomico di Trieste, Via Tiepolo 11, 34131 Trieste, Italy
        \and
        CEA \/ Saclay, Service d'Astrophysique, L'Orme des Merisiers, B\^at. 709, 91191 Gif-sur-Yvette Cedex, France
          \and
            Astrophysikalisches Institut Potsdam (AIP),
            An der Sternwarte~16, 14482 Potsdam, Germany
           \and
            University Observatory Munich, Ludwigs-Maximillians University Munich,
            Scheinerstr. 1, 81679  Munich, Germany
%         \and
%         Max-Planck-Institut f\"ur Radioastronomie (MPIfR), Auf dem H\"ugel 69, 53121 Bonn, Germany , Germany
         \and
         European Southern Observatory (ESO), Karl-Scharzschild-Str.~2, 85748 Garching, Germany
        \and
        Departamento de Astronom\'ia y Astrof\'isica, Pontificia Universidad
        Cat\'olica de Chile, Casilla 306, Santiago 22, Chile
%        \and
%        Department of Astronomy, University of Illinois, Urbana, IL 61801, USA
%        \and
%        Argelander Institute for Astronomy, Bonn University, Bonn, Germany
%        \and
%        Center for Astrophysics and Space Astronomy, University of Colorado, Boulder, CO, 80309, USA
%        \and
%        Physics Department, McGill University, H2T 2Y8 Montreal, Canada ; D-Wave Systems Inc., Burnaby, V5C 6G9, Canada
%        \and
%        Department of Physics, University of California, Berkeley, CA, 94720, USA
%        \and
%        Onsala Space Observatory, Chalmers University of Technology, 43992 Onsala, Sweden
%        \and
%        Lawrence Berkeley National Laboratory, Berkeley, CA, 94720, USA
        \and
        Excellence Cluster Universe, Boltzmannstr.~2, 85748 Garching, Germany
             %University of Alexandria, Department of Geography, ...\\
             %\email{c.ptolemy@hipparch.uheaven.space}
             %\thanks{The university of heaven temporarily does not
             %        accept e-mails}
             }

%   \date{Received September 15, 1996; accepted March 16, 1997}

% \abstract{}{}{}{}{}
% 5 {} token are mandatory

%:Abstract
  \abstract
  % context heading (optional)
  % {} leave it empty if necessary
   {Observations of the formation and evolution of massive galaxy clusters and their matter components provide crucial constraints
   on cosmic structure formation, the thermal history of the intracluster medium (ICM), galaxy evolution, %and
   transformation processes,
   and gravitational and hydrodynamic interaction physics of the subcomponents.}
  % aims heading (mandatory)
   {We characterize the global multi-wavelength properties of the X-ray selected galaxy cluster
   XMMU\,J1230.3+1339 at $z\!=\!0.975$, a new  system  discovered within the XMM-{\it Newton}
   Distant Cluster Project (XDCP). 
   We measure and compare various widely used mass proxies and identify multiple
   cluster-associated components from the inner core region out to the large-scale structure environment.}
  % methods heading (mandatory)
   {We present a comprehensive galaxy cluster study based on a joint analysis of X-ray data, optical imaging and spectroscopy
   observations, %the Sunyaev-Zedovich effect (SZE), 
   weak lensing results, and radio properties for achieving a detailed  %unprecedented 
   multi-component view on a system at $z \sim 1$.}
  % results heading (mandatory)
   {We find an optically very rich and massive system with
   $M_{200}\!\simeq\!(4.2 \pm 0.8)\!\times\!10^{14}\,\mathrm{M_{\sun}}$, $T_{X,2500}\!\simeq\!5.3^{+0.7}_{-0.6}$\,keV, and
   $L^{\mathrm{bol}}_{\mathrm{X,500}}\!\simeq\!(6.5 \pm 0.7)\!\times\!10^{44}$\,erg\,s$^{-1}$.
   We have identified  a central fly-through group close to core passage %, similar to the  'Bullet' cluster', 
   and find marginally
   extended 1.4\,GHz radio emission possibly associated with the turbulent wake region of the merging event.
   %The central region is being disrupted by a fly-through group bullet with a tentative Mach number of $M\simeq 2.1 \pm
   %0.5$...., with extended radio emission from wake region.
   On the cluster outskirts we see evidence for an on-axis
   infalling group with a second Brightest Cluster Galaxy (BCG) and indications for an additional off-axis group accretion event.
   We trace two galaxy filaments beyond the nominal %virial
   cluster radius   and provide a tentative
   reconstruction of the 3D-accretion geometry of the system.
  % We also attempted the first absolute distance measurement at this redshift and find good agreement with concordance model
  % expectations within our error budget.
   }
  % conclusions heading (optional), leave it empty if necessary
   {In terms of total mass, ICM structure, optical richness, and the presence of two dominant BCG-type galaxies, the newly confirmed
   cluster XMMU\,J1230.3+1339 is likely the progenitor of a system very similar to the local Coma cluster, differing by
   7.6\,Gyr of structure evolution. This new system is an ideally suited astrophysical model laboratory for  in-depth %detailed in-situ
   follow-up studies on the aggregation of baryons in the cold and hot phases.
   %provides access to one of the best suited high-z astrophysical laboratories
   %%to observationally address various fundamental questions in cluster physics and cosmology.
%    spanning more than six orders of
%   magnitude in physical scale. These studies could shed new light on the origin of ultra-relativistic electrons and
%   diffuse radio emission in clusters, various galaxy interaction processes in different environments, hydrodynamic merger
%   shocks in the intracluster medium, the origin of the Cool Core - Non-Cool-Core Cluster dichotomy, the accretion modes
%   of (dark) matter, the connection to large-scale structure filaments, and the absolute geometric distance to the cluster.
   %such as the destruction of cool cores, the origin of diffuse cluster radio emission,
   %the accretion modes of (dark) matter, and the site and
   %physical processes of galaxy transformations.
   }


   \titlerunning{A pan-chromatic view of XMMU\,J1230.3+1339 at $z\!=\!0.975$}
   \authorrunning{R. Fassbender et al.}

   \maketitle
%

% >>>>>>>>>> ToDO
% use object command \object for NED

%________________________________________________________________

% page 1-2:  Abstract+Intro
% page 3-5:  Observations+Optical Plots
% page 6-8:  Results X-ray Plots
% page 9-10: Density Plots/Images
% page 11:   Conclusions
% page 12:  Acknowledgements & References

%\clearpage

%============== SECT 1 ===============================================================================================
\section{Introduction}

% Introduce Acronyms:
% 1 page
Massive galaxy clusters at high redshift are unique laboratories to study galaxy evolution in the densest environments,
the thermodynamic properties and chemical enrichment of the hot intra cluster medium (ICM) at large look-back times,
and the structure and growth of the underlying virialized Dark Matter (DM) halos. Moreover, distant galaxy clusters are
among the most promising cosmological probes to shed new light on the properties and evolution of Dark Energy (DE).
This potential has recently been demonstrated based on moderate size X-ray cluster samples out to redshift unity
\citep[e.g.][]{Vik2009b,Allen2008a} and is reflected in the armada of ongoing or planned galaxy cluster surveys in
various wavelength regimes.

Progresses in observational capabilities and search techniques over the past five years have led to an increase in the
number of %known 
spectroscopically confirmed $z\!\ga\!1$ clusters from a handful of objects to now a few dozen systems. On
the X-ray side, new identifications of X-ray luminous clusters are currently largely driven by XMM-{\it Newton} serendipitous
surveys \citep[e.g.][]{Mullis2005a,Stanford2006a,RF2008b,Santos2009a,Schwope2010a} and smaller dedicated 
surveys of a few square
degrees \citep[e.g.][]{Andreon2005a,Bremer2006a,Pierre2006a,Alexis2006a,Alexis2010a,Tanaka2010a}. Optical search methods based on
overdensities of galaxies at similar color \citep[e.g.][]{Gladders2005a} have recently been successfully extended to
the near-infrared (NIR) regime  \citep[e.g.][]{vanBreukelen2006a}, and the mid-infrared (MIR) with {\em Spitzer}\
\citep[e.g.][]{Stanford2005a,Eisenhardt2008a,Wilson2009a,Muzzin2009a,Demarco2010a}. The first detections of galaxy clusters based on
the Sunyaev-Zeldovich effect (SZE) \citep{StanisSPT2009a,Vanderlinde2010a} have opened another promising window for the
selection and study of massive distant systems.

% Search Strategies
%Xray Recently identified high-z clusters X-ray: Mullis, Jee, Stanford \citep{Mullis2005a}
% Optical with SPITZER SpARCS: 3.6mu + deep z: 42deg2 \citep{Wilson2009a,Muzzin2009a,Eisenhardt2008a}

%Recent progress in Sunyaev-Zeldovich effect (SZE) surveys has led to the discovery of the first SZE-selected galaxy
%clusters by the South Pole Telescope \citep{StanisSPT2009a} up to $z\,\sim\,0.9$. At redshifts of $z\,\sim\,1$ and
%beyond SZE surveys have the potential to identify more hot, massive systems in the near future, with similar
%characteristics than the cluster presented here.
%Galaxy Evolution
%Large-scale structure studies: RF 08, Tanaka, Gal 09

%Recent optical studies of $z\,\sim\,1$ clusters revealed that the red-sequence, i.e. the well-defined locus of
%early-type galaxies, at these  %citep{mei,...}
%truncation of red-sequence %citep{DeLucia,Tanaka}
%Large-scale structure %citep{Tanaka, ..., Fassbender08}

A crucial pre-requisite for linking galaxy cluster survey results to cosmological model predictions is reliable and
well quantified mass-observable relations. In this respect, weak gravitational lensing techniques have emerged as 
%one of the most 
a promising tool to achieve a robust calibration of various scaling relations. In the low redshift Universe,
weak lensing (WL) based mass measurements applied to sizable cluster samples have established mass-observable relations
with significantly improved accuracy \citep[e.g.][]{Reyes2008a,Zhang2008a,Marrone2009a}. First weak lensing constraints
on the evolution of scaling relations for X-ray groups are now available \citep[e.g.][]{Leauthaud2009a}, and
significant WL signals have been reported for individual massive clusters out to $z \sim 1.4$
\citep[e.g.][]{Jee2006a,Jee2009a}. However, all mass estimation methods are potentially influenced by different inherent biases \citep[e.g.][]{Meneghetti2010a} emphasizing the importance of an inter comparison of different techniques, in particular for high redshift systems.

Significant progress is yet to be made to establish robust mass-observable relations at $z\sim 1$ and beyond
and to compare vices and virtues of mass proxies derived from optical/IR, X-ray, WL, and SZE observations at these
redshifts. The current limitation is the still persisting small number of known galaxy clusters at $z\!>\!0.9$ that are
sufficiently massive to enable the detection and  cross-comparison of all observational techniques on an individual
cluster basis. Using concordance model parameters, we can estimate the surface density of suitable test objects at $z
>0.9$ with $M_{\mathrm{200}}\!\ga\!4\!\times\!10^{14}\,\mathrm{M_{\sun}}$ to be about one object in a solid angle of 100\,deg$^{2}$
\citep[see e.g.][]{Jee2009a}. %\new{Hans: confirm with LCDM expectations}

%Detailed ICM studies %citep{Rosati,...}
%Cosmological results  %citep{Viklinin,Allen, Ettori}
%Cluster cosmology: Viklinin 09, Allen , Ettori

In this paper we present first details of the newly identified cluster of galaxies XMMU\,J1230.3+1339 at redshift
$z\!=\!0.975$. We perform a %detailed 
cluster characterization based on a joint analysis in X-rays,
optical imaging and spectroscopy, %Sunyaev-Zedovich effect observations, 
and weak lensing. The three main science
objectives of this work 
%first paper on the cluster
are (i) a  multi-wavelength characterization of the physical system parameters, (ii)
the derivation and cross-comparison of a dozen different total mass proxies, and (iii) the identification of substructure-
and large-scale structure components for an assessment of the dynamical state of the system.
An accompanying paper by \citet[Paper\,II hereafter]{Lerchster2010a} provides a more detailed view on  the cluster's galaxy population and the performed weak lensing analysis.

The cluster XMMU\,J1230.3+1339 occupies a special noteworthy location in the sky, as it is located behind our closest
neighbor, the Virgo system. The cluster center is in the direct proximity of the Virgo member NGC4477, about 1.3\,deg
North of M87, corresponding to a projected distance of approximately 400\,kpc, or half of $R_{500}$, at the
Virgo redshift. % of 0.004.

%Special location behind Virgo
%1.3 degree North of M87, corresponding to a projected distance at the Virgo redshift of 390\,kpc.

This paper is organized as follows: Sect.\,\ref{s2_obs} %describes the observations and the data reduction;
introduces the observations, data reduction, and first basic cluster properties;
Sect.\,\ref{s3_results} contains the %cluster properties 
derived results and mass proxies; %from the individual wavelength regimes; 
Sect.\,\ref{s4_multilambda_view} focusses on a global pan-chromatic view and the identification of cluster
associated sub-components; Sect.\,\ref{s4_disc} discusses the dynamical state of the cluster and the comparison of mass
proxies; we conclude in Sect.\,\ref{s5_concl}. Throughout the paper we assume a $\Lambda$CDM cosmology with
$\Omega_\mathrm{m}\!=\!0.3$, $\Omega_{\Lambda}\!=\!0.7$, and $h\!=H_0/(100\,\mathrm{km\,s}^{-1}\,\mathrm{Mpc^{-1}})\!=\!0.7$. Unless otherwise noted, the notation with
subscript $X_{500}$ ($X_{200}$) %denotes
refers to a physical quantity $X$ measured inside a radius, for which the mean total
density of the cluster is 500 (200) times the critical energy density of the Universe $\rho_{\mathrm{cr}}(z)$ at the
given redshift $z$. All reported magnitudes are given in the AB system. At redshift $z\!=\!0.975$ the lookback time is
7.61\,Gyr, one arcsecond angular distance corresponds to a projected physical %comoving
scale of 7.96\,kpc, and the cosmic evolution factor has a value of $E(z)= H(z)/H_0 = [\Omega_{m}(1+z)^3 +
\Omega_{\Lambda}]^{1/2}=1.735$.

% A) Optical Appearance
% full width plot
\begin{figure*}[t]
\sidecaption
    \centering
    \includegraphics[width=14.5cm, clip]{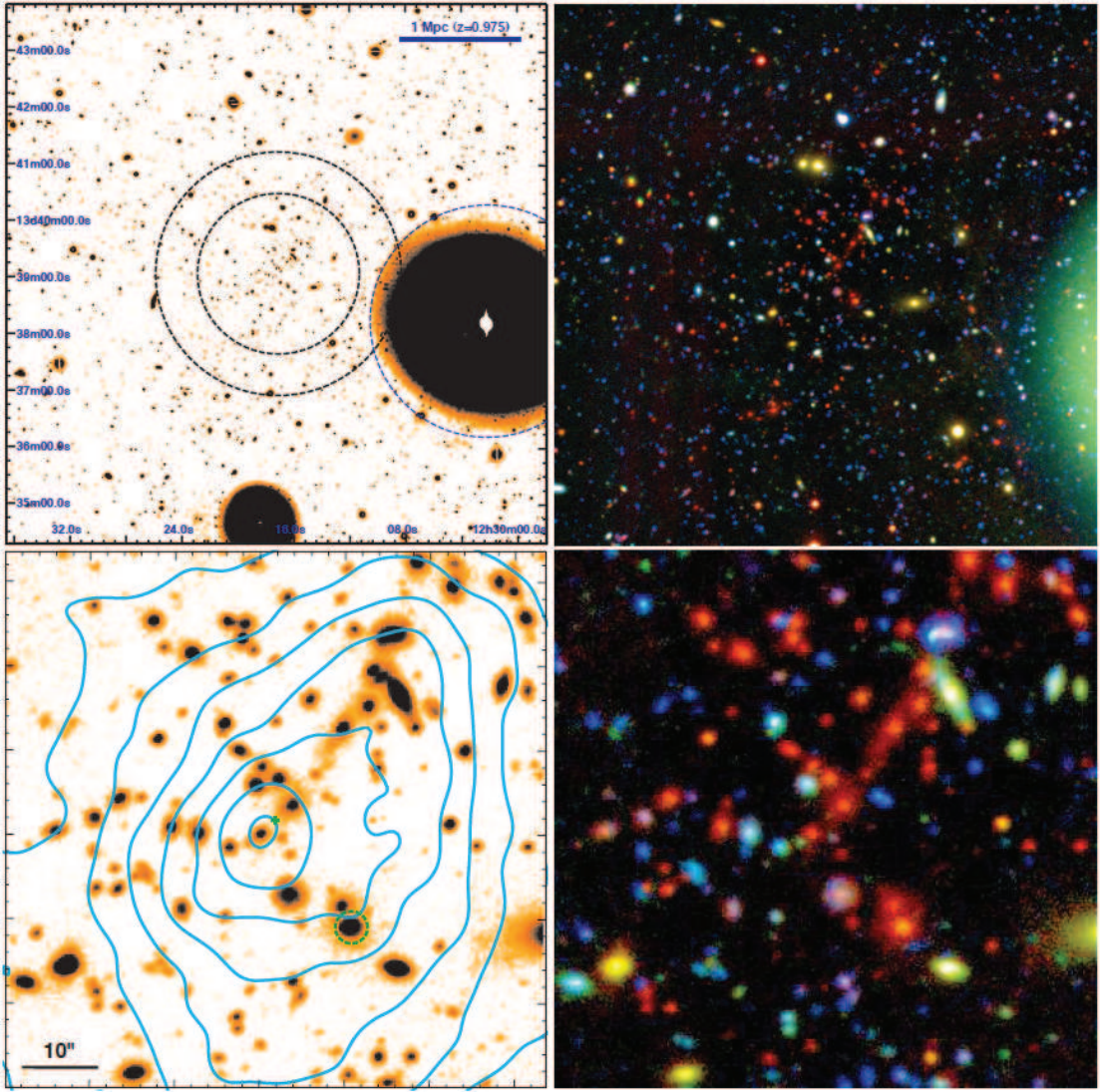}
    %\includegraphics[width=12.0cm, clip]{CombiFig1.ps}
    %\includegraphics[width=14.5cm, clip]{Fig1_all.eps}
    %%\begin{minipage}[b]{14.5cm}
    %%\includegraphics[width=7.25cm, clip]{15204fg1a.ps}
    %%\includegraphics[width=7.25cm, clip]{15204fg1b.eps}
    %%\includegraphics[width=7.25cm, clip]{15204fg1c.ps}
    %%\includegraphics[width=7.25cm, clip]{15204fg1d.eps}
    %%\end{minipage}
      \caption{Optical appearance of the galaxy cluster XMMU\,J1230.3+1339 at $z\!=\!0.975$. {\em Top left:}
      Co-added 9.5\arcmin\,$\times$\,9.5\arcmin \ i\arcmin z\arcmin \, image showing the cluster environment. The %central large
      dashed circles represent the projected cluster radii R$_{200}$ (outer) and R$_{500}$ (inner),
      the physical angular scale at the cluster redshift is given in the upper right corner, and the Virgo galaxy NGC\,4477
      is marked %on the right (blue dashed circle). 
      by the dashed circle on the  right.
      {\em Top right:} RGB color composite %centered on the cluster 
      based
      on the co-added LBT images in i\arcmin z\arcmin \, (red channel), Vr\arcmin \, (green), and UB (blue). The side length of the image corresponds to
      a physical scale of 2\,Mpc\,$\times$\,2\,Mpc (250\arcsec\,$\times$\,250\arcsec), displaying approximately the region inside 
      %the expected %virial
      $R_{200}$.
      % radius of the cluster. 
      The large number of red cluster galaxies are easily visible.
      {\em Bottom left:} Core region of the cluster in i\arcmin z\arcmin \,
      with {\it Chandra} X-ray contours overlaid in cyan and a side length of 560\,kpc (70\arcsec);
      the green dashed circle marks the BCG to the South-West of the nominal X-ray center (green cross).
      %, corresponding to a comoving scale of 560\,kpc. 
      %the  BCG (green dashed circle) is located at a projected distance of 140\,kpc to the South-West of the
      %nominal X-ray center (green cross). 
      {\em Bottom right:} Same sky region as on the left as color composite. }
         \label{fig_OpticalView} %\vfill
\end{figure*}

%\clearpage

% ----------------------------------------- Start Observation Table -----------------------------------------------
\begin{table*}[t]    % h = here ; positioning
\caption{Observation log of the X-ray, optical imaging, and spectroscopic data coverage of XMMU\,J1230.3+1339. } \label{table_obs}

\centering
\begin{tabular}{ c c c c c c }
\hline \hline

Observatory & Instrument & Data Type & Date & Exposure Time & Observation ID  \\
% &  &  & & ksec  &   \\
%& $z$  & $f_{\mathrm{X}}$(0.5--2.0\,keV) & $r_{\mathrm{c}}$  &  $L_{\mathrm{X}}$(0.5--2.0\,keV)  \\ %& $M_{500}$ \\
% & &  Eq. 2000 &  Eq. 2000  &    & erg\,s$^{-1}$cm$^{-2}$ & \arcsec &  erg\,s$^{-1}$  \\ %& $\mathrm{\mathrm{M_{\sun}}}$ \\

\hline

% Xray --> ok
XMM-{\it Newton}  &  EPIC & X-ray  & Jun 2002 & 14\,ksec / 3.9\,h & 0112552101 \\
XMM-{\it Newton}  &  EPIC & X-ray  & Jul 2002 & 13\,ksec  / 3.6\,h& 0106060401 \\
{\it Chandra}  &  ACIS-S & X-ray   & Apr 2008 & 38\,ksec  / 10.6\,h&  9527 \\

\hline
% Optical --> ok
VLT & FORS\,2 & R\,z imaging & 11 Mar 2007 & 1.4\,ksec  / 0.4\,h& 078.A-0265 \\
LBT & LBC & U\,B\,V\,r\arcmin \,i\arcmin \,z\arcmin \, imaging & 28 Feb - 2 Mar 2009 & 37\,ksec / 10.3\,h  &  \\

\hline
VLT & FORS\,2 & MXU spectroscopy & 27 Apr \& 6-7 Jun 2008 & 2.2\,ksec  / 3.9\,h  & 081.A-0312 \\

%\hline
%% SZE --> ok
%APEX & APEX-SZ & mm & 12/15/17 Apr 2009 & 45\,ksec & \\
%% was 27 ksec

\hline
\end{tabular}
\end{table*}
% ----------------------------------------- End Table -----------------------------------------------

%============== SECT 2 ===============================================================================================

%__________________________________________________________________
\section{Observations, data reduction, and basic cluster properties}
\label{s2_obs}

% General Strategy
The cluster of galaxies XMMU\,J1230.3+1339 (see Fig.\,\ref{fig_OpticalView}) was discovered as part of the XMM-{\it Newton}
Distant Cluster Project (XDCP), a serendipitous archival X-ray survey focussed on the identification of very distant
X-ray luminous systems \citep{Mullis2005a,HxB2005a,RF2007Phd}. The first step of the survey strategy is based on the
initial detection of serendipitous extended X-ray sources  as cluster candidates in high galactic latitude XMM-{\it Newton}
archival fields. Extended X-ray sources lacking an optical cluster counterpart in Digitized Sky Survey (DSS) images are
followed-up as distant candidates by optical two-band imaging with the objective to confirm a galaxy overdensity and
obtain a redshift estimate based on the color of early-type cluster galaxies. Subsequent spectroscopic observations of
photometrically identified systems provide the final secure cluster identification as  %three dimensionally
bound objects and accurate redshifts for the  high-z systems. The presented XMM-{\it Newton} and the VLT/FORS\,2 imaging and
spectroscopy data are part of this XDCP survey data set.

Additional follow-up data for a more detailed characterization of the cluster were obtained with the Large Binocular
Telescope (LBT) for deep multi-band imaging . %and with APEX-SZ for the detection of the Sunyaev-Zeldovich effect.
Moreover, high-resolution X-ray data of a field serendipitously  covering the cluster has recently become available in
the {\it Chandra} archive. An overview of the full multi-wavelength data set used for this work is given in
Table\,\ref{s2_obs}. We first present the optical imaging and spectroscopy data and then discuss the analysis of the archival X-ray data.

%We will first present the optical data, then discuss the analysis of the archival X-ray data.
%and finally describe the SZE observations of the cluster.

%................................................
\subsection{Optical follow-up observations}
\label{s2_opt_data}

% Optical Properties

%\begin{figure*} %[t]
%\sidecaption
%   \centering
%   \includegraphics[width=6.0cm]{CMD_Hist.eps}
%   \includegraphics[width=6.0cm]{long_157com_SpectraGrid.ps}
%      \caption{Optical properties of the galaxy cluster XMMU\,J1230.3+1339 {\em Top left:} R$-$z color-magnitude
% diagram of the %cluster environment. {\em Bottom:} Spectra }
%         \label{fig_Optical}
%\end{figure*}

%~~~~~~~~~~~~~~~~~~~~~~~~~~~~~~~~~~~~~~~~~
\subsubsection{VLT Imaging}
\label{s2_VLT_FORS2}

Short-exposure images in the R$_{\mathrm{special}}$ (960\,s, hereafter R) and z$_{\mathrm{Gunn}}$ (480\,s, hereafter z)
filters were acquired in March % on 11 March
2007 with VLT/FORS\,2 in photometric, sub-arcsecond seeing conditions in order to identify the optical counterpart
associated with the extended X-ray source XMMU\,J1230.3+1339. The FORS\,2 imaging data with a field-of-view of
6.8\arcmin$\times$6.8\arcmin \ were reduced in a standard manner following the procedure described in
\citet{Erben2005a}. The 4 (3) 
bias subtracted and flat-fielded frames in %each 
the z (R) band were registered to a common
coordinate system and co-added using {\em SWarp}  and {\em ScAMP} \citep{Bertin2006a}. 
The final stacked images have a measured on-frame
seeing of 0.52\arcsec \ in the z-band and 0.62\arcsec \, in R.

Photometric zero-points were derived from dedicated standard star observations in the R-band and stellar SDSS
photometry in the science field for the z filter. 
%The calibration of the non-standard ESO z-band cut-on filter to the
%SDSS photometric system requires color transformation equations which limits the photometric precision to an
%uncertainty of about 0.03\,mag. 
Photometric catalogs were extracted from the PSF matched images using {\em SExtractor}
\citep{Bertin1996a} in dual image mode with the unconvolved z-band frame as detection image. The 50\% completeness
limits\footnote{The 50\% completeness %limit 
corresponds approximately to a $5\,\sigma$  detection limit (2\arcsec \,apertures).}
%, the 100\% completeness limits are about 0.7\,magnitudes lower.} 
for the two bands were estimated to be
z$_{\mathrm{lim}}(50\%)\simeq$24.7\,mag and R$_{\mathrm{lim}}(50\%)\simeq$25.4\,mag.
%, the 100\% completeness limits are about 0.7\,magnitudes lower.

% The source detection and photometry was performed with {\em SExtractor} \citep{Bertin1996a} in dual imaging mode
%using the z-band as master-image.

\begin{figure}[t]
   \centering
   %\resizebox{\hsize}{!}{\includegraphics[width=\textwidth]{X1230flux.ps}}
   %\resizebox{\hsize}{!}{\includegraphics[width=\textwidth]{X1230flux.ps}}
   %\includegraphics[width=9.0cm, clip]{v6_CMD_Rz_157com.ps}
   %\includegraphics[width=9.0cm, clip]{v4_pub_CMD_Rz_157com.ps}
   %\includegraphics[width=9.0cm, clip]{v3_CMD_Hist_157com.ps}
   %\includegraphics[width=9.0cm, clip]{v6_pub_CMD_Rz_157com.ps}
   \includegraphics[width=9.0cm, clip]{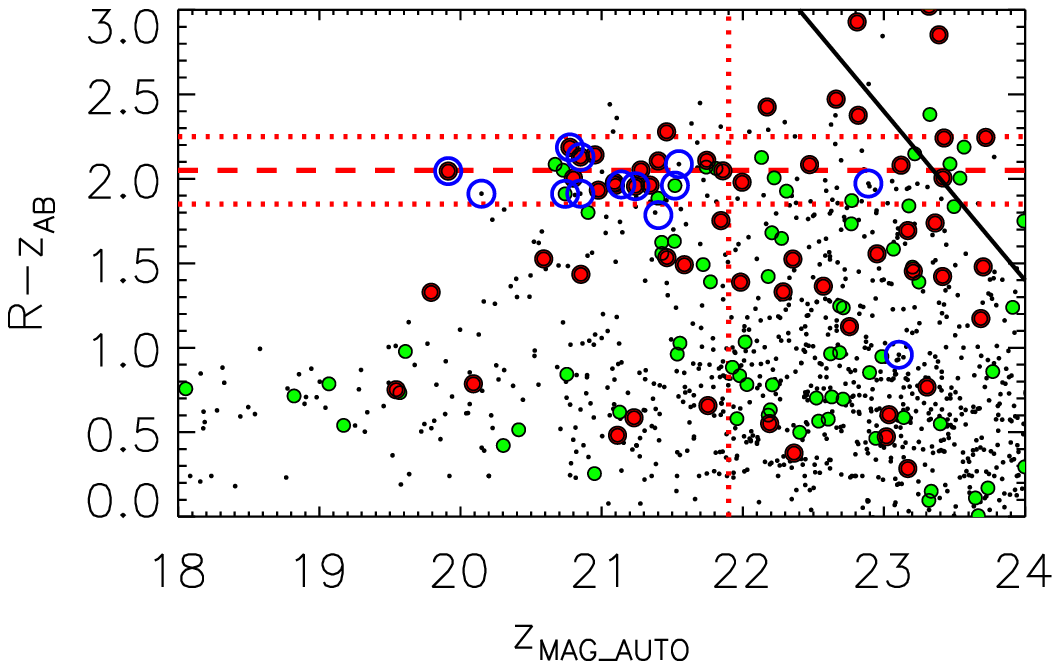}
   \includegraphics[width=9.0cm, clip]{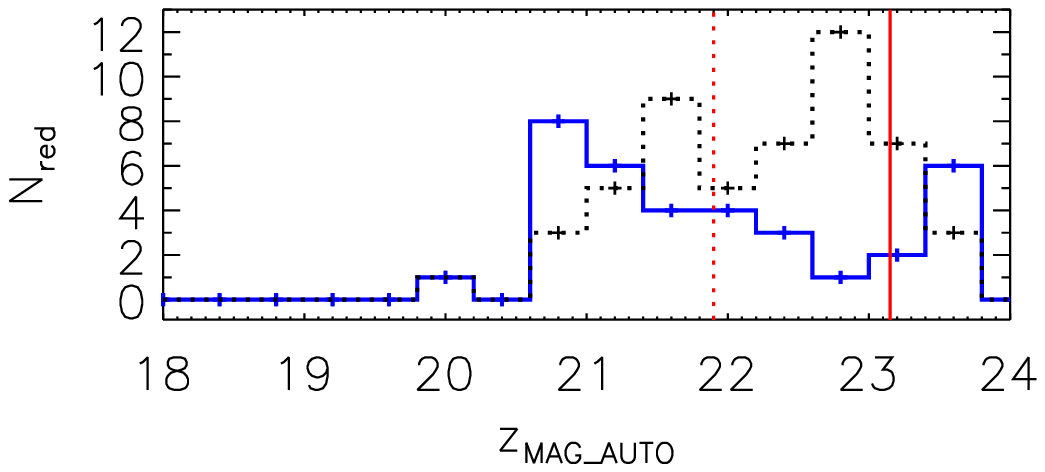}
      \caption{{\em Top:} R$-$z color-magnitude diagram of the cluster environment based on the FORS\,2 imaging data.
      The richly populated cluster red-sequence around the median ridgeline color R$-$z$\sim$2.05 (red dashed line) is clearly visible.
      Filled red circles indicate objects within a 30\arcsec \, radius from
      the X-ray center, green circles correspond %to objects %have cluster-centric distances of 
      to 0.5--1\arcmin, small black dots represent all other
      objects in the field, and open circles mark spectroscopically confirmed cluster members.
Horizontal dotted lines define the applied color cut interval of $\pm$0.2 mag about the median color, while the
dotted vertical line refers to the expected apparent magnitude (m*$\simeq$21.9\,mag) of a passively evolving L* galaxy at the
cluster redshift (in both panels). The 50\% completeness limit is indicated by the black solid line. {\em Bottom:}
Histogram of galaxy counts in the red color interval 1.85$\le$R$-$z$\le$2.25 in bins of 0.4\,mag for galaxies within
$1\arcmin$ \ from the cluster center (blue line).
%The significant drop of galaxies at z$>$21.5\,mag suggests a truncation of the the red-sequence.
The vertical solid line indicates the nominal 50\% completeness limit of the data, while the black dotted line shows the magnitude
distribution for galaxies at larger distances for comparison.}
         \label{fig_CMD}
\end{figure}

For this work we limit the quantitative galaxy color assessment in the cluster environment to this initial FORS\,2 R-
and z-band discovery data set. We constructed the color-magnitude-diagram (CMD) for the observed field
 using the total ({\tt MAG\_AUTO}) z-band magnitude and fixed 2.2\arcsec \,(3.5$\times$\,seeing) %FWHM)
aperture magnitudes for the R$-$z color, both corrected for effects of galactic extinction. The resulting CMD is
displayed in Fig.\,\ref{fig_CMD} after the removal of stellar sources ({\tt CL\_STAR}$\ge$0.95) and objects with saturated
cores. The richly populated cluster red-sequence, i.e. the locus of early-type cluster galaxies in the CMD, clearly stands out
from the background  at a color of R$-$z$\sim$2 and provided an early indication that the galaxy cluster is indeed at
$z\!>\!0.9$ prompting the subsequent spectroscopic follow-up.

We have not attempted to fit a non-zero slope to the cluster red-sequence, since (i) the slope is expected to be small,
(ii) the moderately deep ground-based data does not allow precision photometry down to faint magnitudes, and (iii) the
lack of secure cluster membership and SED-type information along the ridgeline would bias the result due to possible
inclusions of interlopers and non-passive galaxies. For a robust color selection of predominantly cluster galaxies with
a high contrast with respect to foreground and background objects we apply a simple color-cut of $\pm 0.2$\,mag (dotted
lines in Fig.\,\ref{fig_CMD}) around the median bright-end color of R$-$z$\sim$2.05 (red dashed line). This color cut
encompasses 11 out of 13 spectroscopically confirmed cluster members and should hence be a good approximation for the
underlying physical red-sequence of the early-type cluster members, which is expected to be slightly bluer at the faint
end driven mainly by a decreasing average metallicity with increasing apparent magnitude \citep[e.g.][]{Kodama1997a}.

The lower panel of Fig.\,\ref{fig_CMD} shows the histogram of the number of objects along this color-selected strip
1.85$\le$R$-$z$\le$2.25\,mag in bins of 0.4\,mag for galaxies within 1\arcmin \ of the X-ray center (blue solid line)
and galaxies further away (black dotted line), with %z$^{\mathrm{red}}_{\mathrm{lim}}$
the nominal 50\,\% completeness limit at this color indicated by  the red vertical line. The apparent decline of the
number of red ridgeline galaxies towards fainter magnitudes seen in the blue histogram can be attributed to several
possible factors, or combinations of these: (i) the increasing uncertainty for fainter galaxies in the R$-$z color
determination scatters a fraction of red cluster galaxies outside the color cut, (ii) fainter ridgeline galaxies are
preferentially located beyond a cluster-centric distance of 1\arcmin \, (dotted histogram), or (iii) the cluster
red-sequence exhibits a physical deficit of faint galaxies as has been reported for several high-z clusters
\citep[e.g.][]{DeLucia2007b, Tanaka2007a}.
%While the decreasing photometric accuracy at fainter magnitudes (i)
%certainly accounts partially for the observed decline, Fig.\,\ref{fig_RadialDistrib_RSgals} shows indications that also
%option (ii) contributes to the observed effect.

\begin{figure}[th]
   \centering
   \includegraphics[width=9.0cm, clip=true]{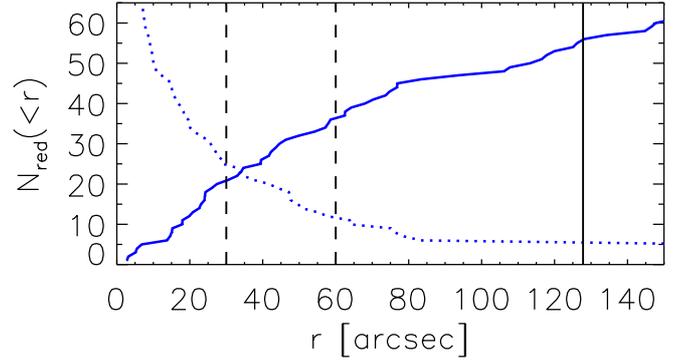}
      \caption{Cumulative radial distribution of color selected  red galaxies (as in Fig.\,\ref{fig_CMD}) as a function of cluster-centric distance
      (blue solid line). Dashed vertical lines indicate the radii used for color coded symbols in Fig.\,\ref{fig_CMD}, the solid line marks the fiducial cluster radius.
      %The flattening of the galaxy number counts beyond 80\arcsec \ is partially due to the masked out area by NGC4477.
      The dotted blue line shows the cumulative
      radial red galaxy number counts normalized to the enclosed area, i.e. the average red galaxy density per square arcminute within the cluster-centric distance r.}
         \label{fig_RadialDistrib_RSgals}
\end{figure}

%The blue solid line in Fig.
Figure\,\ref{fig_RadialDistrib_RSgals} displays the cumulative radial distribution of red galaxies
with the same  color cuts as a function of cluster-centric distance (blue solid line). 
%Whereas the contrast of red galaxies, shown by the
%dotted line as the average surface density within the enclosed area, 
About 20 red galaxies are found beyond a cluster-centric distance of 1\arcmin \, but still within the fiducial cluster radius (solid vertical line, see
Sect.\,\ref{s3_Tx_R200}). 
These galaxies can mostly be attributed to infalling groups, as discussed in
Sect.\,\ref{s3_InfallingGroups}, and could contribute to the faint end of the red population. 
The blue dotted line shows the average surface density of red galaxies within the cluster-centric distance $r$, which is directly related to the galaxy contrast in the enclosed area with respect to the background. This contrast can be optimized for the initial confirmation of the cluster and the identification of its red-sequence  by restricting the galaxy selection to the inner 30\arcsec or 1\arcmin \ (vertical dashed lines),
as was used for the CMD of Fig.\,\ref{fig_CMD}.

For this work, only the particularly richly populated bright end of the red ridgeline  in XMMU\,J1230.3+1339 is highlighted \citep[for
comparison see e.g.][]{Mei2009a}. A more quantitative analysis of the physical cluster red-sequence (and a possible truncation thereof) based in the deeper LBT imaging data is discussed in  Paper\,II. %\citet{Lerchster2010a}.
\subsubsection{LBT Imaging}
\label{s2_LBTimaging}
%Goals:

In order to obtain a more detailed view of the cluster, we initiated deep wide-field imaging observations in six-bands
U\,B\,V\,r\arcmin \,i\arcmin \,z\arcmin\footnote{For our practical purposes, the difference between the photometric reference systems z
and z\arcmin \, is negligible ($\la0.02$\,mag). The derived LBT z\arcmin \, and FORS\,2  z-band magnitudes have been confirmed to be
consistent within the photometric errors. However, for easier distinction we refer to  z-band magnitudes for the
FORS\,2 results and   z\arcmin \, magnitudes for the deeper LBT data.} with the Large Binocular Cameras (LBC) working in
parallel at the 2$\times$8.4\,m Large Binocular Telescope. The data were acquired between 28 February and 2 March 2009
under good, mostly sub-arcsecond, seeing conditions. The net exposure times (per single 8.4\,m telescope) over the
field-of-view of 26\arcmin$\times$26\arcmin \ were 9\,ksec in the U and z'-bands, 5.9\,ksec in B and i\arcmin, 3.6\,ksec in
V, and 3.2\,ksec in the r\arcmin-band.

The LBT/LBC imaging data were reduced in standard manner similar to the treatment in the previous section. For more
information on LBC specific reduction procedures in general we refer to \citet{Rovilos2009a}, and for details on the
photometric  
calibration and object catalogues based on this data set to Paper\,II. % \citet{Lerchster2010a}. % Lerchster et al. (in prep.)
The reduced final co-added image stacks have measured seeing full-width half-maxima (FWHM) of 0.8\arcsec \ in the %red
z\arcmin -band and about 0.9\arcsec \ in all other bands, resulting in  limiting
magnitudes (5$\sigma$ detection in 1\arcsec\,apertures, see Table\,1 of Paper\,II) %(50\% completeness) 
of the co-added frames of: U$_{\mathrm{lim}}\simeq$27.0, B$_{\mathrm{lim}}\simeq$26.9,
V$_{\mathrm{lim}}\simeq$26.3, r\arcmin$_{\mathrm{lim}}\simeq$26.2, i\arcmin$_{\mathrm{lim}}\simeq$25.8, and
z\arcmin$_{\mathrm{lim}}\simeq$25.4. 
% SNR of 5 with 1arcsec aperture

This rich imaging data set will be fully exploited %and described in more detail 
in forthcoming papers (e.g. Paper\,II).
%(Lechster et al.,in prep.; Fassbender et al., in prep.)
The current work uses the deep wide-field LBT imaging data for two main purposes: (i) a quantitative photometric
analysis in the z\arcmin -band (Sect.\,\ref{s3_optical_prop}), which is %about one magnitude 
deeper than the FORS\,2 data and
allows accurate background determinations in independent external regions; and (ii)  a deep optical view in
single-band and color composites as shown in Figs.\ref{fig_OpticalView}, \ref{fig_DynamicalState},
\ref{fig_InfallingGroups}, and \ref{fig_CoreRegion}. In order to maximize the depth of the available data for
visualization purposes, we co-added the i\arcmin \, and z\arcmin -band using an inverse variance weighting scheme to a resulting
combined i\arcmin z\arcmin -frame. This combined band is used as single-band background image with a high contrast of cluster
galaxies and as input for the red-channel of RGB color composite images. Accordingly, a co-added Vr\arcmin-frame was obtained
as input for the green-channel, and a UB-image for the blue-channel.

%and to determine global optical properties for which sufficiently large background regions are required (e.g. richness
%and total light L$_{200}$)
%we use the deep wide-field data mainly for the optical view on the galaxy cluster population (see
%Figs.\ref{fig_OpticalView}, \ref{fig_DynamicalState}, \ref{fig_CoreRegion}).

\begin{figure}[t]
   \centering
   \includegraphics[width=8.8cm, clip=true]{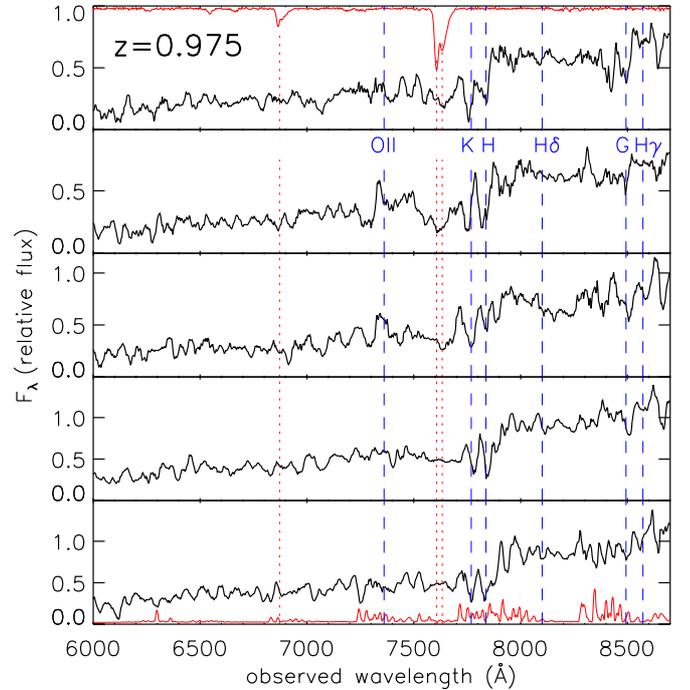}
      \caption{Sample spectra of secure cluster members with the median redshift of $z\!=\!0.975$, smoothed with a 9\,pixel boxcar
      filter. The expected observed positions of prominent spectral features at the median redshift are indicated by blue dashed lines.
      The sky spectrum (bottom) and telluric features (top) are overplotted in red. From top to bottom the spectra correspond to the
      galaxies with object IDs 1-9-4-6-8 in Table\,\ref{tab_specmembers}.}
         \label{fig_Spectra}
\end{figure}

% velocity distribution
\begin{figure}[h]
   \centering
   \includegraphics[width=8.8cm, clip=true]{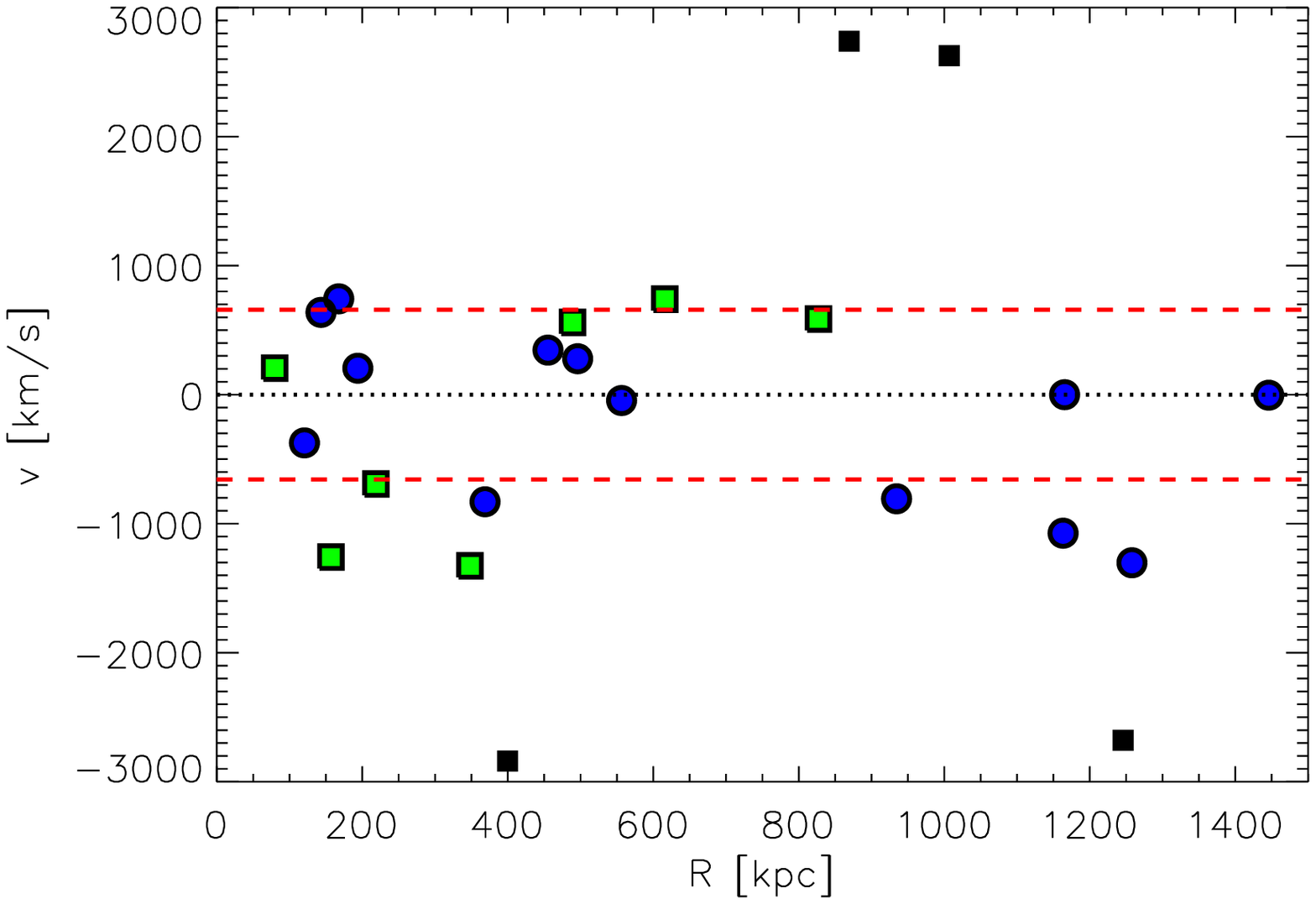}
      \caption{Cluster restframe velocity distribution of spectroscopic cluster members as a function of projected distance
      from the X-ray center.
      Member galaxies of Table\,\ref{tab_specmembers} with secure redshifts are indicated by blue circles, dashed lines
      correspond to the
      formal velocity dispersion estimate.
      % based on the biweight method (blue) and the standard deviation (red).
      %The average errors in the velocity measurement of individual galaxies is
      %about $\sigma_{v}\!\sim\!50$\,km/s.
      For completeness, galaxies with tentative redshifts are also shown as squares, smaller black symbols
      indicate velocities outside the formal 3\,$\sigma$ interval.
      }
         \label{fig_veldistr}
\end{figure}

%\clearpage

%~~~~~~~~~~~~~~~~~~~~~~~~~~~~~~~~~~~~~~~~~
\subsubsection{VLT Spectroscopy}
\label{s2_VLT_Spectr}

For the final redshift determination of the cluster, we obtained spectroscopic observations with  VLT/FORS\,2 using a
single MXU-mode (Mask eXchange Unit) slit-mask centered on XMMU\,J1230.3+1339 for a total net exposure time of
7.85\,ksec (Table\,\ref{table_obs}). Targeted galaxies were selected to be preferentially red galaxies close to the
ridgeline in Fig\,\ref{fig_CMD}. The chosen instrument setup with the 300\,I grism and a slit width of 1\arcsec\, 
provides a wavelength coverage of
6\,000--10\,000\,\AA \ with a resolution of $R\!=\!660$. The data consisted of six individual exposures taken in three
different nights under variable seeing conditions ranging from 0.7\arcsec \ to 1.9\arcsec.
%(program ID: 081.A-0312(B)).

The spectra were reduced with standard techniques using {\it IRAF\/}\footnote{\url{http://iraf.noao.edu}} tasks. In
short, the 2-D spectra were bias subtracted, flatfielded to correct for pixel-to-pixel variations, and co-added to
provide a sufficient signal-to-noise ratio (SNR) for the tracing of weak spectra. 38 1-D spectra were extracted and
sky-subtracted by fitting a low order polynomial %function 
to adjacent background regions along the trace of the spectrum. The
wavelength calibration was achieved by means of a Helium-Argon reference line spectrum observed through the same MXU
mask. The quality of the obtained dispersion solutions was cross-checked by measuring the observed position of several
prominent sky lines yielding typical absolute rms calibration errors below %$\la 
1\,\AA.

For the redshift determination, the reduced and wavelength calibrated spectra were cross-correlated with a range of
galaxy template spectra %of different SED type
using the {\it IRAF\/} package {\it RVSAO\/} \citep{Kurtz1998a}. Out of 38 spectra, we could identify 13 secure cluster
members (see Table\,\ref{tab_specmembers}) with a median cluster redshift of $\bar z\,=\,0.9745 \pm 0.0020$ and a
bootstrapped error estimate. Figure\,\ref{fig_Spectra} displays five sample spectra of cluster members with the most
prominent spectral features labelled. The cross-correlation of absorption-line spectra with a sufficient
signal-to-noise ratio (SNR$\ga$5) yields typical statistical velocity errors of $\pm$50\,km/s for $c\!\cdot\!z$. Adding
the wavelength calibration uncertainty ($\la$1\,\AA \ at 8\,000\,\AA) in quadrature, we estimate the typical
uncertainty for most of the secure individual galaxy redshifts in Table\,\ref{tab_specmembers} to be
$\sigma_{z}\!\sim\!0.0002$

Due to the partially poor seeing conditions, not all redshifts of targeted galaxies could be determined with high
confidence. We have identified 11 additional tentative galaxy redshifts which are close to the median cluster redshift
and %are 
potentially indicate member galaxies.% with a quality flag of $\le$90\% confidence. 
For completeness these objects are
also displayed in Fig.\,\ref{fig_veldistr}, which shows the restframe velocity distribution as a function of
cluster-centric distance further discussed in Sect.\,\ref{s3_vel_dispersion}. % and \ref{s4_masses}.

% Table with individual redshifts:

\begin{table}[t]    % h = here ; positioning
\caption{Secure spectroscopic cluster members of XMMU\,J1230.3+1339 with total z\arcmin -band magnitudes, cluster-centric distances,
and measured redshifts. 
%Typical measurement uncertainties for the individual galaxy redshifts in good quality spectra are $\sigma_{z}\!\sim\!0.0002$. 
} \label{tab_specmembers}

\centering
\begin{tabular}{ c c c c c c }
\hline \hline

ID & RA     & DEC     & z\arcmin \, & d$_{\mathrm{cen}}$  & $z_{\mathrm{spec}}$  \\

   &  J2000 &  J2000  &  mag  &  arcsec             &                       \\

\hline

01 & 12:30:16.41 & +13:39:16.2 &  20.37 &  15.1   &  0.9720   \\
02 & 12:30:16.35 & +13:38:49.5 &  19.90 &  18.0   &  0.9787   \\
03 & 12:30:18.43 & +13:39:06.6 &  21.00 &  21.1   &  0.9794   \\
04 & 12:30:17.76 & +13:39:26.0 &  20.98 &  24.4   &  0.9758   \\
05 & 12:30:20.11 & +13:39:00.9 &  21.89 &  46.3   &  0.9690   \\
06 & 12:30:20.44 & +13:38:38.6 &  20.97 &  57.2   &  0.9767   \\
07 & 12:30:14.95 & +13:39:58.3 &  21.57 &  62.3   &  0.9763   \\
08 & 12:30:18.87 & +13:38:00.0 &  20.09 &  69.9   &  0.9742   \\
09 & 12:30:11.94 & +13:40:33.4 &  20.96 &  117.3  &  0.9692   \\
10 & 12:30:09.58 & +13:40:38.4 &  23.00 &  146.1  &  0.9674   \\
11 & 12:30:21.50 & +13:36:54.1 &  21.86 &  146.4  &  0.9745   \\
12 & 12:30:09.16 & +13:40:49.3 &  21.40 &  158.0  &  0.9658   \\
13 & 12:30:06.11 & +13:40:22.7 &  22.81 &  181.6  &  0.9745   \\

\hline
\end{tabular}
\end{table}

%................................................
\subsection{X-ray Data}
\label{s2_XrayAnalysis}

% X-ray Properties
\begin{figure}[t]
   \centering
   %\resizebox{\hsize}{!}{\includegraphics[width=\textwidth]{X1230flux.ps}}
   %\resizebox{\hsize}{!}{\includegraphics[angle=-90,width=\textwidth]{X1230spec.ps}}
   %\includegraphics[width=8.8cm]{flux_356yray_0112552101.ps}
   %\includegraphics[width=8.8cm]{v2_flux_X1230.eps}
   \includegraphics[width=8.8cm]{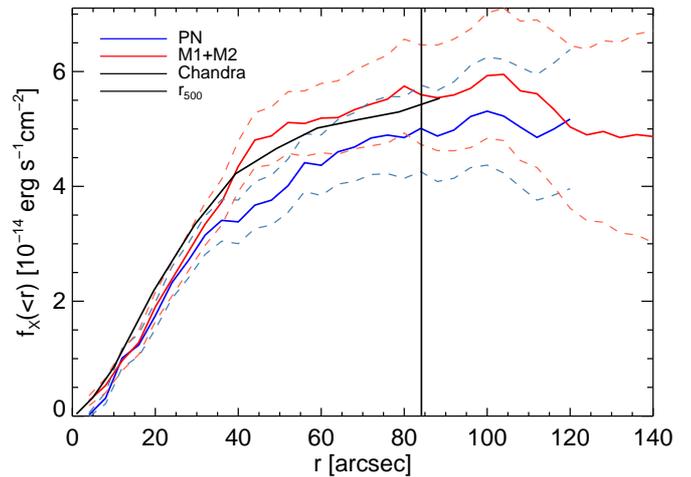}
      \caption{Growth curve analysis for the unabsorbed, background subtracted 0.5-2\,keV X-ray flux as a function of projected cluster radius. The blue solid
      line shows the total 0.5-2\,keV flux inside the projected radius r as measured with the XMM-{\it Newton} PN camera, the red line represents the
      sum of the two MOS instruments. Poisson errors plus 5\% %1\,$\sigma$
      uncertainties in the background determination are displayed by the dashed lines, the vertical line depicts the R$_{500}$ radius.
      The fully consistent {\it Chandra} GCA result is overplotted in black.  The X-ray emission
      can be traced out to about 90\arcsec, where it reaches the plateau value.}
         \label{fig_XrayFlux}
\end{figure}

\begin{figure}[t]
   \centering
\includegraphics[angle=-90,width=8.8cm, clip]{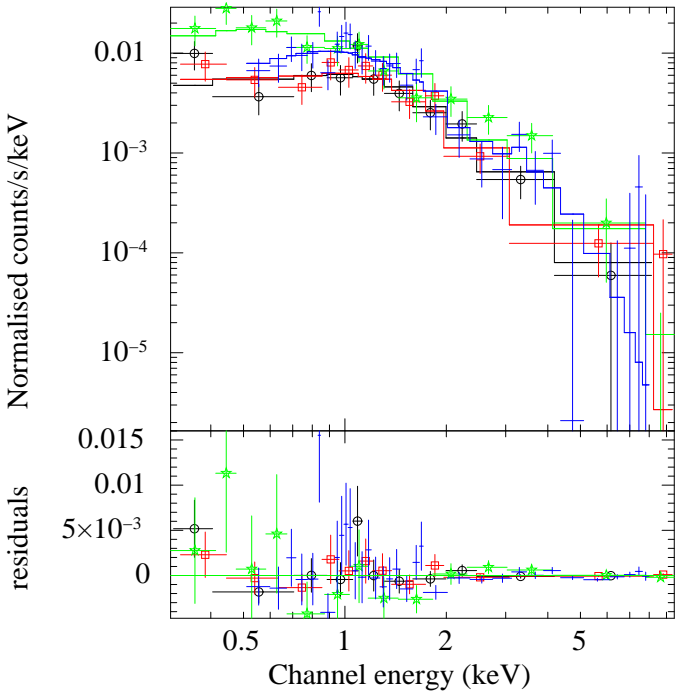}
      \caption{{\it Chandra} and XMM-{\it Newton} X-ray spectra of XMMU\,J1230.3+1339 with a joint fit ICM temperature 
      of $T_{X,2500}\!\simeq\!5.3$\,keV.
%The data were binned for display purposes only. 
The {\it Chandra} spectrum extracted from the ACIS-S3 chip is shown in blue,
and the corresponding data from the three XMM-{\it Newton} instruments in green (PN), black (MOS1), and red (MOS2).}
      %{\em Top:}  XMM-Newton spectrum extracted from an aperture with 71\arcsec \
      %for the PN (green) and two MOS cameras (black and red line). The region around 1.5\,keV was masked out to avoid the prominent
      %Al-K$\alpha$ instrumental line. Residuals with respect to the best fit temperature of 6\,keV %6.44\,keV
      %are shown in the lower part. {\em Bottom:} Corresponding Chandra spectrum extracted from the ACIS-S3 detector within an aperture
      %of 40\arcsec \ and  the a best temperature fit of 5.4\,keV (solid line). A hint of the Fe line is visible at
      %$kT \sim 3.3$\,keV.}
         \label{fig_XraySpec}
\end{figure}

%\clearpage

%\begin{figure}[t]
%   \centering
%   \includegraphics[width=8.8cm]{1230_2bfit.ps}
%      \caption{Chandra surface brightness profile. \new{ preliminary plot} }
%         \label{fig_Chandra}
%\end{figure}

%\clearpage

%~~~~~~~~~~~~~~~~~~~~~~~~~~~~~~~~~~~~~~~~~
\subsubsection{XMM-{\it Newton}}
\label{s2_XMM_Xray}

XMMU\,J1230.3+1339 was initially detected as an extended X-ray source at high significance level in the XMM-{\it Newton}
field with observation identification number (OBSID) 0112552101 at an off-axis angle of 4.3\arcmin \ from the central
target, the nearby galaxy NGC4477 in the Virgo cluster. The original XDCP source detection run, for which 470
XMM-{\it Newton} archival fields were processed, was performed with \texttt{SAS} v6.5 using the tasks \texttt{eboxdetect} and
\texttt{emldetect}. With an extent likelihood of L$_{\mathrm{ext}}\simeq$117, corresponding to a formal probability
that the source extent is due to a statistical Poissonian background fluctuation of
p$_{\mathrm{Poisson}}\!\simeq\!1.5\!\times\!10^{-51}$, the cluster is one of the most significant X-ray sources among
all distant cluster candidates in the XDCP survey. A detailed description of the source detection and screening
procedure can be found in \citet{RF2007Phd}
%(detection maximum likelihood $\simeq$268, extent likelihood $\simeq$117)

For the analysis presented here we re-processed the field with the latest version  \texttt{SAS}\,v8, and additionally
considered a second archival field with OBSID 0106060401 (Table\,\ref{table_obs}), which encloses the cluster source at
a larger off-axis angle of 9\arcmin. After applying a strict two-step flare cleaning process for the removal of high
background periods, a clean net exposure time of 13.3/8.3\,ksec remained for the EMOS/PN cameras in field 0112552101,
and 7.8/5.0\,ksec for the second observation 0106060401.

% XMM Map
In order to optimize the signal-to-noise ratio of low surface brightness regions in the cluster outskirts discussed in
Sect.\,\ref{s3_LSS}, we combined the X-ray images of both XMM-{\it Newton} observations and their corresponding exposure maps
with the \texttt{SAS} task \texttt{emosaic}. This results in more than 1\,100 source photons from XMMU\,J1230.3+1339
in the mosaic image for the used 0.35-2.4\,keV detection band, which provides an optimized SNR for distant cluster
sources \citep{Scharf2002a}. To enable the evaluation of the significance of low surface brightness features in the
presence of chip gaps, we created a signal-to-noise map of the cluster environment. 
%in the following way: (i) in the 
%combined and flatfielded photon count image, we applied
%a local SNR correction to the flux in each pixel according to the effective exposure time in this pixel ($\propto\!\sqrt{t_{\mathrm{exp}}}$), 
%%of the combined and flatfielded
%%exposure time corrected
%%photon count image, 
%(ii) the resulting map is adaptively smoothed, (iii) the background and its rms are measured in 27
%regions in the field devoid of sources, (iv) the median background level is subtracted and the remaining signal
%normalized to the rms value. 
Using this %final 
map, we created a log-spaced contour plot spanning the significance
levels 0.5$\le$SNR$\le$35 used in Fig.\,\ref{fig_DynamicalState} of Sect.\,\ref{s3_LSS}.

% Flux measurement
We applied the growth curve analysis (GCA) method of \citet{HxB2000a} to obtain an accurate 0.5-2\,keV flux measurement
of the cluster as a function of radius.%, as shown in Fig.\,\ref{fig_XrayFlux}. 
Since the effective exposure time and the
point-spread-function (PSF) at the cluster source are significantly better in field 0112552101, we focus the
quantitative analysis on this single observation and use the second field 0106060401 only for a consistency check.
Contaminating regions containing X-ray sources not associated with the cluster were conservatively masked out prior to
the GCA measurement. The X-ray emission of the neighboring S0 galaxy NGC4477 (mostly point sources)
%, dominated by a large number of individual point sources in the galaxy,
only contributes significantly at cluster-centric radii beyond
2\arcmin, i.e. the X-ray analysis %of %the extended emission
of XMMU\,J1230.3+1339 is basically not influenced by the
foreground galaxy.

%most notably the emission of the nearby galaxy NGC4477, were conservatively masked out prior to the GCA measurement.

Figure\,\ref{fig_XrayFlux} displays the GCA results for the cluster emission, which can be traced out to a radius of
about 90\arcsec, i.e. just beyond the nominal radius R$_{500}\simeq 84\arcsec$ as estimated in
Sect.\,\ref{s3_XrayResults}. Within the projected  R$_{500}$ radius we measure an unabsorbed 0.5-2\,keV flux of
f$_{X,500}=(5.14 \pm 0.54)\times 10^{-14}$\,erg\,s$^{-1}$cm$^{-2}$. For completeness we also show the fully consistent
{\it Chandra} GCA result in Fig.\,\ref{fig_XrayFlux} (black line).
%Going out to the plateau value nominal virial radius R$_{200}\simeq 135\arcsec$,
%we find the plateau value (horizontal line) of the GCA of
%f$_{X,200}=7.78 \pm 0.82$\,erg\,s\,cm$^{-2}$ \new{revise}.

%Comparison to Chandra
% no LX yet, need T for k correction

% Temperature
For the temperature determination of the cluster we followed the general procedure described e.g. in
\citet{Pratt2010a}. As for the flux determination, we focused the spectral analysis on field 0112552101, since the
effects of the broadened XMM-{\it Newton} PSF %point-spread function 
at larger off-axis angle outweigh the gain of additional
source photons.
%Since the effects of the broadened XMM-Newton point-spread function at this off-axis angle outweigh the gain of
%additional source photons, we also discarded the latter observation from the analysis.
%Data were processed with SASv 8 as described in e.g., Croston et al.(2008).
We extracted an X-ray spectrum of the source from two circular regions of radius 40\arcsec \ and 71\arcsec \ centered
on the X-ray peak. The smaller 40\arcsec \ aperture is signal-to-noise optimized in conjunction with the {\it Chandra}
spectral analysis, the larger aperture corresponds to the maximum radius for which reliable results can be derived with
XMM-{\it Newton}\footnote{The source-flux-to-total-background-flux ratio (0.5-2\,keV) is about 3:2 for the 40\arcsec \ aperture and 1:2 for the 71\arcsec \ aperture.}. The presence of extended thermal foreground emission from the Virgo cluster results in an increased
background level\footnote{The Virgo contribution to the total background is about 80\%.}, which we assume to be homogeneous over the cluster scale of interest of about 2\arcmin. To account
for this extra foreground component, we have tested several methods with both local and external background spectrum
determinations. The most robust results were obtained for the 40\arcsec \ aperture using a local background spectrum
extracted from a nearby uncontaminated region after the masking of point sources. With the metal abundance fixed to
Z=0.3\,Z$_\odot$, we derive a best fitting core region temperature for XMMU\,J1230.3+1339 of
$5.28^{+0.90}_{-0.79}$\,keV, based on a single temperature MEKAL
model, a minimally binned spectrum ($\ge$1 cts/bin), and C-statistics. 
%unbinned spectrum and C-statistics. 
The XMM-{\it Newton} spectral fit and the
residuals are displayed in Fig.\,\ref{fig_XraySpec} (green, red, and black lines).

For the larger 71\arcsec \ aperture, we measured a slightly higher X-ray temperature of $5.6^{+1.4}_{-1.1}$\,keV
applying the same local background subtraction procedure. Since this larger extraction region exhibits a lower signal-to-noise
ratio and is hence more prone to background uncertainties, we cross-checked this temperature trend with an alternative
spectral fitting method using an external background spectrum. % (see below). 
In addition to the physically motivated model of the
cosmic X-ray background (CXB) consisting of two unabsorbed MEKAL models plus a power law with fixed index $\gamma\!=\!1.4$ \citep[see][]{Lumb2002a,DeLuca2004a}, we considered  a further thermal MEKAL component %with kT${_\mathrm{Virgo}}=1.88$\,keV  
of the %to compute the 
total background spectrum in order to account for the local foreground emission of  Virgo at the position of XMMU\,J1230.3+1339.
From this four-component model we obtained a best fitting local Virgo temperature of kT${_\mathrm{Virgo}}=1.88$\,keV consistent with  previous ASCA measurements of this region \citep{Shibata2001a}.
%consisting of two unabsorbed MEKAL models plus a power law with fixed index $\gamma =
%1.4$ and the absorption fixed to the Galactic value (N$_{\mathrm{H}}\!=\!2.6\!\times\!10^{20}$\,cm$^{-2}$)  in the
%direction of the cluster.
This %best fitting 
four-component %background 
model, with a renormalization appropriate to the ratio of the surface area of the extraction regions, 
was then added %as an extra component 
to the fit of the cluster spectrum, resulting in an  %best fitting 
X-ray temperature based on this external background model  
of $6.4^{+1.7}_{-1.2}$\,keV. %, with a reduced $\chi^2=1.0$.

%We modeled the background  

%The resulting best fitting X-ray temperature
%based on this external background model is $6.4^{+1.7}_{-1.2}$\,keV, with a reduced $\chi^2=1.0$.
%The XMM-Newton
%spectral fit and the residuals are displayed in the top panel of Fig.\,\ref{fig_XraySpec}.

\begin{figure*}[t]
\begin{center}
\includegraphics[width=8.0cm,clip=true]{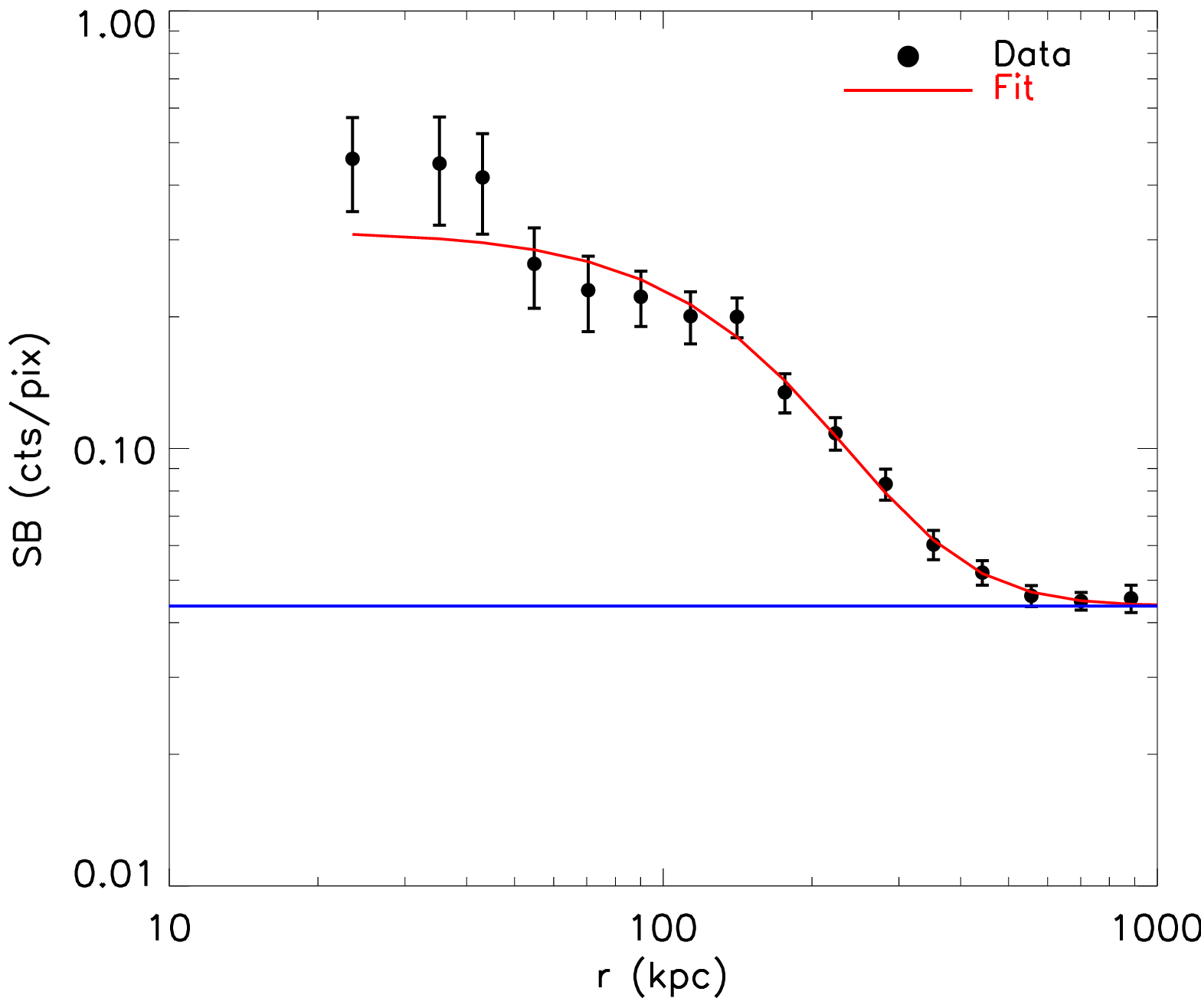}
\includegraphics[width=8.0cm,clip=true]{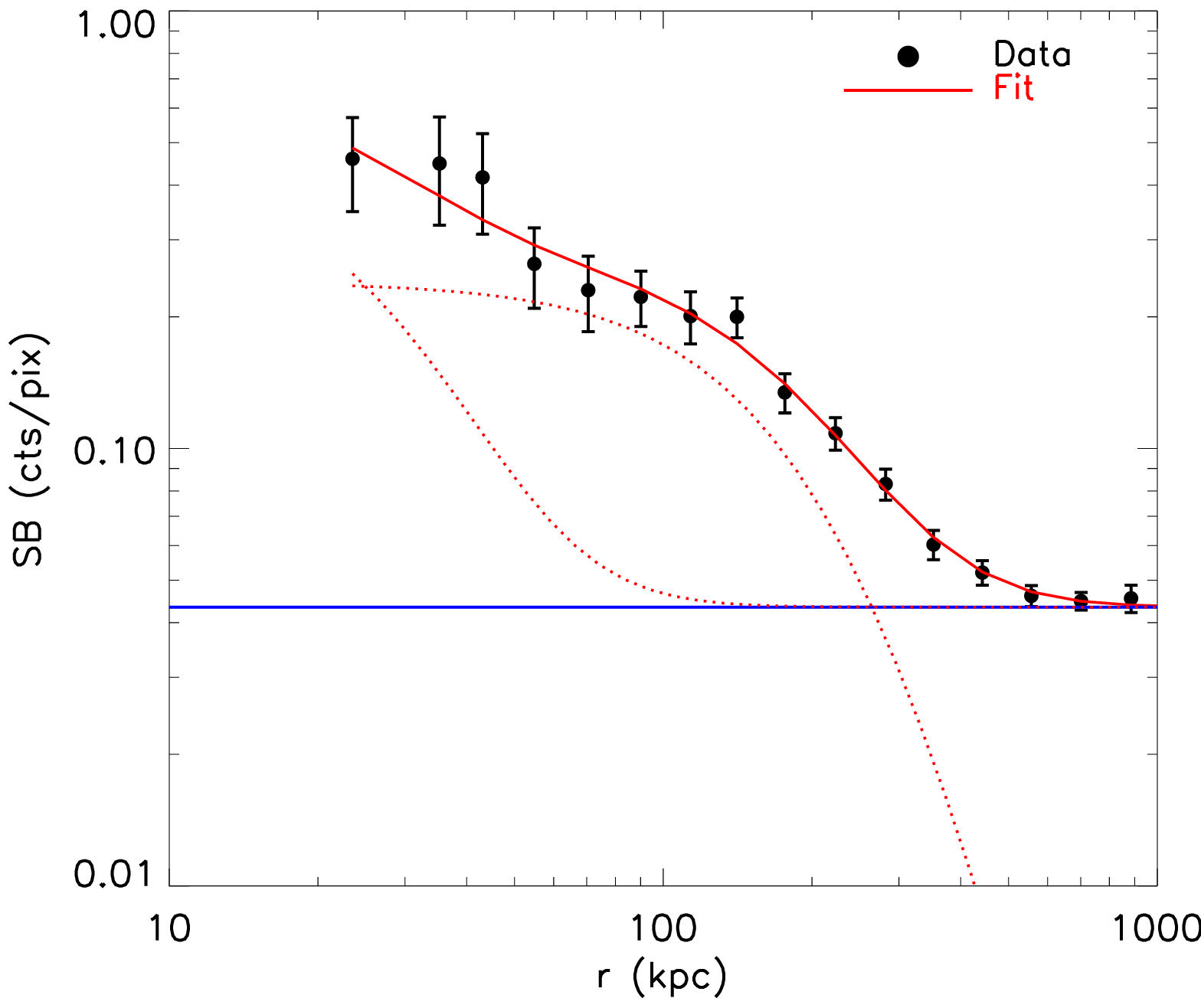}
\caption{{\it Chandra} surface brightness profile of XMMU\,J1230.3+1339 (black data points) with the best fitting single
({\em left}) and two-component ({\em right}) beta models (red solid lines). The background level is indicated by the
blue line. The dotted lines in the right panel represent the individual fit components, the inner one includes the
background and the outer one is additive. 
%\new{Plot might be better without background??}
}
 \label{fig_SBprof}
\end{center}
\end{figure*}

\subsubsection{{\it Chandra}}
\label{s2_Chandra}
%The cluster was additionally serendipitously observed in a 38\,ksec Chandra observation (ID ???) of the target NGC4477,
%which has %recently become public. ...
%SB profile

Additional X-ray coverage of XMMU\,J1230.3+1339 has recently become publicly available in the {\it Chandra} archive, where
the cluster was serendipitously observed almost on-axis in a 38.2\,ksec pointing targeted on NGC4477 (Observation ID
9527, see Tab.\,\ref{table_obs}). The field was observed in {\tt VFAINT} mode, with the cluster being located on the
central ACIS-S3 detector at an off-axis angle of approximately 2\arcmin.

%Our target was observed in one Chandra pointing of the galaxy NGC4477 (Observation ID: 9527, PI: Sarazin)
%with an exposure of 38.2 ksec observed with ACIS-S3 in the {\tt VFAINT} mode. The cluster is well within the
%chip, at an off-axis angle of approximately 2~arcmin.

We performed a standard data reduction using the {\tt CIAO v4.1} software package, with a recent version of the
Calibration Database ({\tt CALDB v4.1.1}). Since the observation was taken in  {\tt VFAINT} mode, we ran the task {\tt
acis$\_$process$\_$events} which provides an improved flagging of probable background events by distinguishing real
X-ray photon events and events most likely associated with cosmic rays. With this procedure the ACIS particle
background can be significantly reduced.
%The exposure is cleaned with the {\tt vfaint} option to reduce the background.
The data is filtered to include only the standard event grades 0, 2, 3, 4 and 6.  We confirmed that there were no
flickering pixels with more than two events contiguous in time and checked visually for hot columns remaining after the
standard cleaning. The resulting total net exposure time after this cleaning procedure is 37.68\,ksec.

We repeated the growth curve analysis for {\it Chandra}  in the soft 0.5-2.0\,keV band (Fig.\,\ref{fig_XrayFlux}), which
traces the cluster emission out to similar radii as in the single XMM-{\it Newton} field.
After subtraction of the background of $b\!=\!0.043\!\pm\!0.008$\,cts/pixel measured in an external region %we measure %571 net counts within r=80\arcsec \ and
634 net cluster counts within r=2\arcmin \ remain. The resulting unabsorbed {\it Chandra} fluxes are in full agreement with the
specified XMM-{\it Newton} values as shown by the black line in Fig.\,\ref{fig_XrayFlux}.

The main advantage of the high-resolution {\it Chandra} X-ray data is the ability to resolve small-scale ICM structures in
the inner cluster region and to obtain precise radial surface brightness (SB) profiles down to the central core. We
measured the binned azimuthally-averaged surface brightness profile %in bins of fixed radius that were
and performed a fit with the widely used approximation of the single isothermal $\beta$ model \citep{Cavaliere1976a},
tracing the ICM out to about R$_{500}$.
%(Cavaliere \& Fusco-Femiano 1978), 1976 is better
We fitted the functional form  $S(r) = S_{0} (1+(r/r_{c})^2) ^{-3\beta+0.5} + b$ for the radial SB profile $S(r)$ with
 free parameters $S_{0}$ for the central SB, $r_{c}$ for the core radius, $\beta$ for the slope, and a constant $b$ for
 the background using a Levenberg-Marquardt least-squares minimization. The radial profile and the best fitting
 single $\beta$ model are shown in the left panel of Fig.\,\ref{fig_SBprof}, with resulting parameters
 $S_{0}\!=\!0.28\pm0.04$\,cts/pix\footnote{Using 2x2 binned pixels with an effective scale of 0.984\arcsec per pixel. }, 
$\beta\!=\!0.837\pm0.471$, $r_{c}\!=\!215\pm110$\,kpc, %$b=0.04$\,cts/pix,
and a reduced-chi square of 0.70. Beyond a projected radius of about 150\,kpc, the model is a very good representation
of the data. The reasons for the slight deviations in the core region at $r\la 50$\,kpc and the flat shoulder at $r
\sim\!100$\,kpc %with a subsequent SB jump,
can be understood from a more detailed look on the cluster core region (Sect.\,\ref{s3_XCoreProper}) and the dynamical
structure in Sect.\,\ref{s3_GroupBullet} and Fig.\,\ref{fig_CoreRegion}.

%\new{Joana: can you get a realistic lower bound for $\beta$, this is very important for Sect.3.1.2 and
%even more for \ref{s4_cosmo_check}. For large errors, they should be assymetric (see e.g. Reese+02, p28 left panel). I
%would expect something like $\beta=0.84 ^{+47}_{-30}$, when the upper limit is kept.}
% --> this is used for error in DA

%profile with a good level of accuracy up to 2 arcmin radius, that is roughly R200. We computed the azimuthally-averaged
%surface brightness profile in bins of fixed radius that we fitted with the widely used approximation of the single
%isothermal $\beta$ model (Cavaliere \& Fusco-Femiano 1978):
%\begin{equation}
% S(r) = S_{0} (1+(r/r_{c})^2) ^{-3\beta+0.5} + b
%\label{equ_sbeta1}
%\end{equation}
%\noindent where $S_{0}$, $r_{c}$, $\beta$ and $b$ correspond to the central surface brightness, core radius, slope and
%constant background, respectively. The fitting procedure was performed with a Levenberg-Marquardt least-squares
%minimization.
%The fitting parameters we obtain are the following: $\beta$=0.802 $\pm$ 0.427, $r_{c}$=205 $\pm$ 102 kpc,
%and a reduced-chi square of 0.913.

In order to obtain an improved  %a qualitatively more appropriate
model representation %fit 
for the innermost bins of the radial profile we also performed a two-component $\beta$ model fit, with a single
$\beta$ coefficient, but  allowing for two independent core radii $r_{c1}$ and $r_{c2}$

\begin{equation}\label{e2_double_beta_model}
% $S(r) = S_{1} (1+(r/r_{c1})^2) ^{-3\beta+0.5} + S_{2} (1+(r/r_{c2})^2) ^{-3\beta+0.5}$ \, .
S(r) = S_{1} \left[1+ \left(\frac{r}{r_{c1}}\right)^2 \right]^{-3\,\beta+ \frac{1}{2}} + S_{2} \left[1+
\left(\frac{r}{r_{c2}}\right)^2 \right] ^{-3\,\beta+ \frac{1}{2}} + b \ .
\end{equation}

\noindent The resulting fit parameters for this model are $\beta\!=\!1.0$ (the maximum allowed value),
$S_{1}\!=\!0.42\pm0.51$\,cts/pix, 
%\new{Joana: error correct?, changed from 0.51??} --> large error correct 
$r_{c1}\!=42\pm\!34$\,kpc, $S_{2}\!=\!0.24\pm0.04$\,cts/pix, and $r_{c2}\!=\!272\pm32$\,kpc with a reduced-chi square equal to 0.45
%$r_{c1}=265 \pm 26$\,kpc, and $r_{c2}=26 \pm 12$\,kpc with a reduced-chi square equal to 0.548.
As shown in the right panel of Fig.\,\ref{fig_SBprof}, the core region of the SB distribution at $r\la\!100$\,kpc is now
qualitatively better modelled. %, as can be expected when introducing additional free parameters.

We also measured the surface brightness concentration for the cluster using the parameter
$c_{\mathrm{SB}}=(\mathrm{SB}[r<40 \,\mathrm{kpc})])/(\mathrm{SB}[r<400\,\mathrm{kpc}])$ as defined by
\citet{Santos2008a}, which allows an improved structural characterization of high-z clusters compared to methods
requiring very high photon statistics. For XMMU\,J1230.3+1339 this parameter is found to be $c_{\mathrm{SB}}=0.07$, which places
the system in the formal category of Non-Cool Core (NCC) clusters, in contrast to the more centrally concentrated
objects in the moderate and strong Cool Core Cluster (CCC) class at high redshift \citep{Santos2008a}.

%It is well-known that when a cluster presents a central excess emission and thus is termed a cool core (ref), a
%single-$\beta$ model is often inappropriate to describe its core region, requiring the use of a double-$\beta$ model.
%While this conclusion is clear for nearby clusters where the count statistics are high and the cluster angular size is
%large, for high-z cluster this situation is ambiguous because the photon statistics is usually not sufficient to firmly
%discriminate between a single or a double $\beta$ function (e.g. Ota \& Mitsuda 2004, Ettori et al. 2004, (Santos et
%al. 2008)).

%To investigate the nature of the high SB value in the core we measured the surface brightness concentration,
%$c_{SB}$, a parameter defined as csb=sb(r$<$40 kpc)/sb(r$<$)400 kpc (see Santos for more details) and optimized
%for high-z clusters, that allows us to discriminate between strong cool-cores, moderate cool-cores and non cool-cores.
%We obtain $c_{SB}$ equal to 0.07, indicating that this is a non cool-core, hence the peaked profile could be
%due to noise or could be an indication of merger activity.

For the visual representation of the two dimensional X-ray surface brightness structure
(Figs.\,\ref{fig_OpticalView},\ref{fig_MultiLambda_DensityMaps}, and \ref{fig_CoreRegion}), we adaptively smoothed the
{\it Chandra} soft band image and created an X-ray contour map with eight log spaced surface brightness levels spanning the
significance range of 1-70\,$\sigma$ \ above the background rms. It should be noted that the cluster does not contain
any detectable point sources within $R_{500}$ at the current {\it Chandra} depth. The structural and spectral analysis of the
cluster properties should thus be unbiased, also for XMM-{\it Newton} with its lower spatial resolution.

%Since the total number of Chandra photons for the cluster is comparable to the deeper of the two XMM-exposures,
We performed a spectral temperature fit in a signal-to-noise optimized aperture of 40\arcsec \, following the procedure
described in \citet{Tozzi2003a}. Applying the same general procedure as for XMM-{\it Newton} with the local background
measured in an on-chip region with %a 
60\arcsec \ diameter, we obtain a best fitting  {\it Chandra} core temperature of
$5.40^{+1.27}_{-0.94}$\,keV with the abundance fixed at $0.3\,Z_\odot$. The spectrum and model fit is displayed in blue in
%the lower panel of 
Fig.\,\ref{fig_XraySpec}, which also shows an indication of the Fe line at $kT\!\sim\!3.3$\,keV. Leaving
the metal abundance as a free parameter, we can derive a weak {\it Chandra}-only core region constraint of
$Z=0.46^{+0.38}_{0.28}\,Z_\odot$.

\section{Results and mass estimates}
\label{s3_results}

In the following section we will present results derived from the X-ray measurements and the optical imaging and
spectroscopic data. %, and the SZE observations. 
A special focus will be given to a range of observables that are commonly
used as proxies for the cluster mass. All final $M_{200}$ cluster mass estimates (ME) based on a given observable are
labelled with a number (e.g. ME\,1) for the purpose of easier referencing, intermediate results are indicated by an
additional letter (e.g. ME\,1a, ME\,1b). All used mass-observable scaling relations are stated explicitly  for clearness, transformed into our assumed cosmology and the used cluster mass definition $M_{200}$, where applicable.

%................................................
\subsection{Global X-ray properties}
\label{s3_XrayResults}

We will start with the X-ray determined bulk properties of the intracluster medium in order to establish some first
robust fundamental characteristics of the cluster. For all subsequent discussions, we define the `center-of-mass' of
the ICM X-ray emission determined with XMM-{\it Newton} (RA=12:30:16.9, DEC=+13:39:04.3, see Tab.\,\ref{table_results}) as
the reference center of the cluster, and measure all cluster-centric distances relative to this point.

%~~~~~~~~~~~~~~~~~~~~~~~~~~~~~~~~~~~~~~~~~
\subsubsection{X-ray temperature}
\label{s3_Tx_R200}

The {\it Chandra} %($5.43^{+1.25}_{-0.96}$\,keV)
and XMM-{\it Newton} %($6.44^{+1.68}_{-1.20}$\,keV)
results on the ICM temperature in the 40\arcsec \ aperture core region are of comparable statistical quality and are
fully consistent with each other.
%are well consistent within the specified errors. Having two independent measurements
%with comparable uncertainties at hand, we can readily combine the results into a joint best temperature estimate for
%XMMU\,J1230.3+1339 of kT$_{\mathrm{X,com}}=5.94^{+1.05}_{-0.77}$\,keV.
%, using standard error propagation for the combined uncertainties.
In order to further decrease the uncertainties, we have performed a joint {\it Chandra}/XMM-{\it Newton} spectral fit to these
data %. We obtain  a best joint temperature
for the region within $40\arcsec\!\cong\!R_{2500}$ (see below). 
%finding  %XMMU\,J1230.3+1339 of
We find a final joint X-ray temperature of 
kT$_{\mathrm{X,40\arcsec}}\!\simeq\!\mathrm{kT}_{\mathrm{X,2500}}\!=\!5.30^{+0.71}_{-0.62}$\,keV (see Fig.\,\ref{fig_XraySpec}) using the same method
as in Sect.\,\ref{s2_XrayAnalysis} with a fixed metallicity of Z=0.3\,Z$_\odot$, minimally binned data, and C-statistics. Leaving the metallicity as a free parameter, we can constrain the iron abundance in the cluster core to
Z$_{2500}=0.43^{+0.24}_{-0.19}$\,Z$_\odot$, i.e. with a 2.3$\,\sigma$ detection significance.

%With XMM-Newton we see indications
The  XMM-{\it Newton} data analysis indicates %yields 
that the global ICM temperature of XMMU\,J1230.3+1339 is higher when the
measurement aperture is increased. With the currently available X-ray data set this is a tentative trend, as the
measurement errors also increase. However, to allow for an evaluation of possible biases, we take the average of the
71\arcsec \ temperatures (local and external background) as a second reference temperature
$\mathrm{kT}_{\mathrm{X,71\arcsec}}\simeq 6.0^{+1.6}_{-1.2}$\,keV. Since ideally one would  like to have a measurement
of $\mathrm{T}_{\mathrm{X,500}}$, i.e. within the full $R_{500}$ region, the latter value in the 6\,keV range could indeed
be a better approximation of the underlying global temperature of the cluster.

% 13% error
This global X-ray temperature is the fundamental parameter of the intracluster medium and is closely linked to the
depth of the underlying potential well of the cluster through the Virial theorem.
% and thus the cluster as a whole.
We use the jointly determined best (core) X-ray temperature $\mathrm{T}_{\mathrm{X,2500}}$ with an uncertainty of about
$\pm$13\% to derive robust size measures of the cluster, which is an important pre-requisite for the measurement of
most cluster related quantities.
% and has in fact already been used e.g. for the flux determination.
We are using the R-T scaling relations of \citet{Arnaud2005a}, which have been calibrated using a local cluster sample
based on high-quality XMM-{\it Newton} data\footnote{Temperatures in \citet{Arnaud2005a} are determined in the region [0.1--0.5]$R_{200}$. The effect of using $T_{X,2500}$ for this work should be small and is accounted for by the discussed systematics.}. 
Using the R-T scaling relation for hot systems ($T_{\mathrm{X}}>3.5$\,keV)
$R_{200}=(1714 \pm 30) \times E^{-1}(z) \times (kT/5\,\mathrm{keV})^{0.50 \pm 0.05}$\,kpc, we obtain for
XMMU\,J1230.3+1339 an estimated $R_{200}=1017^{+68}_{-60}$\,kpc.
%, which should closely resemble the virial radius of the system.
For $R_{500}$, using the normalization $1129 \pm 17$\,kpc of the R-T relation, the estimated radius is
$R_{500}=670^{+45}_{-39}$\,kpc, and similarly  $R_{2500}=297^{+20}_{-17}$\,kpc (normalization $500 \pm 5$\,kpc).
Standard error propagation was applied in all cases to evaluate the uncertainties. In terms of projected distance in
the plane of the sky at the cluster redshift, these representative length scales correspond to angular distances of
[$R_{200}$, $R_{500}$, $R_{2500}$]=[127.8\arcsec, 84.2\arcsec, 37.3\arcsec]. With respect to the higher temperature
estimate $T_{\mathrm{X,71\arcsec}}$, all derived radii would increase by about 6\%, i.e. possible systematic effects
are mostly small in comparison to other uncertainties.
%135.3\arcsec \ and 89.1\arcsec \ for $R_{200}$ and $R_{500}$, respectively.

As the next step, we can derive a first mass estimate of the cluster by using the
M-T relation of \citet{Arnaud2005a}, %which was
calibrated with the same local cluster sample as above. Using again the best joint temperature
$\mathrm{T}_{\mathrm{X,2500}}$, we obtain from the relation

\begin{equation}\label{e3_Tx_M200}
    M^{T_X}_{200}= \frac{(5.74 \pm 0.30)}{E(z)} \times \left(\frac{kT}{5\,\mathrm{keV}} \right)^{1.49 \pm 0.17} \times 10^{14}\,\mathrm{M_{\sun}}
\end{equation}

%$M_{200}=(5.74 \pm 0.30) \times E^{-1}(z) \times (kT/5\,\mathrm{keV})^{1.49 \pm 0.17} \times 10^{14}\,M_{\sun}$
\noindent an estimate of the approximate total %virial
mass of the cluster of $M_{200}=3.61^{+0.77}_{-0.68} \times 10^{14}\,\mathrm{M_{\sun}}$ (ME\,1a). For 
%the second commonly used mass measure 
$M_{500}$ the M-T relation, with the normalization of $(4.10 \pm 0.19)\times 10^{14}\,\mathrm{M_{\sun}}$, yields
$M_{500}=2.58^{+0.54}_{-0.48} \times 10^{14}\,\mathrm{M_{\sun}}$.

The use of the higher temperature estimate $T_{\mathrm{X,71\arcsec}}$ translates into a 20\% higher mass of $M_{200}
\simeq 4.34 \times 10^{14}\,\mathrm{M_{\sun}}$ (ME\,1b). We take this positive offset of $0.73 \times 10^{14}\,\mathrm{M_{\sun}}$ into
account as a potential systematic error (denoted in brackets) and arrive at a final X-ray temperature-based mass
estimate of $M^{T_X}_{200}=3.61^{+0.77(+0.73)}_{-0.68} \times 10^{14}\,\mathrm{M_{\sun}}$ (ME\,1).

% derive M, from scaling relation
One of the main objectives of this work is the comparison of total cluster mass estimates, most commonly approximated
by $M_{200}$, as derived from various mass-observable relations. For scaling relations calibrated to yield only
$M_{500}$, we apply the ratio of the normalizations of the respective  M-T relations (see above) $f_{500 \rightarrow
200}=M_{200}/M_{500} \simeq 1.40$ to obtain a total mass estimate $M_{200}$. In the same way, $f_{2500 \rightarrow
200}$ is given by $f_{2500 \rightarrow 200}=M_{200}/M_{2500} \simeq 3.21$.

%~~~~~~~~~~~~~~~~~~~~~~~~~~~~~~~~~~~~~~~~~
\subsubsection{Hydrostatic mass estimate}

Under the assumption of an ideal isothermal gas in hydrostatic pressure equilibrium (HE), the following widely used
approximation for the total mass profile can be derived \citep[e.g.][]{Hicks2008a,Rosati2009a}

\begin{eqnarray}\label{e2_HSmass_profile}
M^{\mathrm{HE}}_{\mathrm{tot}}(<\!r) \simeq \frac{3\, k_{\mathrm{B}}}{G\,\mu\,m_{\mathrm{p}}} \cdot T \cdot \beta \cdot \frac {r
\cdot \left( \frac{r}{r_{\mathrm{c}}} \right)^{2}}{1 + \left( \frac{r}{r_{\mathrm{c}}} \right)^{2}} \\ \nonumber \simeq
1.13 \cdot \beta \cdot \left( \frac{T}{1\,\mathrm{keV}} \right) \left( \frac{r}{1\,\mathrm{Mpc}}\right) \frac {\left(
\frac{r}{r_{\mathrm{c}}} \right)^{2}}{1 + \left( \frac{r}{r_{\mathrm{c}}} \right)^{2}} \times 10^{14} \,\mathrm{M_{\sun}} \ ,
\end{eqnarray}

\noindent where $k_{\mathrm{B}}$ is the Boltzmann constant, $G$ the gravitational constant, $m_{\mathrm{p}}$ the proton
mass, and $\mu\!\simeq\!0.59$ the mean molecular weight in atomic mass units. By using the joint X-ray temperature
$\mathrm{T}_{\mathrm{X,2500}}$, and the single beta model fit parameters from {\it Chandra} ($\beta\!\simeq\!0.84$,
$r_{\mathrm{c}}\!\simeq\!215$\,kpc), we obtain the hydrostatic mass estimates for XMMU\,J1230.3+1339 of  $M_{500}\!\simeq\!3.1\!\times\!10^{14}\,\mathrm{M_{\sun}}$ and $M_{200}\!\simeq\!4.9\!\times\!10^{14}\,\mathrm{M_{\sun}}$ (ME\,2a), whereas the latter value is
slightly extrapolated beyond the last reliable data point of the measured X-ray SB profile.
% ^{\mathrm{extrapol}}

Note that only about 11\% (18\%) of the total mass $M_{200}$ ($M_{500}$) is actually enclosed within one core radius
$r_{\mathrm{c}}$, within which the single beta model exhibits some deviations from the data possibly due to the onset of
cooling (see Sect.\,\ref{s3_XCoreProper}) and significant dynamical activity (see Sect.\,\ref{s3_GroupBullet}).
% leads to deviations from the best fitting single beta model and the assumption of hydrostatic equilibrium.
Between one core radius and $r\!\ga\!R_{500}$ the model describes the radial surface brightness profile %extremely well,
with good accuracy, so that an extrapolation out to $R_{200}$ is justified.
% and the resulting $M_{200}$ should be a good mass approximation.
The dominant source of uncertainty here is the $\sim$56\% error on the $\beta$ parameter due to a degeneracy with the
fitted core radius over the accessible data range. Taking this and the temperature and radii errors into account we
arrive at a final hydrostatic equilibrium mass estimate of $M^{\mathrm{HE}}_{200} = (4.9 \pm 2.8) \times 10^{14} \,\mathrm{M_{\sun}}$ (ME\,2).
%Instead of extrapolating the radial profile beyond the data out to $R_{200}$, a more robust way for a hydrostatic
%approximation of the total mass is the use of the previously determined
%mass ratio factor $f_{500 \rightarrow 200}$, which yields $M_{200}\simeq f_{500 \rightarrow 200} \cdot M_{500} \simeq
%4.9^{+2.7}_{-2.6} \times 10^{14}\,M_{\sun}$.

%~~~~~~~~~~~~~~~~~~~~~~~~~~~~~~~~~~~~~~~~~
\subsubsection{X-ray luminosity}
\label{s3_Lx_lum}

From an observational point of view, the X-ray luminosity $L_{X}$ is the ICM bulk property which is easiest to
determine, as it only requires  flux and redshift measurements. For this reason, the X-ray luminosity is of crucial
importance for most large-scale X-ray surveys and provides the link to other ICM properties and the total cluster mass
\citep[e.g.][]{Reiprich2002a} via $L_{X}$ scaling relations.

For XMMU\,J1230.3+1339 we determine a restframe 0.5-2\,keV cluster luminosity of $L^{0.5-2\,\mathrm{keV}}_{X,500}=(1.92
\pm 0.20) \times 10^{44}$\,erg/s within the accessible aperture $R_{500}$.
% and $L^{0.5-2\,\mathrm{keV}}_{X,200}=2.10 \pm 0.21 \times 10^{44}$\,erg/s for the total soft band cluster luminosity.
The bolometric X-ray luminosity can be determined by applying a temperature dependent bolometric correction factor
$f^{\mathrm{bol}}_{0.5-2\,\mathrm{kev}}$ to the soft band luminosity \citep[e.g.][]{HxB2004a}, in our case
$f^{\mathrm{bol}}_{0.5-2\,\mathrm{kev}}\!\simeq\!3.38$, %This 
yielding a total bolometric [0.01-100\,keV] luminosity for the
cluster of $L^{\mathrm{bol}}_{X,500}\!=\!f^{\mathrm{bol}}_{0.5-2\,\mathrm{kev}}\!\cdot\!L^{0.5-2\,\mathrm{keV}}_{X,500} =
(6.50 \pm 0.68)\! \times\!10^{44}$\,erg/s.
%, currently surpassed only by a handful objects at this redshift or beyond.
% X2235, Warps Cl1415, X1229, 2XMM0830 (Georg)

Recently the use of a core-excised X-ray luminosity has been shown to reduce the scatter in the $L_{X}$ scaling
relations by a factor of 2 \citep{Pratt2009a}, simply by avoiding the central cluster region and its related complex
physical processes. 
%Following \citet{Pratt2009a}, 
For XMMU\,J1230.3+1339 we measure a core-excised  luminosity in an aperture of
[0.15-1]$R_{500}$ of $L^{0.5-2\,\mathrm{keV}}_{X,[0.15-1]500}\!=\!(1.53 \pm 0.21)\!\times\!10^{44}$\,erg/s.

\citet{Pratt2009a} also provide the latest $L_{X}$ scaling relations calibrated with the representative local REXCESS
sample \citep{HxB2007a}. Applying the self-similar $L$-$T$ relation\footnote{Derived with the BCES orthogonal fit method
and corrected for Malmquist bias.}
%with an assumed  evolution
$E^{-1}(z)\cdot L^{\mathrm{bol}}_{X,500} = (7.13 \pm 1.03) (kT/5\,\mathrm{keV})^{3.35 \pm 0.32}\times 10^{44}$\,erg/s,
we would predict an X-ray temperature for the cluster of $\simeq$4.1\,keV, i.e. a value approximately 2\,$\sigma$ below
our spectroscopic temperature measurement. The main reason for this discrepancy is the assumption of self-similar
evolution of the X-ray luminosity, i.e. $L_{X} \propto E(z)$ at fixed temperature. While some empirical results at
$z\!\la\!1$ are consistent with this evolutionary scenario \citep[e.g.][]{Vik2002a,Kotov2005a,Maughan2006a}, other studies
extending towards higher redshifts find little or no evolution \citep[e.g.][]{OHara2007a,Rosati2002a}, or even negative
trends with redshift \citep{Ettori2004a}. XMMU\,J1230.3+1339 by itself would be consistent with either of the latter
two scenarios. Since the evolutionary trends of the X-ray luminosity at high redshifts ($z>0.8$) are not well
constrained yet, we assume as a starting point a no-evolution scenario, which would predict $T_{X}\!\simeq\!4.9$\,keV for
the given luminosity, and treat a possible L$_X$ redshift evolution as a systematic uncertainty. 
The use of a no-evolution scenario for the $L$-$T$ relation instead of the self-similar prediction is also empirically supported by the latest compilation of high-z cluster measurements (Reichert et al., in prep.).

With this assumption on the evolution of %of a non-evolving  
$L_{X}$, we can derive an additional mass estimate of the cluster based on the
$L$-$M$ relation of \citet{Pratt2009a}
\begin{equation}\label{e3_LX_M}
  \frac{L^{\mathrm{bol}}_{X,500}}{E^{\frac{4}{3}}(z) } = (1.38 \pm 0.12)\left( \frac{M_{500}}{2\times 10^{14}\,\mathrm{M_{\sun}}}\right)^{2.08 \pm 0.13} \times
10^{44}\, \mathrm{erg/s} \, ,
\end{equation}

%$E^{-4/3}(z)\cdot L^{\mathrm{bol}}_{X,500} = (1.38 \pm 0.12)(M_{500}/2\times 10^{14}\,M_{\sun})^{2.08 \pm 0.13} \times
%10^{44}$\,erg/s,
\noindent
corrected for one factor of $E(z)$ to take out the self-similar redshift scaling\footnote{This approach is somewhat ad hoc at this point, but provides a significantly improved agreement with other high-z cluster measurements.} (in $L$-$T$). We thus obtain a mass
estimate of
$M_{500} \!\simeq\!(2.96 \pm 0.41)\!\times\!10^{14}\,\mathrm{M_{\sun}}$ from %a non-evolving
the L-M relation, which yields $M_{200}\!\simeq\!f_{500 \rightarrow 200} \cdot M_{500}\!\simeq\!(4.1 \pm 0.6)\!\times\!10^{14}\,\mathrm{M_{\sun}}$ (ME\,3a). However, the dominant uncertainty in this luminosity-based mass proxy is the discussed
weak constraint on the luminosity evolution. We take this current lack of knowledge into account by adding a 23\%
systematic error, based on the decreased mass estimate of the self-similar evolution model.
%predicted mass difference compared to a self similar evolution model,
The final luminosity-based total mass estimate is hence $M^{L_X}_{200}\!\simeq\!(4.1 \pm 0.6 \,(\pm 1.0))\!\times\!10^{14}\,\mathrm{M_{\sun}}$ (ME\,3). Using the core excised X-ray luminosity $L^{0.5-2\,\mathrm{keV}}_{X,[0.15-1]500}$ with the
corresponding scaling relation yields a consistent mass estimate within 8\%, implying that the central region of
XMMU\,J1230.3+1339 does not exhibit any particular feature that would drive a deviation.

%Using the core-excised $L^{0.5-2\,\mathrm{keV}}_{X,[0.15-1]500}$ instead yields the same value within 10\%.

%~~~~~~~~~~~~~~~~~~~~~~~~~~~~~~~~~~~~~~~~~
\subsubsection{Core properties of the ICM}
\label{s3_XCoreProper}

We now have a more quantitative look at the  properties of the intracluster medium in the densest central core region.
The line-of-sight emission measure $\mathrm{EM}\!=\!\int
n^2_{\mathrm{e}}\,dl\!=\!4\,\pi\,(1\!+\!z)^4\,S(r)/\Lambda(T,z)$ gives direct observational access to the local
electron density  $n_{\mathrm{e}}$, deduced from the deprojected measured surface brightness profile $S(r)$ as a
function of %projected
cluster-centric distance $r$  in the soft X-ray band, and the redshifted and absorption corrected cooling function
$\Lambda(T,z)$. For an assessment of the central density of the cluster, we use the double beta model surface
brightness fit (right panel of Fig.\,\ref{fig_SBprof}), since the inner beta model component dominates the densities
within the central 40\,kpc. After the deprojection of the %2D
SB model fits (S$_1$, r$_{c1}$, S$_2$, r$_{c2}$, $\beta$)
of Sect.\,\ref{s2_Chandra}, we obtain a total %central
electron density of  $n_{\mathrm{e},0} \simeq (2.3 \pm 1.4) \times 10^{-2}$\,cm$^{-3}$ in the very center, and
$n_{\mathrm{e},20\,\mathrm{kpc}} \simeq (1.7 \pm 1.0) \times 10^{-2}$\,cm$^{-3}$ at a cluster-centric radius of
20\,kpc.
%The errors were estimated from variations of the radial binning of the SB profile and the subsequent changes
%of the central fitting parameters, resulting in up to 30\% higher densities with a very fine (but noisy) binning
%scheme.
%\new{Hans: ne from sum of models, ok?; error estimate ok?}

The central density and temperature of the ICM determine the cooling time scale $t_{\mathrm{cool}}$ in the cluster core
\citep{Sarazin1986a}, i.e. the time for which the thermal energy of the gas would be radiated away at the current
cooling rate %\new{check: taken from Pratt and Arnaud 2003, different from thesis; ne-nH conversion?}

\begin{equation}\label{e3_cooling_time}
t_{\mathrm{cool}} \simeq 2.9 \times 10^{10} \ \mathrm{yr} \cdot \left(\frac{n}{10^{-3}\,\mathrm{cm}^{-3}}\right)^{-1}
\left(\frac{k\,T}{1\,\mathrm{keV}}\right)^{1/2} \ . %\simeq 3.4 \, \mathrm{Gyr} \ .
\end{equation}

\noindent From this expression we derive a central cooling time scale\footnote{When considering only the single $\beta$-model fit parameters, this cooling time scale increases by a factor of 3.} for XMMU\,J1230.3+1339 of $t_{\mathrm{cool,0}}\!=\!(2.9 \pm 1.8)\,$Gyr and at the 20\,kpc radius $t_{\mathrm{cool,20\mathrm{kpc}}}\!=\!(4.0 \pm 2.4)\,$Gyr. This time scale
is to be compared to the age of the Universe at the redshift of the cluster ($t \simeq 5.85$\,Gyr), or for more
realistic cluster formation scenarios, the time span since $z=3$ ($t \simeq 3.74$\,Gyr) or $z=2$ ($t \simeq
2.63$\,Gyr).
%\new{Hans: tcool from Sarazin shorter than from your program, ok?}

The central electron density $n_{\mathrm{e},0}$ provides an alternative test for the presence of a cool core. Following
the CCC definition of \citet{Pratt2009a} ($n_{\mathrm{e},0}\!>\!4 \times 10^{-2}\cdot E^2(z)$\,cm$^{-3}$),
XMMU\,J1230.3+1339 is consistently classified as a Non-Cool-Core cluster, %as was already seen from
%consistent 
in concordance with the surface brightness concentration measure $c_{\mathrm{SB}}$ in Sect.\,\ref{s2_Chandra}. However, the
cluster's core region is in an interesting transition phase, where the dense gas has basically been going through a
single full cooling time scale since redshift 2-3. %, which can be considered as a realistic ICM formation epoch.
Hence, the central core region could indeed contain gas which has already cooled considerably, %in support of
consistent with our tentative finding of an increasing average X-ray temperature with radius in Sect.\,\ref{s3_Tx_R200}.

%The measurement of the central surface brightness concentration $c_{SB}=0.07$ in Sect.\,\ref{s2_XrayAnalysis} formally
%placed the cluster in the non-cool core regime.

%This apparent contradiction \new{?revise} can be explained by assuming that the observed ongoing merging event
%discussed in Sect.\,\ref{s3_GroupBullet} flattens out the central surface brightness profile (see
%Fig.\ref{fig_MultiLambda_DensityMaps}), and hence destroys the onset of a cool core. This scenario is predicted in
%\cite{Burns2008a} and can be observed in situ in XMMU\,J1230.3+1339.

%~~~~~~~~~~~~~~~~~~~~~~~~~~~~~~~~~~~~~~~~~
\subsubsection{Gas mass}
\label{s3_GasMass}

The total gas mass $M_{\mathrm{gas}}$ of clusters is obtained from the integrated radial gas density profile.
%For XMMU\,J1230.3+1339 we can savely measure $M_{\mathrm{gas}}$ out to $R_{500}\simeq 90\arcsec$.
For XMMU\,J1230.3+1339, both the deprojected single beta model ($n_{\mathrm{e},0}\simeq 0.71 \times
10^{-2}$\,cm$^{-3}$, $\beta \simeq 0.84$, $r_{\mathrm{c}} \simeq 215$\,kpc) and the double beta model yield consistent
results for the gas mass of $M_{\mathrm{gas,500}} \simeq (3.0 \pm 0.9)\times 10^{13}\,\mathrm{M_{\sun}}$, where the estimated
error is based on the uncertainties of the latter model in combination with the allowed temperature range.
%\new{Hans: 30\% error realistic?}

%we measure $M^{1\,\beta}_{\mathrm{gas,500}} \simeq (2.7 \pm 1.6)\times 10^{13}\,M_{\sun}$
%and for the double beta model parameters the consistent value $M^{2\,\beta}_{\mathrm{gas,500}} \simeq (3.3 \pm
%0.5)\times 10^{13}\,M_{\sun}$ is derived.\new{average? or take one}

Using the mean local gas mass fraction for similarly massive clusters of $\bar{f}_{\mathrm{gas},500}\simeq 0.10 \pm
0.02$ \citep{Pratt2009a}, we readily obtain a fourth total mass estimate of $M^{\mathrm{gas}}_{200} \simeq f_{500 \rightarrow
200}\cdot M_{\mathrm{gas,500}} / \bar{f}_{\mathrm{gas},500}  \simeq (4.2 \pm 1.5) \times 10^{14} \mathrm{M_{\sun}}$ (ME\,4).
% 30% error + 20% error --> 36% error

With the knowledge of the final combined %(independent) 
best estimate for the total cluster mass $M^{\mathrm{best}}_{200}$ as
discussed in Sect.\,\ref{s4_masses}, we can turn the argument around and estimate the cluster gas mass fraction for
XMMU\,J1230.3+1339 for the assumed cosmology. This consistency check yields $f_{\mathrm{gas},500}\!\simeq\!M_{\mathrm{gas,500}} / (M^{\mathrm{best}}_{200}/ f_{500 \rightarrow 200}) \simeq 0.100 \pm 0.035$ in good agreement
with the average local value.
% assume Mbest = 4.3 +-20%
% error: 30% error on Mgas + 20% error on Mbest --> 36% error

%~~~~~~~~~~~~~~~~~~~~~~~~~~~~~~~~~~~~~~~~~
\subsubsection{ Y$_{\mathrm{X}}$ }

As the final {\it Chandra} and XMM-{\it Newton} derived property, we determine the X-ray version of the Compton-Y parameter
$Y_{\mathrm{X}}\!=\!M_{\mathrm{gas}}\!\cdot\!T_{\mathrm{X}}$. This quantity reflects the total thermal energy of the
ICM and is predicted to be a robust, low scatter mass proxy from simulations \citep[e.g.][]{Kravtsov2006a,Nagai2007a}
which has recently been well confirmed by observations \citep[e.g.][]{Arnaud2007a,Vik2009a,Arnaud2010a}.

%\new{revise}
From the measurements of the global X-ray temperature and the gas mass within the radius $R_{500}$, we obtain
$Y_{\mathrm{X,500}} = (1.6 \pm 0.5 \, (+0.2))\times 10^{14} \,\mathrm{M_{\sun}}$\,keV, where again a systematic error is
considered for a possible temperature bias (Sect.\,\ref{s3_Tx_R200}). With the latest calibration of the M-Y$_{X}$
scaling relation from \citet{Arnaud2010a} we find

\begin{eqnarray}\label{e3_MYx}
M_{500} = \frac{10^{14.567 \pm 0.010}}{E^{\frac{2}{5}}(z)} \left( \frac{Y_{X}}{2\times
10^{14}\,\mathrm{M_{\sun}}\,\mathrm{keV}}\right)^{0.561 \pm 0.018} \,\mathrm{M_{\sun}} \\ \nonumber \simeq  (2.6 \pm 0.8 \, (+0.3))
\times 10^{14}\,\mathrm{M_{\sun}} \ ,
\end{eqnarray}

\noindent and hence $M^{Y_X}_{200}\simeq f_{500 \rightarrow 200} \cdot M_{500} \simeq (3.6 \pm 1.1  \, (+0.4)) \times
10^{14}\,\mathrm{M_{\sun}}$ (ME\,5).

%................................................
\subsection{Optical properties}
\label{s3_optical_prop}

We now turn to the optical cluster properties as derived from the VLT/FORS2 imaging and spectroscopic data and the
wide-field multi-band LBT imaging observations.

The rich galaxy population of XMMU\,J1230.3+1339 is easy to recognize when inspecting the top panels of
Fig.\,\ref{fig_OpticalView}, either as a galaxy overdensity measure in the single band (left panel), or in color space
(right), where the red early-type cluster galaxies clearly stand out from the foreground/background. An interesting
note is the fact that the angular size of the foreground Virgo galaxy NGC\,4477 is almost identical to the cluster
total radius.
%, due to the angular scale ratio of $\simeq$86 between $z=0.0045$  (0.093\,kpc/\arcsec) and  $z=0.975$ (7.96\,kpc/\arcsec).
At a projected cluster-centric distance of about 700\,kpc towards the West, the optical light of
the NGC\,4477 halo starts to significantly contaminate the colors of background galaxies, and at $>$850\,kpc in this
Western direction the region is basically masked out by the foreground elliptical. The proximity of the cluster to
NGC\,4477 hence has two effects on the optical analysis: (i) galaxy counts have to be corrected for the geometrically
covered area in the outer regions, and (ii) the measured colors of galaxies overlapping with the outer halo of
NGC\,4477 are likely biased towards the blue.

%R200 scaling summary:

Throughout this Sect.\,\ref{s3_optical_prop} we are facing the task to appropriately re-scale measurement apertures
that were defined in the local Universe to be applicable at $z\!\sim\!1$. This is not a straight-forward exercise, since
(i) one has to distinguish fixed-size physical apertures from relative cluster apertures that evolve with redshift, and
(ii) this evolution depends on the reference density relative to which all cluster quantities are defined. For the
overdensity definition with respect to the critical density $\rho_{\mathrm{cr}}(z)$, used in this work, the radius
scales in the self-similar model for a cluster at fixed mass as $r\!\propto\!E^{-2/3}(z)$ \citep[e.g.][]{Voit2005a}, i.e.
it changes by %shrinks by
a factor of 1.44 from $z=0$ to the cluster redshift. On the other hand, a definition relative to the mean
density $\rho_{\mathrm{mean}}(z)$ implies a scaling of $r\!\propto\!(1+z)^{-1}$, i.e. a decrease by 1.98, and with
reference  to the top-hat collapse density $\Delta (z)$ radii scale as $r\!\propto\!E^{-2/3}(z)\cdot \Delta^{-1/3} (z)$,
equivalent to a shrinking by about a factor of 1.67 at the cluster redshift. %\new{Hans: check} 
For the following
discussion we choose an intermediate radius re-scaling approach consistent with \citet{Calrberg1997a} 
$r\!\propto\!E^{-1}(z)$, i.e. where appropriate we apply a radius correction factor of $f_{\mathrm{resc}}= 1/ E(z) \simeq 0.58$.
Since the total enclosed mass close to the outer radius scales approximately as $M_{\mathrm{tot}}(<r) \propto r$ (see
e.g. Equ.\,\ref{e2_HSmass_profile}), the use of different re-scaling approaches  for the measurement aperture of
optical quantities can lead to systematic total mass shifts of $\pm$15-20\%. Working with fixed physical apertures, on
the other hand, could %easily 
result in an overestimation of the total cluster mass by  up to a factor of 2 at $z \sim 1$, when
using optical observables such as richness or total luminosity.

%~~~~~~~~~~~~~~~~~~~~~~~~~~~~~~~~~~~~~~~~~
\subsubsection{Richness measures}
\label{s3_opt_richness}

% Richness estimate: Abell's original richness definition R of counting the number of cluster galaxies within the
% projected distance 1.5\,h$^{-1}$\,Mpc in the magnitude interval [m$_3$,m$_3$+3], i.e. the galaxies within two
% magnitudes of the third brightest galaxy m$_3$.
%R=0: 30-49 R=1: 50-79 R=2: 80-129 galaxies

%R2: top 30\% of Abell's statistical sample top 20\% of all Abell clusters
%Coma cluster (from Kaiser 1986): R=2, mass about 10E15 Msun, velocity dispersion of about 1000km/s M/L=400 h
%R=1 clusters about 5E14Msun

%N$^{\mathrm{gal}}_{200}$ \citep{Koester2007a}
%Scaling Relations \citep{Reyes2008a}

As starting point, we determine the classical optical richness measure, the Abell richness $R^{\mathrm{rich}}$
\citep{Abell1958a}, defined as the number of cluster galaxies within the projected Abell radius $R_{\mathrm{A}} \!\simeq\!2.14\,h^{-1}_{70}$\,Mpc and with apparent magnitudes in the interval [$m_3,m_3+2$], where $m_3$ is the third brightest
cluster galaxy.
%After these general remarks on aperture re-scaling, we return to the richness estimates of XMMU\,J1230.3+1339.
We use the LBT/LBC z\arcmin -band as the reddest optical wide-field imaging filter, providing the highest
cluster-to-foreground/background contrast and allowing background estimations in the same field. The magnitude interval
for the Abell richness in our case is [$m_{3,z},m_{3,z}+2$]=[20.74, 22.74], and for the re-scaled  Abell radius we
assume $R_{\mathrm{A,resc}}\simeq f_{\mathrm{resc}} \cdot R_{\mathrm{A}} \simeq 1240$\,kpc, corresponding to an angular
scale of 156\arcsec.

%, rescaled according to the expected background density evolution as in Sect.\,\ref{s3_Tx_R200}, is
%$R_{\mathrm{A}(z)}\simeq R_{\mathrm{A}} / E(z) \simeq 1240$\,kpc corresponding to 156\arcsec.

We determined the statistical background/foreground galaxy counts in this magnitude interval in 17 external background
regions with a radius equivalent to $R_{500}$  %90\arcsec \ radius  ($\simeq R_{500}$)
each, yielding a median z\arcmin -band field galaxy density of $8.4 \pm 1.5$\,arcmin$^{-2}$.
%$(9.5\pm 1.4)$\,arcmin$^{-1}$.
For the cluster region within $R_{500}$, which is not influenced by NGC\,4477, we obtain a galaxy overdensity of $(74
\pm 9)$. The richness estimate for the scaled Abell radius $R_{\mathrm{A,resc}}$ is then $(109 \pm 17)$, after applying
a 10\% correction ($\sim$3 galaxies) for the geometrically covered area in the outer ring beyond $R_{500}$. This
implies that XMMU\,J1230.3+1339 falls in the Abell richness class $R^{\mathrm{rich}}\!=\!2$, defined as having 80-129 counted galaxy
members, and would thus rank among the top 20\% of the richest clusters even in the local Abell catalog.

% V4 Doku in X1230_VLT_Doku.txt
% V2:
% Note: from average density, get 202.2 background object in Abell region; measured are 292.9, yielding 90.7
% different from analysis in docu
% error: squrt(21.2/7.1 arcmin)x7.8 = 13.51
% 458 out of 2712 = 16.8% clusters in Abell catalog have richness class 2 or higher

A more modern and better calibrated richness measure is $N_{200}$ \citep{Koester2007a,Reyes2008a}, the number of E/S0
ridgeline member galaxies within $R_{200}$, fainter than the BCG and brighter than $0.4\,L*$. This richness definition
has been used for the  MaxBCG catalog of \citet{Koester2007b} of optically selected groups and clusters from the Sloan
Digital Sky Survey (SDSS) in the redshift range $0.1 < z < 0.3$.
%The application of this and other locally defined optical quantities at $z \sim 1$ requires cosmological corrections
%that account for the discussed scaling behavior of the cluster radius. For luminosity based quantities as below,
%passive evolution for the considered galaxy population and a $K$-correction term has to be taken into additionally
%account.
We follow \citet{Reyes2008a} for a self-consistent procedure to determine $N_{200}$, which has been used and calibrated
with a different definition of $R'_{200}$, evaluated relative to the mean density $\rho_{\mathrm{mean}}$ of the
Universe.
%considering the cluster matter overdensity relative to the mean density of the Universe instead of the critical density.
For the selection of E/S0 ridgeline galaxies we use the color cuts R$-$z=$2.05 \pm 0.2$ as shown in
Fig.\,\ref{fig_CMD}, with a comparable color width as used for the  MaxBCG catalog. The considered z-band magnitude
range is then [19.9\,mag, 22.9\,mag], %19.9\,mag$\le$z$\le$22.9\,mag --> FORS2 magnitudes
confined by the measured BCG magnitude  and a SSP (Simple Stellar Population) model magnitude of a 0.4\,L* galaxy at
the cluster redshift (see below). For the iterative process, we first determine the number of galaxies fulfilling the
color and magnitude cuts within a fixed projected radius of 1\,$h^{-1}$\,Mpc $\simeq$1.43\,Mpc yielding
$N_{\mathrm{1\,h^{-1}\,Mpc}}\!=\!56$. From the very similar scaled final search radius $R^{*}_{200}\! \simeq\!0.223\,
N^{0.6}_{\mathrm{1\,Mpc}}\!\cdot\!f_{\mathrm{resc}} \,h^{-1}_{70}\,\mathrm{Mpc}\!\simeq\!1.44\,\mathrm{Mpc}$\footnote{Note that the optically defined radius $R^{*}_{200}$ for the determination of $N_{200}$ and $L_{200}$ is different from the $T_X$ derived radius $R_{200}$ used throughout most of the paper.}, we obtain the final richness measure $N_{200}\!\simeq\!57$, which again includes a small correction (3 galaxies) for the region covered by NGC\,4477.

A total mass estimate based on this richness estimate is then obtained from the calibrated  MaxBCG scaling relation of
\citet{Reyes2008a}. Due to different radius definitions, the cluster mass $M'_{200}$ resulting from this scaling
relation is systematically higher than $M_{200}$ used in this work. In the local Universe $M'_{200}$ is about 40\%
larger, which reduces to approximately 8\% at $z\!\sim\!1$, due to the fact that $\rho_{\mathrm{mean}}(z)$ converges to
$\rho_{\mathrm{cr}}(z)$ with increasing redshift. For consistency with our $M_{200}$ mass definition we apply a
correction factor $f^{\mathrm{mean}}_{\mathrm{cr}}(z\!\sim\!1)\!\simeq\!0.92$ %\new{Hans: cross-check}  
and convert the mass units to $h^{-1}_{70} \mathrm{M_{\sun}}$

%compared to the one used in this work. Correcting for this, we obtain a richness based total mass estimate of $M_{200}
%\simeq 2.4 \pm 1.1 \times 10^{14}\,M_{\sun}$. \new{revise}

\begin{samepage}

\begin{eqnarray}\label{e3_N200_M}
    M^{N_{200}}_{200} = f^{\mathrm{mean}}_{\mathrm{cr}} \cdot (2.03 \pm 0.11) \cdot \left(\frac{N_{200}}{20}\right)^{1.16 \pm 0.09} \times 10^{14} \mathrm{M_{\sun}} \\ \nonumber
    \simeq (6.3 \pm 0.7) \times 10^{14}\,\mathrm{M_{\sun}} \ .
\end{eqnarray}

\end{samepage}

\noindent In terms of systematic uncertainties, we consider an additional $\pm$15\%  error for the aperture re-scaling,
and $\pm$20\% for the selected color cuts and possible foreground/background contamination in the richness estimate.
Furthermore, all optical scaling relations in \citet{Reyes2008a} are derived from stacked cluster quantities, i.e. they do not contain the information on the scatter of individual clusters with respect to the best fitting relation. To account for this intrinsic scatter, we add a   conservative minimum of $\pm$20\%  systematic mass uncertainty to all optical mass proxies \citep[see e.g.][]{Popesso2005b}.
With a combined  possible systematic mass offset of $\pm$32\%, the final richness based total mass estimate is hence
$M^{N_{200}}_{200} \simeq (6.3 \pm 0.7 \, (\pm 2.0)) \times 10^{14}\,\mathrm{M_{\sun}}$ (ME\,6).

%~~~~~~~~~~~~~~~~~~~~~~~~~~~~~~~~~~~~~~~~~
\subsubsection{Brightest Cluster Galaxy}
\label{s3_BCG}

We address the characteristics of the Brightest Cluster Galaxy (BCG) of XMMU\,J1230.3+1339 and compare them to
predictions from Simple Stellar Population (SSP) models. The BCG can be readily identified in the
color-magnitude-diagram in Fig.\,\ref{fig_CMD} as the galaxy with total z-band magnitude
z$^{\mathrm{BCG}}_{\mathrm{tot}}=19.90$ and a color of R$-$z$\simeq$2.04. In the lower left panel of
Fig.\,\ref{fig_OpticalView} the BCG is marked by a green dashed circle at a projected offset from the nominal X-ray
center of about 18\arcsec \ corresponding to 140\,kpc. Furthermore, the BCG (ID02 in
Table\,\ref{tab_specmembers}) seems not to be at rest in the cluster reference system, but rather exhibits a velocity
of about $+600$\,km/s with respect to the median redshift. The lower panels of Fig.\,\ref{fig_OpticalView} also reveal
that the BCG has a very close companion galaxy with a consistent color in the projected view, which could be an
indication for ongoing or upcoming merging activity. We can summarize, that XMMU\,J1230.3+1339 has a dominant BCG in
the cluster core with properties in the CMD (magnitude, color, luminosity gap) very similar to most local clusters, but
additional characteristics (cluster center offset, cluster velocity offset, close companion) that point towards a
dynamical BCG state which is yet to reach a final equilibrium configuration at the minimum of the cluster potential
well.

Interestingly, we can identify a second galaxy with BCG properties on the outskirts of the cluster, from now on
referred to as BCG2 for distinction. This galaxy can be seen in the Southern half of Fig.\,\ref{fig_OpticalView} (top
right panel) and in Fig.\,\ref{fig_InfallingGroups}. %(top panel). 
The BCG2 spectrum is shown in the lower panel of
Fig.\,\ref{fig_Spectra} and further properties are specified in Table\,\ref{tab_specmembers} (ID08). With a total
magnitude of z$^{\mathrm{BCG2}}_{\mathrm{tot}}=20.1$, the BCG2 is only 0.2\,mag fainter than the core BCG and has a
slightly bluer color in comparison. Due to the projected distance of about $70\arcsec\!\simeq\!550$\,kpc from the cluster
center, the BCG2 appears in the CMD of Fig.\,\ref{fig_CMD} as a spectroscopically confirmed (open circle) non-core member
(small black dot). The cluster restframe velocity of the BCG2 is consistent with zero, giving rise to the conclusion
that the BCG2 has just entered the $R_{500}$ region of the cluster approximately along the plane of the sky, which will
be further discussed in Sect.\,\ref{s3_InfallingGroups}.

%The application of this and other locally defined optical quantities at $z \sim 1$ requires cosmological corrections
%that account for the discussed scaling behavior of the cluster radius. For luminosity based quantities as below,
%passive evolution for the considered galaxy population and a $K$-correction term has to be taken into additionally
%account.

The comparison to stellar population synthesis models allows us to probe in more detail the BCG properties and to
compare them with observations in the local Universe. We have computed a grid of SSP models using PEGASE2
\citep{Fioc1997a} with different formation redshifts and metallicities for relating them to the observed redshift
evolution of passively evolving galaxies \citep[for details, see][]{RF2007Phd}. The observed R$-$z color is well
matched by a model with a single star formation burst at $z_f\!\simeq\!5$ and slightly supersolar metallicity for the BCG
and about solar metallicity for BCG2. For $z\!\sim\!1$, our models further predict an apparent magnitude of $z\!\simeq\!21.9$\,mag for a passive L* galaxy, implying that the  BCG has a luminosity of $\simeq 6.5\,$L* (or m*-2).

For the comparison of passive galaxy luminosities at higher-z to local measurements such as MaxBCG, two different
approaches could in principle be followed. The first option is to perform the measurements in the same observed band, r
in the case of MaxBCG, and then account for passive evolution and a K-correction term of order 2\,mag using the model
predictions. The observationally preferable second approach uses the best suited band for the measurement, z in our
case, then applies significantly smaller evolution and K-correction terms, and finally transforms the local values
into the desired reference band using the model. Here we follow the latter method in order to take advantage of the
$\sim$2 magnitudes brighter passive cluster galaxies in the z-band compared to r or R.

We determine the absolute z-magnitude by applying the distance modulus equation $m_{\mathrm{BCG}}(z)-M_{\mathrm{BCG}}\!=\!25+5\log \left(d_{\mathrm{lum}}\, [\mathrm{Mpc}]\right)+K_{\mathrm{BCG}}(z)$. With
z$^{\mathrm{BCG}}_{\mathrm{tot}}=19.90$, the luminosity distance $d_{\mathrm{lum}}=6403.6$\,Mpc, and a K-correction
of K$_{\mathrm{BCG}}\simeq 0.82$ we obtain $M^{\mathrm{z}}_{\mathrm{BCG}} = -24.95 \pm 0.03$\,mag for our assumed concordance
cosmology. Using the absolute magnitude of the sun $M^{\mathrm{z}}_{\sun} = 4.51$ \citep{Blanton2003b} %\new{Daniele: check} 
we can derive the BCG luminosity as $L^{\mathrm{z}}_{\mathrm{BCG}}\simeq 6.1 \times 10^{11}\,\mathrm{L_{\sun}}$.

% SSP Model Mags:
% R$-$z(AB) = R$-$z(Vega) - 0.326
% at 0.975, zf=5: 1Z: 1.91  ; 3Z: 2.25
% AB 1Z: m*_z(z=1)=21.9, m*_z(z=0.975)=21.806 , m*_R(z=1)=23.8  , m*_R(z=0.975)=23.7
% --> BCG = m* - 1.906
% 3Z: m*_z(z=1)=22.3, m*_z(z=0.975)=22.2
% K-corrections: z=0.975-->0: R_k=2.13 ; Z_k=0.82
% K-corrections: z=0.975-->0.25: R_k=2.13-0.316=1.81 ; Z_k=0.82-0.234=0.59
% 0  |    0.1910  |    0.2340

% r-band mag at z=0.25
% use M_R (sun) = 4.63 (http://www.ucolick.org/~cnaw/sun.html)

% From Blanton+2003, APJ 592
%There are small differences between the system output by the SDSS pipelines and our best estimate of the true AB
%system, amounting to Dm ¼ mAB  mSDSS ¼ 0:042, 0.036, 0.015, 0.013, and 0.002 in the u, g, r, i, and z bands
% yields the absolute solar AB magnitudes: M;0:1u ¼ 6:80 ; M;0:1g ¼ 5:45 ; M;0:1r ¼ 4:76 ; M;0:1i ¼ 4:58 ; Mz=4.51
% Sun absolute magnitude in z-band: M_zsun=4.51 (Blanton+2003, APJ 592); M_Rsun=4.63

The BCG luminosity is known to correlate  weakly  with the cluster mass \citep{Lin2004a}. Although found to be a fairly
poor tracer of $M_{200}$, \citet{Reyes2008a} provide a calibrated $L_{\mathrm{BCG}}$-$M$ scaling relation for BCG r-band
luminosities evaluated and K-corrected at $z\!=\!0.25$. With the discussed correction term
$f^{\mathrm{mean}}_{\mathrm{cr}}$ and luminosities and masses transformed to $h^{-2}_{70}\,\mathrm{L_{\sun}}$ and
$h^{-1}_{70}\,\mathrm{M_{\sun}}$ we arrive at the following relation %\new{Daniele: check}

%\new{check}
\begin{equation}\label{e3_LBCG_M}
    M^{L_{\mathrm{BCG}}}_{200} = f^{\mathrm{mean}}_{\mathrm{cr}} (1.53 \pm 0.10) \left(\frac{L^{r}_{\mathrm{BCG}}}{1.02 \times 10^{11}
    \mathrm{L_{\sun}} }\right)^{1.10 \pm 0.13} \times 10^{14} \mathrm{M_{\sun}} \ .
\end{equation}

\noindent In order to apply this relation, we need to passively evolve the BCG to $z\!=\!0.25$ and then evaluate its r-band
luminosity ($M^{r}_{\sun}\!=\!4.76$) yielding $L^{\mathrm{r}}_{\mathrm{BCG}}\!\simeq\!1.7 \times 10^{11}\,\mathrm{L_{\sun}}$. 
%\new{Daniele: check used R-band offsets instead of r} 
From Equ.\,\ref{e3_LBCG_M} we obtain the mass estimate $M_{200}\!\simeq\!(2.4 \pm 0.4)\!\times\!10^{14}
\mathrm{M_{\sun}}$ (ME\,7a).
%, which after correction to our total mass definition yields $M_{200} \simeq 2.1 \times 10^{14} M_{\sun}$.

So far, we assumed that the BCG's stellar population is passively dimming without a change in stellar mass. If we
instead assume a non-evolving absolute magnitude, i.e. the dimming of stellar light is compensated by stellar mass
acquisition, then the r-band luminosity at $z\!=\!0.25$ is 70\% higher ($L^{\mathrm{r}}_{\mathrm{BCG}}\!\simeq\!2.9 \times
10^{11}\,\mathrm{L_{\sun}}$) resulting in a mass estimate of $M_{200} \simeq (4.4 \pm 0.7) \times 10^{14} \mathrm{M_{\sun}}$ (ME\,7b).
For consistency with \citet{Reyes2008a}, we use the non-evolving luminosity as reference mass proxy, and treat the offset to
the result including passive evolution as a systematic error. With an additional $\pm$10\% systematic for the SSP model
evolution uncertainty and $\pm$20\% for the intrinsic scatter, we arrive at the final cluster mass estimate of $M^{L_{\mathrm{BCG}}}_{200}\!\simeq\!(4.4 \pm 0.7 (^{+1.0}_{-2.2}))\!\times\!10^{14} \mathrm{M_{\sun}}$ (ME\,7).

%correct by 30\%

%bit too red, but within 2 sigma of calibration color error
% two cluster with redder zero point in \citep{Mei2009a} real or data
%Truncation consistent with downsizing picture, RS built up from bright magnitudes first
%truncation in surrounding groups of 1252, but not the main cluster truncation at Ks(AB)=22 = 20.1 Vega \citep{Tanaka2007a}

%~~~~~~~~~~~~~~~~~~~~~~~~~~~~~~~~~~~~~~~~~
\subsubsection{Optical luminosity}
\label{s3_Opt_Lumin}

The next optical  observable and useful   % characteristic and 
mass proxy %observable
%we would like to measure
is the cluster luminosity $L_{200}$,
defined as the summed r-band luminosity of all red ridgeline galaxies in $N_{200}$,  K-corrected to $z\!=\!0.25$
\citep{Reyes2008a}. Following the procedures of the last two Sects.\,\ref{s3_opt_richness} and \ref{s3_BCG}, we derive
a total ridgeline z-band luminosity inside $R^{*}_{200}$ of $L^{\mathrm{z}}_{200}\!\simeq\!9.2\!\times\!10^{12}\,\mathrm{L_{\sun}}$ corresponding to $15.1\!\cdot\!L^{\mathrm{z}}_{\mathrm{BCG}}$ . In contrast to the BCG, where the accretion of additional stellar mass  can potentially compensate
the dimming of its predominantly passive stellar population, the natural choice for the treatment of the total
ridgeline light, which should be closely linked to the total stellar mass in the cluster, is the assumption of
passively evolving luminosities. The expected evolved and transformed bulk luminosity in the r-band at $z=0.25$ is
therefore $L^{\mathrm{r}}_{200}\!\simeq\!2.5\!\times\!10^{12}\,\mathrm{L_{\sun}}$. The corresponding scaling relation (in units of
$h^{-2}_{70}\,\mathrm{L_{\sun}}$ and $h^{-1}_{70}\,\mathrm{M_{\sun}}$) %\new{Daniele: check}

% L200,  also profile? --> no

\begin{equation}\label{e3_L200_M}
    M^{L_{200}}_{200} = f^{\mathrm{mean}}_{\mathrm{cr}} (2.51 \pm 0.24) \left(\frac{L^{r}_{200}}{8.16\times 10^{11} \mathrm{L_{\sun}}}\right)^{1.40
    \pm 0.19} \times 10^{14} \mathrm{M_{\sun}}
\end{equation}

\noindent yields a total cluster mass estimate of $M_{200}\!\simeq\!(11.1 \pm 2.8)\!\times\!10^{14} \mathrm{M_{\sun}}$ (ME\,8a). The
systematic uncertainties now add up to about $\pm 34$\%, based on the discussed contributions from aperture re-scaling
(15\%), color cut selection (20\%), the SSP evolution model (10\%), and the intrinsic scatter assumption (20\%). The final $L_{200}$ mass proxy for the cluster
is hence $M^{L_{200}}_{200}\!\simeq\!(11.1 \pm 2.8 \, (\pm 3.7))\!\times\!10^{14} \mathrm{M_{\sun}}$ (ME\,8).

While the passively evolved BCG luminosity pointed towards a low cluster mass estimate (ME\,7a), the total $L_{200}$
light (ME\,8) would give rise to the opposite conclusion of an extremely high total mass. This discrepancy is to be
attributed to the highly populated bright end of the red galaxy ridgeline (see Fig.\,\ref{fig_CMD}), which apparently
places XMMU\,J1230.3+1339 in a low mass-to-light ratio regime (see Sect.\,\ref{s3_ML_ratio}) with respect to the
underlying MaxBCG cluster sample used for the calibration of the $L_{200}$ relation.

%~~~~~~~~~~~~~~~~~~~~~~~~~~~~~~~~~~~~~~~~~
\subsubsection{Optimal optical mass tracer}

%correct by 30\%

\citet{Reyes2008a} also tested the possibility of an improved optical mass tracer by considering power law combinations
of either the richness $N_{200}$ or the total luminosity $L_{200}$ with the BCG luminosity $L_{\mathrm{BCG}}$. The
resulting optimal optical mass tracer with an improved scatter of the form (in units of $h^{-2}_{70}\,\mathrm{L_{\sun}}$ and
$h^{-1}_{70}\,\mathrm{M_{\sun}}$)
%\new{Daniele: check}

%Best combined mass tracer:
\begin{eqnarray}\label{e3_best_optM}
    M^{\mathrm{opt}}_{200} =
    f^{\mathrm{mean}}_{\mathrm{cr}} (1.81 \pm 0.11) \left(\frac{N_{200}}{20}\right)^{1.20 \pm 0.09} \\ \nonumber
    \cdot \left(\frac{L_{\mathrm{BCG}}}{\bar{L}_{\mathrm{BCG}}(N_{200})}\right)^{0.71 \pm 0.14}   \times 10^{14}
    \mathrm{M_{\sun}}
\end{eqnarray}

\noindent points towards an anti-correlation of the richness with the BCG luminosity at fixed cluster mass. Here
$\bar{L}_{\mathrm{BCG}}(N_{200})$ is the mean local BCG luminosity at a given richness $\bar{L}_{\mathrm{BCG}}(N_{200})\!=\!3.14\!\cdot\!N^{0.41}_{200}\!\times\!10^{10} \mathrm{L_{\sun}}$. Using the determined parameters for XMMU\,J1230.3+1339 with the
non-evolved BCG luminosity $L^{\mathrm{r}}_{\mathrm{BCG}}\!\simeq\!2.9\!\times\!10^{11}\,\mathrm{L_{\sun}}$ and $N_{200}\!\simeq\!57$ we obtain
an optimal optical mass estimate of $M^{\mathrm{opt}}_{200}\! =\!(8.6 \pm 1.3 \, (^{+2.8}_{-3.9}))\!\times\!10^{14}
\mathrm{M_{\sun}}$ (ME\,9). The systematic uncertainties here include the richness systematics of Sect.\,\ref{s3_opt_richness}
and  the passive evolution offset discussed in Sect.\,\ref{s3_BCG}, which would decrease the total mass by 32\%.

%................................................
\subsubsection{Mass-to-light ratio}
\label{s3_ML_ratio}

With our available LBT wide-field imaging data, we have an independent way to confirm the high optical luminosity of
the cluster using the statistical background subtraction method. This approach does not include a color cut, i.e. also
galaxies outside the red ridgeline region are counted in a statistical sense. We focus here on the uncontaminated
$R_{500}$ region of the cluster and determine the background luminosity in the 17 external regions %of the same size
as in Sect.\,\ref{s3_opt_richness}, taking all galaxies with $m\!\ge\!m_{\mathrm{BCG}}$ into account. This way we measure
a total background subtracted apparent z\arcmin -band magnitude of $m^{\mathrm{tot,z}\arcmin}_{500,\mathrm{ap}}\!=\!16.79$\,mag within the $R_{500}$ aperture, corresponding to a restframe
luminosity of $L^{\mathrm{tot,z}\arcmin}_{500,\mathrm{ap}} = (1.06 \pm 0.16)\times 10^{13} \mathrm{L_{\sun}} $ (see Sect.\,\ref{s3_BCG}). The integrated apparent magnitude of the cluster is comparable to the total foreground/background light contribution, implying %meaning
that about half of all detected z\arcmin -band photons in the $R_{500}$ region originate from XMMU\,J1230.3+1339. Only about
6\% of this luminosity is contributed by faint galaxies with magnitudes $m^{\mathrm{z}\arcmin}>22.9$. The red ridgeline galaxies of
Sect.\,\ref{s3_Opt_Lumin} account for approximately 61\% of the galaxy light with $m^{\mathrm{z}\arcmin}\le 22.9$ in the $R_{500}$
region. 

Although the system is at  $z\!\sim\!1$, the contrast of the cluster's z\arcmin -band light with respect to the integrated background is sufficiently high to allow an evaluation of  the spatial light distribution in the core region and some surrounding substructures without any particular pre-selection of member galaxies. As before, we merely mask out stars and galaxies brighter than the BCG ($m\!<\!m_{\mathrm{BCG}}$) in the sky-subtracted z\arcmin-band image and apply Gaussian smoothing with a 180\,kpc (22\arcsec) kernel over the field. The resulting observed z\arcmin-band  light distribution of the cluster field is shown in the lower left panel of Fig.\,\ref{fig_MultiLambda_DensityMaps}, displayed with 6 linear spaced contours corresponding to significance levels  of $2\sigma$-$12\sigma$ above the background. %In this representation 

%Using the final total mass estimate of the cluster (Sect.\,\ref{s4_masses}), we can derive the z'-band mass-to-light
%ratio in the $R_{500}$ region as %\new{revise}

Next, we can derive the total z\arcmin-band mass-to-light ratio within the $R_{500}$  aperture using the final best mass estimate of the cluster (see Sect.\,\ref{s4_masses})

\begin{equation}\label{e3_ML_ratio}
    \left( \frac{M}{L^{\mathrm{z}\arcmin}} \right)_{500,\mathrm{ap}} \simeq \frac{f_{\mathrm{3d} \rightarrow \mathrm{ap}} \cdot M^{\mathrm{best}}_{200}}{f_{500 \rightarrow 200} \cdot L^{tot,z\arcmin}_{500,\mathrm{ap}}} =
(46.7 \pm 11.3) \frac{\mathrm{M_{\sun}}}{\mathrm{L_{\sun}}} \ .
\end{equation}

\noindent 
Here $f_{\mathrm{3d} \rightarrow \mathrm{ap}}\!=\!M_{500,\mathrm{ap}}/ M_{500}\!\simeq\!1.65$ is the ratio of the aperture mass $M_{500,\mathrm{ap}}$ within  a cylinder with radius $R_{500}$ and the three dimensional spherically enclosed mass $M_{500}$, following the projection formula 
provided in \citet{Jee2005b} for the derived single $\beta$-model parameters of the cluster.
In order to compare this rather small value with locally determined mass-to-light ratios of clusters, we have
to take evolution effects into account, since the optical luminosity is expected to undergo significant
changes over a time span of 7.6\,Gyr due to the increasing mean age of the underlying stellar populations.
%Under the assumption that light traces mass, the discussed cluster radius
%evolution will increase the total cluster mass in the same way as the total light and should hence not %significantly
%alter the mass-to-light ratio. On the other hand, t
%The absolute optical luminosity is expected to undergo significant
%changes over a time span of 7.6\,Gyr due to the increasing mean age of the underlying stellar populations. 
A suitable approximation for this change in the integrated luminosity is the passive evolution model, since the total light is
dominated by galaxies on the red ridgeline or very close to it in color space. This model predicts a dimming of the
absolute z\arcmin-magnitude by 0.7\,mag, equivalent to a decrease of the luminosity by a factor of 1.9 at $z\!=\!0$,
resulting in an %increased
evolved mass-to-light-ratio of  $(M/L^{\mathrm{z}\arcmin})^{z=0}_{500,\mathrm{ap}}\!\simeq\!89\,(\mathrm{M_{\sun}}/\mathrm{L_{\sun}})$. The transformation into the
V-band leads to an additional increase to $(M/L^{\mathrm{V}})^{z=0}_{500,\mathrm{ap}}\!\simeq\!184\,(\mathrm{M_{\sun}}/\mathrm{L_{\sun}})$, i.e. to a mass-to-light ratio
% rather normal
%M/L ratio %within the
in the lower half %part 
of the  $M/L$ parameter region spanned by other massive local clusters \citep[see e.g.][]{Popesso2007a}.
% Lz/LV = 2.12 at z=0

%Combining both effects of the standard evolution scenario results in an increased mass-to-light-ratio of a factor of
%2.77 by $z=0$, i.e. $(M/L^{z'})^{z=0}_{500} \simeq 83 (M_{\sun}/L_{\sun})$.

%the total cluster mass according to the self-similar model $M(z=0) \simeq E^{2/3}(z)\cdot M_{z} \simeq 1.44 \cdot
%M^{\mathrm{best}}_{200}$.

%M/L ratio profile out to R=800kpc
%M(r) from X-ray
%L(r) from z-band

%~~~~~~~~~~~~~~~~~~~~~~~~~~~~~~~~~~~~~~~~~
\subsubsection{Velocity dispersion}
\label{s3_vel_dispersion}

Under the assumption of virial equilibrium a dynamical mass estimate can be obtained from a measurement of the
cluster's radial velocity dispersion $\sigma_r$ \citep[e.g.][]{Finn2005a,Kurk2009a}

\begin{equation}\label{e3_vel_dispersion}
M_{200} \simeq \frac{3 \,\sigma^2_r \,R_{200}}{G} \simeq \frac{1.7}{E(z)}
\left(\frac{\sigma_r}{1000\,\mathrm{km\,s^{-1}}}\right)^3 \times 10^{15} \, h^{-1}_{70} \, \mathrm{M_{\sun}} \ .
\end{equation}

\noindent We determined the velocity dispersion for XMMU\,J1230.3+1339 using the `robust' estimator of
\citet{Beers1990a} and \citet{Girardi1993a} applied to the 13 secure cluster members\footnote{This method yields a
slightly lower estimated cluster redshift of $z_{\mathrm{cl}}=0.9737$ compared to the median redshift $\bar{z}=0.9745$
of Sect.\,\ref{s2_VLT_Spectr}, but still well within the 1\,$\sigma$ uncertainty. The first value is likely biased low
due to coherently infalling sub-structure (see Sect.\,\ref{s3_CosmicFilaments}), so that the median value is preferable
as cluster-centric reference point.} in Table\,\ref{tab_specmembers} and found $\sigma_r \simeq 658 \pm
277$\,km\,s$^{-1}$ (Fig.\,\ref{fig_veldistr}) in good agreement with the classical dispersion measure of
\citet{Danes1980a}. 
%The uncertainty was estimated using Jackknife resampling of the data \new{Mike: check???}. 
From Equ.\,\ref{e3_vel_dispersion} we derive a formal dynamical cluster mass estimate of $M^{\sigma_r}_{200}\!\simeq\!2.8^{+5.2}_{-2.3}\!\times\!10^{14} \mathrm{M_{\sun}}$ (ME\,10).

% z_bi = 0.9737

Although the statistics of available galaxy member redshifts is low with corresponding large uncertainties, the general situation for a first dynamical mass estimate based on about a dozen redshifts  is typical for most optical/infrared selected distant galaxy clusters \citep[e.g.][]{Stanford2005a,Eisenhardt2008a,Wilson2009a,Muzzin2009a}. % and is hence worth a closer look. 
While at local redshifts dynamical mass estimators  can provide robust measurements  for relaxed clusters even for rather small %redshift 
spectroscopic samples   \citep[e.g.][]{Biviano2006a},  high redshift systems are particularly prone to be affected by potentially significant biases.
In the case of XMMU\,J1230.3+1339 the fairly low velocity dispersion estimate  compared to similarly rich and massive local clusters becomes plausible in the light of the large-scale structure and component analysis in Sect.\,\ref{s3_LSS}. 
More than half of the available redshifts are likely to be attributed to galaxies residing in coherently  infalling structures, i.e. they are not virialized tracer particles of the underlying potential well.
%In the case of XMMU\,J1230.3+1339,
%4/13 secure redshifts are associated with galaxies located beyond the nominal $R_{200}$ radius of the cluster
%(Fig.\,\ref{fig_veldistr}) and 
What is effectively measured is hence a combination of the `fingers-of-God effect', corresponding to the virialized
tracer particles the mass estimate is based upon, and the prolate ellipsoid  in redshift space due to the coherent
infall of the surrounding large-scale structure \citep[e.g.][]{Guzzo2008a}. The effect of this latter systematic
velocity dispersion bias depends on the angle of the targeted bulk matter flows with respect to the plane of the sky, which can suppress the measured velocity dispersion in the case of coherently infalling substructures mainly in the transversal direction (see Fig.\,\ref{fig_3Dreconstr}), or boost up the dynamical mass estimates for radial coherent flows.

\subsubsection{Spatial galaxy distribution}

As the final optical result discussed in this work we derive a spatial density map of the red ridgeline galaxies
selected with the previously applied R$-$z color cut as shown in Fig.\,\ref{fig_CMD}. The density map was constructed
by marking the world coordinate positions of the 88 selected galaxies over the FORS\,2 field-of-view and the subsequent
application of an adaptive smoothing filter. % with a varying kernel size of
The resulting spatial red ridgeline galaxy distribution  is displayed in the upper left panel of
Fig.\,\ref{fig_MultiLambda_DensityMaps} with log spaced color cuts and contour levels analogous to the representation
of the X-ray surface brightness in the upper right panel. Going from the lowest to the highest contour level
corresponds to a galaxy density increase by a factor of 26, with the peak density coinciding with the X-ray center marked
by the white cross. Within the $R_{500}$ region of the cluster, two additional galaxy concentrations towards the
North-West and South-East from the center are obvious. Along the same axis, galaxy overdensities can be traced out to
beyond the nominal cluster radius. These components and features will be further discussed in the multi-wavelength view
of the cluster in Sect.\,\ref{s4_multilambda_view}.

\subsection{Weak lensing results}
\label{s3_WL}
% own subsection

The projected total mass surface  density is observationally accessible through the method of weak gravitational
lensing. We have used the deep multi-band LBT imaging data of Sect.\,\ref{s2_LBTimaging} to perform a full weak lensing
analysis in the field of XMMU\,J1230.3+1339, which is presented in detail in the accompanying Paper\,II. 
%paper of \citet{Lerchster2010a}.
%Lerchster et al. (in prep.).
For obtaining a complete multi-wavelength view on the cluster, we summarize here some relevant results of the analysis for this work.

\begin{figure*}[t]
   \centering
    \includegraphics[width=16.0cm, clip]{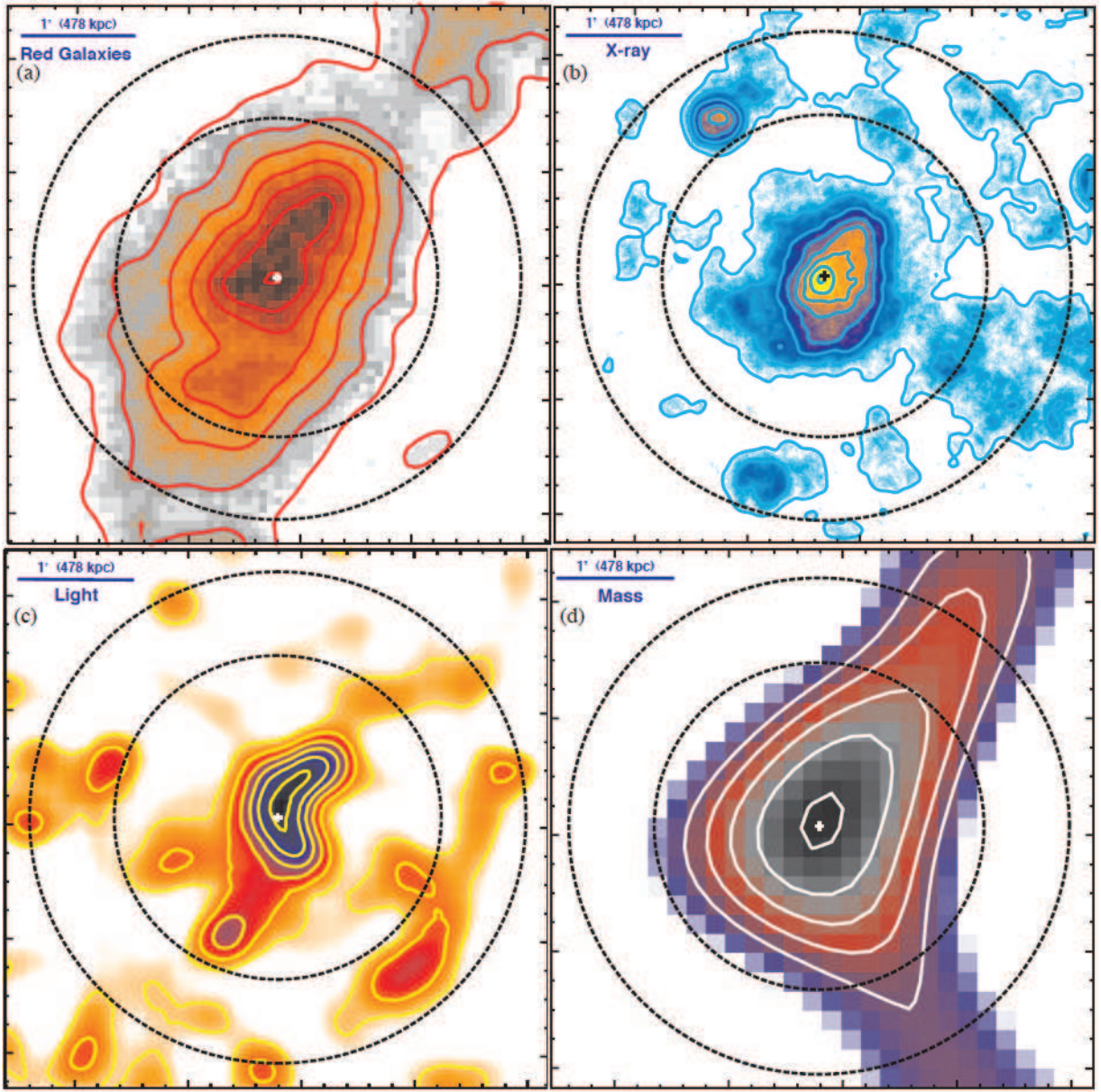}
  %%% \includegraphics[width=16.0cm, clip]{ss15204fg9.eps}
    % \includegraphics[width=8.0cm, clip]{v5_R200_RSgals.ps}
    %\includegraphics[width=8.0cm, clip]{v5_R200_Chandra.ps}
    %\includegraphics[width=8.0cm, height=8.0cm, clip]{v5_R200_zLum.ps}
    %\includegraphics[width=8.0cm, clip]{v5_R200_WL.ps}
%   \begin{overpic}[width=8cm, clip=true]{15204fg9a.ps}
%    \put(2,86){(a)}
%    \end{overpic}    
%   \begin{overpic}[width=8cm, clip=true]{15204fg9b.ps}
%    \put(2,86){(b)}
%    \end{overpic}    
%   \begin{overpic}[width=8.0cm, height=8.0cm, clip]{15204fg9c.ps}
%    \put(2,86){(c)}
%    \end{overpic}    
%   \begin{overpic}[width=8cm, clip=true]{15204fg9d.ps}
%    \put(2,86){(d)}
%    \end{overpic}    
      \caption{A pan-chromatic view of XMMU\,J1230.3+1339 seen through projected surface density maps  of different
      physical quantities within the cluster's $R_{200}$ region (outer dashed circle). All panels show the same $4.8\arcmin\!\times\!4.8\arcmin$ sky region
      in standard orientation (North is up, East to left), the nominal cluster center (central
      cross) and $R_{500}$ (inner dashed circle) are also indicated.
      {\em Top left (a):} Density map of color-selected red galaxies (log-spaced contours).
      {\em Top right (b):} {\it Chandra} X-ray surface brightness map (log-spaced contours). 
      %ping the projected square of the ICM gas density (log-spaced contours).
      {\em Bottom left (c):} Total z\arcmin-band  light distribution map (linearly-spaced contours).
      %SZE temperature decrement map as a measure of the integrated ICM pressure along the line-of-sight (linearly-spaced contours).
      {\em Bottom right (d):} Weak lensing mass density map (linearly-spaced contours).
      %signal-to-noise map is proportional to the total projected mass, i.e. predominantly the Dark
      %Matter distribution (linearly-spaced contours). \new{add Latex labelling: a, b, c, d}
      }
         \label{fig_MultiLambda_DensityMaps}
\end{figure*}

With a redshift of $z\!\sim\!1$, XMMU\,J1230.3+1339 is currently at the feasibility limit for weak lensing studies
using ground-based imaging data. However, the available LBT data allowed the detection of a significant weak lensing
signal based on the shape distortions of background galaxies, which were %identified  through
separated from foreground objects by means of multi-band photometric redshifts.
%multi-band photometric redshifts for the reliable identification of galaxies more distant than cluster.
The resulting weak lensing signal-to-noise map, which is proportional to the projected total mass surface density, is
displayed in the lower right panel of Fig.\,\ref{fig_MultiLambda_DensityMaps}. %Similarly to the SZE map, 
The five linearly spaced contour levels span the significance range from half the peak value (1.75\,$\sigma$) to the maximum at
3.5\,$\sigma$. The location of this mass density maximum coincides with the X-ray center and the peak of the galaxy
overdensity. Furthermore, the general cluster elongation in %along 
the SE-NW direction %axis 
is also clearly visible in the weak
lensing map. The detected extension towards the North-West is significant to better than 2\,$\sigma$, whereas the
South-Western  feature is to be considered  tentatively due to the close proximity to the contaminating foreground
galaxy NGC\,4477.

% B) Density Maps
% full width plot

For the reconstruction of the total spherical 3D-mass of the cluster we fitted two parametric mass models to the
observed tangential shear profile  (see Paper\,II for details). %.\citep[see][for details]{Lerchster2010a}. 
For an assumed singular isothermal sphere
(SIS) density profile we determine the corresponding best fitting velocity dispersion of the cluster to be
$\sigma^{\mathrm{SIS}}_r\!\simeq\!1271 \pm 255$\,km\,s$^{-1}$. 
% $\sigma^{\mathrm{SIS}}_r \simeq 1034 \pm 155$\,km\,s$^{-1}$. 
Using $R_{200}$ of Sect.\,\ref{s3_Tx_R200} as outer
cluster boundary we obtain a SIS model\footnote{Based on a fit to 10 data points out to 10\arcmin.} mass estimate (ME\,11) of  
%\new{Mike: check; derived from analytic SIS model}

\begin{eqnarray}\label{e3_SIS}
    M^{\mathrm{SIS}}_{200} \simeq 4.65 \cdot \left( \frac{\sigma^{\mathrm{SIS}}_r}{1000\,\mathrm{km\,s^{-1}}} \right)^2
    \left( \frac{R_{200}}{1\,\mathrm{Mpc}} \right)  \times 10^{14} \, \mathrm{M_{\sun}} \\ \nonumber
    \simeq (7.6 \pm 3.3\,(+3.0)) \times 10^{14} \, \mathrm{M_{\sun}} \ .
     %\simeq (5.1 \pm 1.6 \, (+2.3)) \times 10^{14} \, M_{\sun} \ .
\end{eqnarray}

\noindent As a second parametric model we considered a Navarro-Frenk-White (NFW) density profile \citep{Navarro1997a}
with a concentration parameter of $c\!=\!4.0$ and a scale radius of $r_s\!=\!354$\,kpc\footnote{For $h_{70}$, slightly scaled
from the values given in Paper\,II %Lerchster et al. (2010) 
for $h_{72}$.}. These NFW parameters were determined from the measured
shear components, in good agreement with the expected values from simulations of halos of similar mass and redshift by
\citet{Bullock2001a} and \citet{Dolag2004a}. 
This best fit NFW model to the weak lensing data yields an enclosed total mass within $R_{200}$
of $M^{\mathrm{NFW}}_{200}\!\simeq\!(6.9 \pm 3.3\,(+2.2)) \times 10^{14} \, \mathrm{M_{\sun}}$ (ME\,12).
%,well consistent with the SIS model.
% finalize, and scale by 1/0.972

For consistency and robustness we used the X-ray determined value for $R_{200}$ as outer cluster boundary. However,
this way the weak lensing mass estimate is not fully independent. Indeed, the self-consistent direct  determination of  $R_{200}$  from the best fitting NFW profile yields a significantly larger radius of $R_{200}^{\mathrm{NFW}}\!\simeq\!1415$\,kpc.
%direct estimation of the $R_{200}$ radius from 
%$\sigma^{\mathrm{SIS}}_r$ using the relation in \citet{Calrberg1997a} 
%yields a significantly larger radius
%($R^{\mathrm{SIS}}\!\simeq\!1475$\,kpc), and similarly for the self-consistent determination using the NFW profile
%($R^{\mathrm{NFW}} \simeq 1415$\,kpc).
%, which is linearly linked to the total mass.
To account for this potential positive offset in the outer cluster radius, we considered the resulting mass differences
as additional systematic uncertainty.

%Using the mean value of the two parametric models we thus arrive at a final weak-lensing based total mass estimate of
%$M^{\mathrm{WL}}_{200} \simeq (5.2 \pm 1.5 \, (+1.0)) \times 10^{14} \, M_{\sun}$ (ME\,12).
%\new{revise}

%\newpage

% ----------------------------------------- Start Figs -----------------------------------------------
% 1 column width: max=88mm
% Optical Properties

%\clearpage

%\clearpage

% C) Dynamic State with Thumbnail images of the identified components
% column width plot
\begin{figure*}[t]
   \centering
   % velocity info plot
   \includegraphics[width=18cm, clip=true]{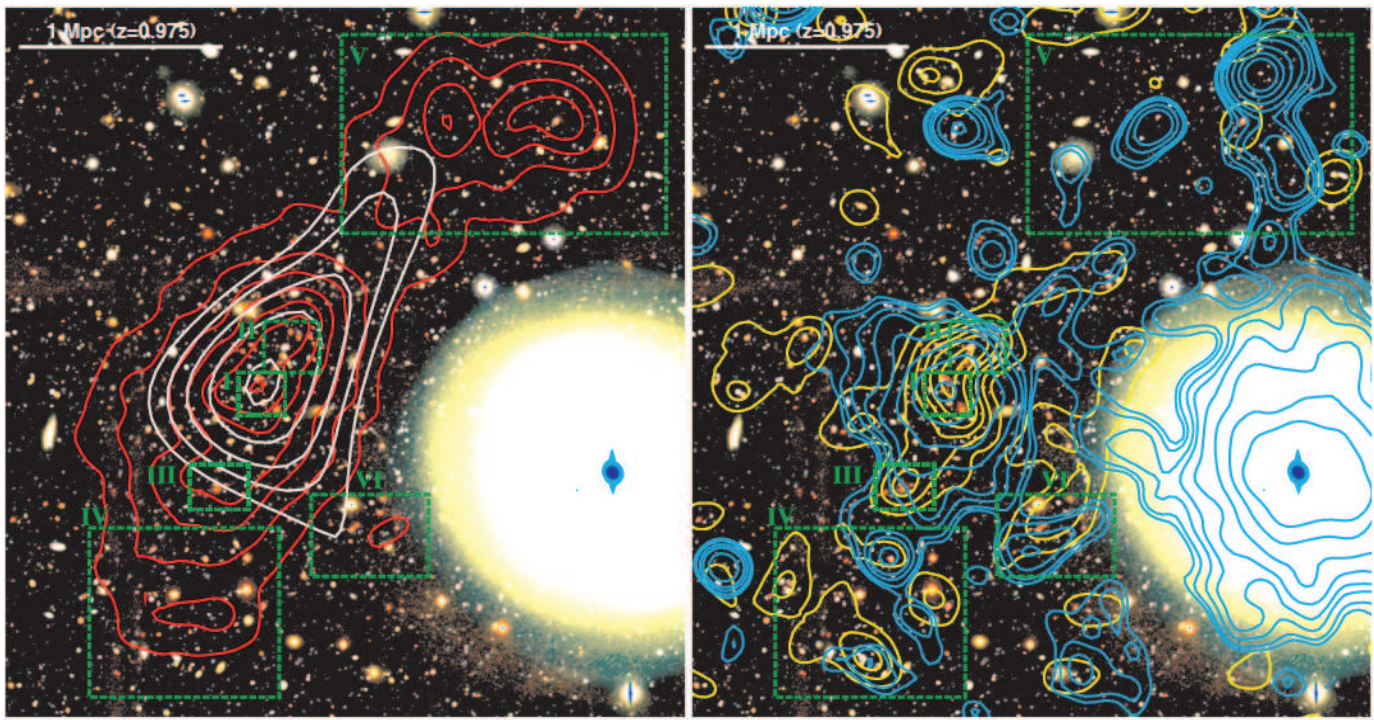}
      \caption{Cluster components and large-scale structure environment of XMMU\,J1230.3+1339 in 
      a 7.0\arcmin\,$\times$\,7.3\arcmin \ LBT color
      composite  view with different density contours overlaid. Six identified cluster components are marked 
      by dashed green boxes labelled with
      roman numerals. % {\em Top:} 7.5\arcmin\,$\times$\,7.5\arcmin \ LBT color composite image in standard orientation (North up, East to left) with red-sequence galaxy density (red) and
      {\em Left:} Contours show the galaxy densities in red and the weak lensing total mass density in white. {\em
      Right:} Same field with the %SZE temperature decrement signal overlaid 
      z\arcmin-band light distribution in yellow and XMM-{\it Newton} X-ray contours in cyan.} %dark blue.}
         \label{fig_DynamicalState}
\end{figure*}

%\clearpage

%\clearpage

%\begin{figure*}
%   \sidecaption
%   \includegraphics[width=5.8cm]{X1230flux.ps}
%   \includegraphics[angle=-90,width=5.8cm]{X1230spec.ps}
%      \caption{X-ray properties of}
%         \label{fig_Xray}
%\end{figure*}
% ----------------------------------------- End Figs ----------------------------------------------

%============== SECT 4 ===============================================================================================
%______________________________________________________________

\section{Multi-wavelength view on large-scale structure and small-scale physics}
\label{s4_multilambda_view}

\noindent We now combine the results derived for the different cluster components (i.e. galaxies, ICM, and Dark Matter) 
%from the individual wavelength regimes 
in order to obtain a detailed pan-chromatic view of XMMU\,J1230.3+1339.

%................................................
\subsection{Comparison of projected density maps}

Figure\,\ref{fig_MultiLambda_DensityMaps} displays the four derived spatial density maps (Sect.\,\ref{s3_results})
within the cluster's $R_{200}$ region (outer dashed circle), corresponding to an angular image side length of 4.8\arcmin. The
maps are based on independent data sets or techniques and visualize different physical projected properties and matter components.
The top panels show the log-spaced density contours of the red ridgeline galaxies (left, a), based on the VLT/FORS\,2 data
of Sect.\,\ref{s2_VLT_FORS2}, and the {\it Chandra} observed (Sect.\,\ref{s2_Chandra}) X-ray surface brightness distribution
(right, b). The bottom panels display  the linearly-spaced contours  of 
%, at lower spatial resolution, the linearly-spaced contours of the APEX-SZ
%(Sect.\,\ref{s2_APEX_SZ}) temperature decrement measurements 
the total projected light distribution in the z\arcmin-band (Sect.\,\ref{s3_ML_ratio}) on the left (c) and the weak lensing signal (Sect.\,\ref{s3_WL}) on the right (d) both derived from the deep LBT/LBC imaging data of Sect.\,\ref{s2_LBTimaging} 
(see Paper\,II for details).  %\citep[see also][for details]{Lerchster2010a}. 
In terms of
physical projected cluster quantities, the WL signal is proportional to the total mass surface density (d), the 
%SZE temperature decrement  maps the integrated gas pressure $P$ along the line-of-sight 
observed z\arcmin-band flux represents the restframe B/V total stellar light (c), the X-ray surface
brightness  is a measure of the projected gas density squared $n_e^2$\ (b), and the galaxies can be
approximately considered as tracer particles of the underlying gravitational potential well (a).

The first observation is that the nominal cluster center (central cross), measured as the `center-of-mass' of the X-ray
emission detected with XMM-{\it Newton} (Sect.\,\ref{s3_XrayResults}), shows a very good correspondence to the peak signal
within a few arcseconds in all projected density maps. %(SZE map with larger spatial uncertainty). 
This implies that the
galaxies, the ICM gas, and the dark matter component are linked to the same (projected) center of their corresponding
density profiles. Note that this concordant cluster center for XMMU\,J1230.3+1339 does {\em not} coincide with the BCG
location (see lower left panel of Fig.\,\ref{fig_OpticalView}), which is often used as fiducial reference center in
optical cluster studies.

A second common feature in all maps is an elongation or extension in the SE-NW direction, which indicates the apparent
main axis for the cluster assembly and the connection to the large-scale structure filaments of the surrounding cosmic
web. Just NW of the center, all projected density maps (at lower resolution for the WL signal) show another clear extension 
feature in the core region, which is  identified with an ongoing merging event discussed in Sects.\,\ref{s3_LSS}\&\ref{s3_GroupBullet}.
%Clear projected density extensions just NW of the center are visible in the galaxy map (a), the X-ray surface
%brightness (b), and the weak lensing signal  at lower resolution (c). This feature is identified with an ongoing
%merging event discussed in Sects.\,\ref{s3_LSS}\&\ref{s3_GroupBullet}. 
The galaxy (a) and total matter overdensities (d) in the NW direction can even be traced to beyond the nominal $R_{200}$ cluster radius, indicative for the presence
of a LSS filament. In the opposite SE direction, the galaxy map (a) reveals a further density peak just inside the inner
$R_{500}$ radius, also reflected in the light map (c),  and another extension to beyond the cluster radius. 
%Interestingly, this density extension is well
%matched in the SZE map (c), although at low SNR, which only allows a tentative statement with the current data. 
%The same 

 Towards the South-Western (SW) direction from the system center, an additional but less pronounced  likely cluster feature is present in all maps. In contrast to the conclusions on the overall shape of XMMU\,J1230.3+1339 along the identified main
NW-SE axis, which was not  not influenced by NGC\,4477 (see Fig.\,\ref{fig_DynamicalState}),  the proximity to the bright foreground galaxy in the SW region  now hinders the  component analysis of the cluster in the far background and  weakens the significance %in different ways.
due to the  deteriorated achievable shape and color measurement accuracy of background galaxies and possible residual contributions to the total light and X-ray emission in this region. As a result, the measured %underlying 
South-Western red galaxy density is to be regarded as lower limit %can be expected to be underestimated 
(a), while the  X-ray emission  (b, see also the deeper XMM SNR map in Fig.\,\ref{fig_DynamicalState}, right) and total light distribution (c) could be slightly increased, and a tentative character applies to the SW  feature in the weak lensing map (d).

%Due to the deteriorated achievable shape measurement accuracy of background galaxies, a tentative character applies to the SW feature in the weak lensing map (d). The local SW peak in the stellar light distribution (c) and the X-ray emission 
%(see the deeper XMM SNR map in Fig.\,\ref{fig_DynamicalState}, right)  
% 
%The proximity to NGC\,4477 deteriorates the achievable shape measurement accuracy of background galaxies, and also biases
%the color determination of potential cluster member galaxies,
%
%A preliminary character applies to the interpretation of the SW feature in the weak lensing map (d), where the %. Here the
%proximity to NGC\,4477 deteriorates the achievable shape measurement accuracy of background galaxies, and also biases
%the color determination of potential cluster member galaxies, which can result in an underestimation of the true red
%galaxy density in map. 
%However, the conclusions on the overall shape of XMMU\,J1230.3+1339 and the identified main
%NW-SE axis are not influenced by NGC\,4477, as can be seen in Fig.\,\ref{fig_DynamicalState}.
%

%................................................
\subsection{Cluster components and large-scale structure}
\label{s3_LSS}

We turn the focus on the global view and the characterization of the dynamical state of  XMMU\,J1230.3+1339 based on
the available multi-wavelength data set. Figure\,\ref{fig_DynamicalState} displays an optical color composite with a
larger %$6.8\arcmin\!\times\!7.6\arcmin$  
7.0\arcmin\,$\times$\,7.3\arcmin \, field-of-view and the different projected density contours overlaid.  The weak
lensing signal (white) and the galaxy densities (red) are shown in the left panel, and the %SZE temperature decrement
z\arcmin-band light distribution (yellow) and the XMM-{\it Newton} X-ray emission (cyan) on the right.  % (dark blue)
For a better qualitative evaluation of low surface
brightness X-ray emission in the cluster outskirts the combined XMM-{\it Newton} data is used (Sect.\,\ref{s2_XMM_Xray})
resulting in an improved sensitivity at the cost of lower spatial resolution compared to {\it Chandra}. The lowest four cyan %blue
contours correspond to significance levels of the X-ray emission of (0.5, 1, 2, 3)\,$\sigma$ \, above the background,
whereas the central cluster contour displays a 35\,$\sigma$ %\, 
significance.

We have identified a total of six different cluster components or associated structures in the multi-wavelength maps,
with identification labels I-VI in order of decreasing significance: the cluster core (I), an ongoing merger event
towards the NW of the center (II), two infalling groups on the cluster outskirts (III \& VI), and two filament
extensions beyond the nominal cluster radius towards the South-East (IV) and the North-West (V).

% seen as a       `fly-through group bullet'; (III) an radially infalling
%      group with a second dominant galaxy (BCG2) shown in the {\em bottom left panel}; (IV) a South-Eastern extension
%      beyond the nominal cluster virial radius; (V) a North-Western filament; (VI) and an off-axis infalling group with
%      associated weak X-ray emission, see also {\em bottom right panel}.

The central cluster region, indicated by box I in Fig.\,\ref{fig_DynamicalState}, encompasses the projected density
peaks of all the main matter components, i.e. the Dark Matter dominated total mass density (white contours), the hot ICM
gas component %(blue and yellow)
(cyan),  and the cold baryon component in form of the total stellar light (yellow) or %and 
the spatial galaxy density (red). 
The region also includes the spectroscopically confirmed BCG
(ID\,02 in Table\,\ref{tab_specmembers}) at a radial distance of 18\arcsec \, from the nominal cluster center (bottom
panels of Fig.\,\ref{fig_OpticalView}). For the further discussion, we follow the flow of accreted matter onto the main
cluster halo from the largest scales towards the central regions.

%................................................
\subsubsection{Cosmic web filaments}
\label{s3_CosmicFilaments}

The North-Western galaxy filament (box V) is traced to a radial distance of more than twice $R_{200}$, %. The
where the current
limit is set by the FORS\,2 FoV, i.e. the actual overdensity of red galaxies might extend even further.  The projected
galaxy density appears to have a minimum close to $R_{200}$ and then increases again towards the local maximum in the
filament of region (V) (left panel of Fig.\,\ref{fig_DynamicalState}). Three spectroscopic cluster members (IDs
09,10,12 in Table\,\ref{tab_specmembers}) are located in the cluster-filament transition zone, %region, 
all are blue shifted
with respect to the median redshift with cluster-centric velocities in the range [-800,-1\,300]\,km/s, which is
indicative for a line-of-sight bulk flow velocity of the filament structure of $\simeq-1\,000$\,km/s. In terms of the
weak lensing mass density, the signal of an extended mass distribution is measured
% to outside the nominal cluster radius
well into the cluster-filament transition regime at a signal-to-noise level of more than two. From the X-ray side,
there are some indications for low surface brightness extensions from the cluster core towards the NW filament (right
panel of Fig.\,\ref{fig_DynamicalState}), and several X-ray point sources in box V, which may or may not be %(un)
related to the cluster environment. 
%Also the $1.5\,\sigma$ temperature decrement signal on the Western edge of box V (yellow
%contour) is currently too weak to allow a physical interpretation.

The opposite galaxy extension towards the South-East (IV) is less pronounced and reaches about 0.5\,Mpc beyond
$R_{200}$. %The $1.5\,\sigma$  SZE decrement contour indicates a possible elongations in this directions, while the weak
%lensing mass map does not show a significant signal. 
The association of the three X-ray sources with corresponding light overdensities in box IV to the
filament can currently not be probed with the limited spectroscopy available, which includes one secure member galaxy
(ID 11 in Table\,\ref{tab_specmembers}) matching the median redshift, i.e. with vanishing cluster-centric line-of-sight
velocity. The tentative interpretation of a galaxy filament along the plane of the sky is fostered by the similar
cluster-centric velocity of %the 
BCG2 (ID 08) as part of the infalling galaxy group III just inside $R_{500}$.

\begin{figure}[t]
   \centering
   \includegraphics[width=7.8cm, clip=true]{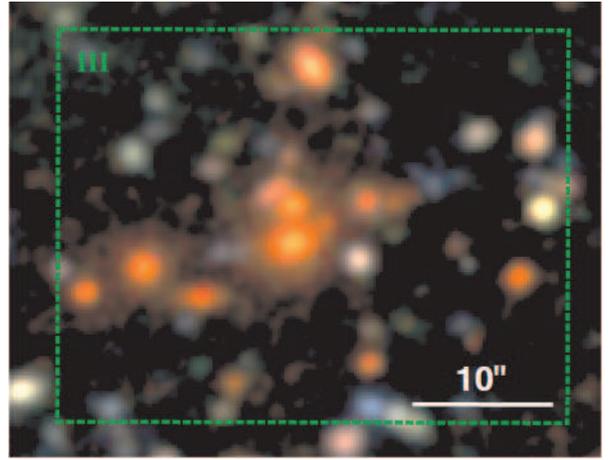}
      \caption{Closeup view on the infalling BCG2 group (III). % shown without any contours.      
      % ({\em top panel}) , and
      %component VI ({\em bottom panel}) on the cluster outskirts, displayed with XMM-Newton (blue)
      %and galaxy density contours (red).
       }
         \label{fig_InfallingGroups}
\end{figure}

%................................................
\subsubsection{Infalling groups on the cluster outskirts}
\label{s3_InfallingGroups}

%The 
BCG2 (see Sect.\,\ref{s3_BCG}) appears to be the central galaxy of an intact %and still bound 
group environment  approaching the cluster center from the SE. 
%The top panel of 
Figure\,\ref{fig_InfallingGroups} displays a closeup view on this group
component (III), which still seems %appears 
to be a bound sub-system without %and shows no 
signs of strong interactions with the main cluster components yet. 
The very low cluster-centric velocity of %the 
BCG2 points at a radial infall orbit towards the
center of XMMU\,J1230.3+1339 along the plane of the sky with a 
current cluster-centric distance of about 560\,kpc. The BCG2 group is clearly
visible as a local peak in the red galaxy density (a) or total light (c) map %(top left panel 
of Fig.\,\ref{fig_MultiLambda_DensityMaps} and
%is located along the direction of the tentative elongation in the SZE temperature map and the 
a less pronounced
distortion in the XMM-{\it Newton} X-ray contours (right panel of Fig.\,\ref{fig_DynamicalState}). At the current mass and
spatial resolution of the weak lensing map, the BCG2 group does not produce a local mass peak feature in the weak
lensing contours, suggesting that the associated total mass of the BCG2 group is small compared to the main cluster
halo.

A second infalling group candidate at a slightly larger projected cluster-centric distance of about 900\,kpc is marked
as component VI in Fig.\,\ref{fig_DynamicalState}. 
%and as closeup view in the lower panel of Fig.\,\ref{fig_InfallingGroups}. 
This South-Western %likely
cluster component is currently the least secure one, as spectroscopic confirmation is still lacking and the discussed 
proximity to NGC\,4477 may impose a bias on the projected quantities.
Nevertheless, %three of the projected 
all four density maps show features that point towards an off-axis infall scenario of
matter also from the SW direction, but currently unconstrained along the radial %redshift 
direction. Besides the %discussed
tentative mass feature in the weak lensing map (white contours) in the direction of component VI, XMM-{\it Newton} detects X-ray emission
within this region on the $3\,\sigma$ significance level (cyan). This X-ray emission seems to be extended, 
%(lower panel of Fig.\,\ref{fig_InfallingGroups}), 
but due to the low-surface brightness a further characterization is currently not
feasible. The general, almost tangential direction with respect to the cluster center is also outlined by the coincident total light distribution of this region (yellow) and the density of red galaxies (red), %coincident with the X-ray emission. 
%However, the measured surface density of red galaxies
%shown by the red contour in region VI 
which is certainly biased low compared to the true density as a result of color
contamination from the blue foreground emission on the outskirts of NGC\,4477.
%of the background galaxies measured through the blue foreground emission on the outskirts of NGC\,4477.

%galaxy transformation processes, discussed in \citep{Moran2007a}

%................................................
\subsubsection{Central merging activity}

Moving back into the core region of XMMU\,J1230.3+1339, we now turn to the cluster component II, which extends to the
North-West of the nominal center at a projected cluster-centric distance of 100-400\,kpc.  This component features
prominently in the galaxy density (a) and total light (c) maps of Fig.\,\ref{fig_MultiLambda_DensityMaps} as linear NW prolongation 
%of the galaxies 
close to the central peak value. %(second central contour, top left panel). 
The {\it Chandra} X-ray surface brightness
map (b) reveals an associated wedge-like %cone-shaped 
region with an almost constant SB, followed by a rapid
outward SB decline at a projected distance of about 250\,kpc. This surface brightness plateau was already visible in the
azimuthally averaged radial SB profile of Fig.\,\ref{fig_SBprof} and can also be recognized at lower spatial resolution
in the XMM-{\it Newton} X-ray contours of Fig.\,\ref{fig_DynamicalState}. The weak lensing map (d) provides a view on the total
mass distribution at a further reduction of resolution, but still reveals a significant elongation of the projected
mass in the direction of component II.

The overall geometry and features in the projected density maps are %very 
similar to the merger configuration in the well studied 
`Bullet Cluster' 1E\,0657-56 at a lower redshift of $z\!\simeq\!0.3$
\citep[e.g.][]{Markevitch2002a,Clowe2006a,Bradac2006a}. In comparison, the observable signs of merging activity in
XMMU\,J1230.3+1339 occur at smaller angular scales of $\la$1\arcmin \, and correspondingly shorter projected physical
distances of $\la$500\,kpc, which implies either a steeper viewing angle with respect to the plane-of-the sky or a
merger event closer to cluster core passage.

All currently available redshifts within component II are blue-shifted with respect to the median zeropoint. Besides
one secure member (ID\,01 in Table\,\ref{tab_specmembers}) at cluster-centric line-of-sight velocity of
$\simeq-400$\,km/s, four more tentative redshifts were identified closer towards the NW front at larger blue-shifted
values of about [-700,-1300,-1300,-2800]\,km/s, shown as squares in the lower left quadrant of Fig.\,\ref{fig_veldistr}.
With the currently available spectroscopy, we can establish a negative line-of-sight bulk velocity of the central
merger component II, with a tentative best estimate for the center-of-mass velocity of the merging group of
$v_r\simeq(-800\pm 500)$\,km/s.

%\begin{figure}[t]
%   \centering
%   \includegraphics[width=7.8cm, clip=true]{v3_Pub_CIII.ps}
%   %\includegraphics[width=6.8cm, clip=true]{v3_Pub_CVI.ps}
%   %\includegraphics[width=8.8cm, clip=true]{MachCone_28.5deg.ps}
%      \caption{Closeup view on the infalling BCG2 group (III). % shown without any contours.      
%      % ({\em top panel}) , and
%      %component VI ({\em bottom panel}) on the cluster outskirts, displayed with XMM-Newton (blue)
%      %and galaxy density contours (red).
%       }
%         \label{fig_InfallingGroups}
%\end{figure}
%

% ----------------------------------------- Start Figs ----------------------------------------------
% E) 3D Reconstruction
% full width plot
\begin{figure}[t]
   \centering
    \includegraphics[width=9cm, clip=true]{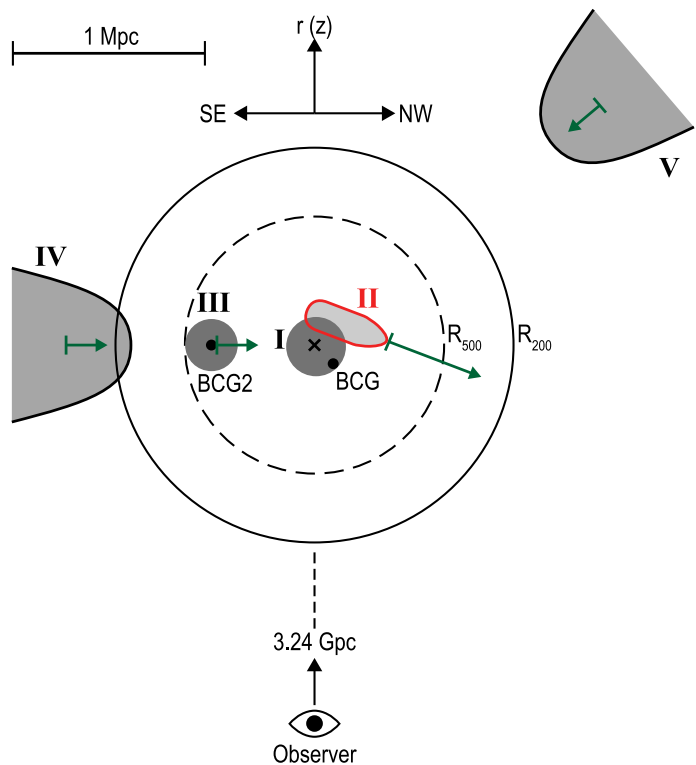}
      \caption{Sketch of the reconstructed 3D accretion geometry for the cluster environment. The horizontal direction shows a
      cut through the main SE-NW cluster axis, whereas the tentative radial (i.e. redshift space) distances  are plotted along the vertical axis for
      components I-V. The nominal cluster center is indicated by the central cross, the positions of the BCG and BCG2 by dots, and
      the reference radii $R_{200}$ and $R_{500}$ by the solid and dashed circles.
      The physical scale in the cluster reference frame is given in the upper left corner and the value in %stated 3.24\,Gpc in
      the lower part refers to the comoving distance to the cluster in the assumed cosmology. 
      Green arrows indicate  schematically the approximate traversed
      distance within 200\,Myr for linearly extrapolated trajectories under the assumption of matter infall velocities of 
      2500\,km/s for the central merging group II  (see Sect.\,\ref{s3_GroupBullet}) and fiducial reference values of 1000\,km/s for the outer components (III,IV,V).
      %1000\,km/s 
      %for the outer components (III,IV,V) and 2500\,km/s for the central merging group II.    
      }
         \label{fig_3Dreconstr}
\end{figure}

\subsection{3D mass accretion geometry}
\label{s4_3dGeometry}

Using the identified cluster components in the 2D multi-wavelength maps of %the last 
Sect.\,\ref{s3_LSS} and the available
spectroscopic information, we can reconstruct a tentative model of the three-dimensional accretion geometry of the
cluster environment of XMMU\,J1230.3+1339. In Fig.\,\ref{fig_3Dreconstr}, the horizontal directions corresponds to a
cut along the cluster's principal SE-NW axis, and the vertical direction indicates the tentatively identified radial line-of-sight 
configuration consistent with the currently available limited spectroscopic information for the individual components
I-V. Component VI is not considered here since the orientation along the redshift space direction is not constrained at
the moment.

The cluster core (component I) with the nominal center (cross) and the offset BCG (black dot) was most likely passed by
the group bullet (II) within the last 100\,Myr at a small but non-zero impact parameter %and 
under a relatively small
angle with respect to the plane-of-the-sky (see Sect\,\ref{s3_GroupBullet}). 
Judging from the undistorted galaxies in the immediate vicinity of the
nominal cluster center, and the concordant peak location in the density maps of the galaxy-, light-, ICM-, and total mass
distribution, the indications do not support the scenario of a direct hit (i.e. zero impact parameter) of the seemingly
intact cluster center. The green arrow connected to component II visualizes the approximate traversed distance of an
assumed linearly extrapolated bullet trajectory at a relative group velocity of 2500\,km/s within 200\,Myr. This
distance of $\simeq$0.5\,Mpc emphasizes the relatively short timescale of $\sim$100\,Myr of a fierce interaction of the
cluster core with the infalling group, and the implied rarity of these events to be `caught-in-the-act'.

The BCG2 group (component III) has recently entered the $R_{500}$ region of the main cluster from the South-East and
does not yet show visual signs of interactions with the main halo. The spectroscopic information indicates a radial
orbit with a minimal impact parameter. Assuming a  reasonable fiducial current infall velocity of 1000\,km/s, we obtain a 
traversed distance
of about 200\,kpc in 200\,Myr (green arrow), and a likely direct core impact within   a timescale of roughly $\sim$0.5\,Gyr. For similar
 assumed fiducial  bulk flow velocities  \citep[e.g.][]{Pivato2006a}  from the identified galaxy filaments IV and V, the timescales for matter infall to the cluster
core are approximately 1-2\,Gyr. From this first model of the 3D configuration of XMMU\,J1230.3+1339 and its
large-scale structure environment, the cluster is likely to be in a continuing active mass accretion and growth phase
for several Giga years to come.

%.........Sect 4.4 (former 4.3) .......................................
\subsection{The central group `bullet'}
\label{s3_GroupBullet}

Merging events constitute a crucial role in the galaxy cluster growth history within the hierarchical structure
formation framework. However, the in situ capture of merging events close to the cluster core passage is rare due to
the %expected
 relatively short core interaction and crossing timescale of the order of 100\,Myr. Since merger events are
important laboratories for a rich number of further detailed investigations on Dark Matter properties, shock physics of
the ICM, galaxy interactions and disruptions, and relativistic particle acceleration, we summarize the current
observational status of the new high-z group `bullet' in XMMU\,J1230.3+1339 here in order to establish a tentative
multi-wavelength reference model of the merger configuration, which is to be tested and extended with upcoming deeper %more detailed
observations.

Figure\,\ref{fig_CoreRegion} shows the currently most detailed optical view on the dynamically active region of interest
to the North-West of the nominal cluster center, with {\it Chandra} X-ray contours and the position of a VLA/FIRST radio
source overlaid.

%................................................
\subsubsection{Merging features in the ICM}

The merging process of a sub-cluster with the main halo is expected to occur at peak velocities corresponding to
typical Mach numbers of $M\!=\!v/c_s\!\simeq\!1$-$3$, where $c_s$ is the velocity of sound in the main cluster and $v$ is the
relative velocity of the colliding gas fronts \citep[e.g.][]{Markevitch2007a}. Interestingly, the expected typical
value is obtained, when interpreting the opening angle with respect to the symmetry axis of the observed wedge-like
structure in the X-ray surface brightness map as Mach cone. The dashed yellow lines in Fig.\,\ref{fig_CoreRegion}
represent the approximate observed wedges of constant surface brightness with an opening angle of $\varphi\!\simeq\!28\degr$, 
which yields $M\!=\!1/\sin(\varphi)\!\simeq\!2.1$. The currently available {\it Chandra} data does not have sufficient
depth for detailed diagnostics of the wedge-like X-ray surface brightness discontinuity, which would require
measurements of the gas density and temperature jumps across the front. However, using this typical Mach number of
$M\!\simeq\!2.1$ as a first estimate for the relative velocities, we can obtain a first approximation of the orientation
angle with respect to the plane of the sky. From the ICM temperature $T_X\!\simeq\!5.3\,\mathrm{keV}\!\simeq\!6.2\!\times\!10^{7}$\,K we derive a sound velocity of $c_s\!\simeq\!1480 \cdot (T/10^8\mathrm{K})^{1/2}\,\mathrm{km/s}\!\simeq\!1160$\,km/s, and
hence a relative velocity between merging group and cluster of $v_{\mathrm{rel}}\!\simeq\!2400$\,km/s. The combination
with the radial line-of-sight velocity component estimate of $v_r\!\simeq\!-800$\,km/s yields a first approximation for
the angle of the merger axis with respect to the plane-of-the-sky of $\simeq$ 20\degr. %, i.e. .

In analogy to the `Bullet Cluster' 1E\,0657-56, the observed wedge structure in XMMU\,J1230.3+1339 is likely a `cold
front' in the `stripping stage', i.e. a contact discontinuity on the boundary of the colder gas of the merging group
and the hotter cluster ICM expected to form close to the core passage point \citep[e.g.][]{Markevitch2007a}. The
associated bow shock, as seen in 1E\,0657-56, is expected to be detectable only on the outbound trajectory, i.e. it
might not have formed prominently yet at the particular merger stage in XMMU\,J1230.3+1339. The tentative
interpretation of a bullet-like merging event close to core passage with an observable 
`cold front' (i.e. with an outward rising temperature) would also be
consistent with the indications of an increasing ICM temperature at radii beyond the SB wedge (Sect.\,\ref{s2_XMM_Xray}).
% and the absence of features in the SZE map at the bullet location, as the pressure is quasi-continuous across a cold front
%while the temperature rises outwards. 
Moreover, in the early `stripping stages' the SB wedge opening angle could be
closer connected to the actual Mach cone, which is not the case any more for cluster 1E\,0657-56 due to the decreasing
bullet size with time and a corresponding increase in the apparent opening angle \citep{Markevitch2007a}.
A more detailed X-ray analysis of the merging physics in XMMU\,J1230.3+1339 based on deeper {\it Chandra} observations will be presented in a forthcoming paper.

%................................................
\subsubsection{Interactions of the galaxy component}

\begin{figure}[t]
   \centering
   \begin{overpic}[width=9cm, clip=true]{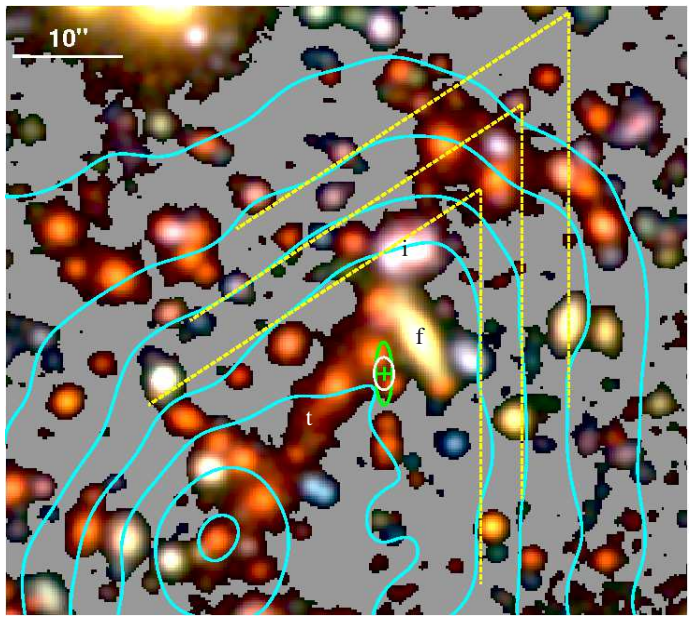}
   %\includegraphics[width=9cm, clip=true]{GrayBack60_MachCone_28.5deg.ps}
   % \put(66,1){\rotatebox{45}{\Huge\textbf{SCSA}}}
    \put(60,40){f}
    \put(58,53){i}
    %\textcolor{white}(44,28){t}
    \put(44,28){\textcolor{white}{t}}
    \end{overpic}
      \caption{%Detailed 
      Zoom on the dynamically active North-Western core region of the cluster. The $62\arcsec\!\times\!55\arcsec$
      LBT/LBC optical color composite displays the Northern part of
      core component (I) and a full view on region (II) identified in Fig.\,\ref{fig_DynamicalState}. The {\it Chandra} X-ray surface
      brightness contours (cyan) exhibit a pronounced cone-like structure with an approximate opening angle of $56\deg$ (yellow lines).
%The central red galaxies of the cone region show indications for trailing tidally stripped material (t). 
Marginally extended 1.4\,GHz radio emission is detected along the approximate symmetry axis of the configuration, displayed by the
green cross for the location, the white spatial error ellipse, and the green ellipse indicating the observed source
shape and extent. 
Character labels refer to candidate events of tidal stripping (t), galaxy-galaxy interaction (i), and a %non-cluster 
foreground galaxy (f). 
To enhance the contrast of low surface brightness features in the optical image, a smoothing with a
0.7\arcsec \ Gaussian kernel was applied to the data, and the black background was remapped to gray scale.}
         \label{fig_CoreRegion}
\end{figure}

Indications for seeing a snapshot of an early merger stage close to core passage %compared to  1E\,0657-56  
also arise from a closer look at the galaxy component in Fig.\,\ref{fig_CoreRegion}. The front end of the X-ray wedge
exhibits a surprisingly coincident alignment with a dense cloud of galaxies (upper right quarter of
Fig.\,\ref{fig_CoreRegion}), seemingly the leading front of the merging group's galaxy content. The practically
collisionless galaxy component has thus not yet been spatially separated from the fluid-like ICM component, as in the case of
1E\,0657-56.

Several galaxies in the central part of Fig.\,\ref{fig_CoreRegion} show signs of ongoing galaxy transformation
processes. The object with the bright core (label i) at the inner tip of the wedge  North of the yellow foreground spiral (label f)
%(see also lower right panel of Fig.\,\ref{fig_OpticalView}) 
 is a candidate event  for the %could be the result of a 
collision of two gas rich galaxies,  based on the strong distortions of the inner isophotes and the blue color. The
photometric redshift $z_{\mathrm{phot}}\!\simeq\!1.0$ of Paper\,II %\citet{Lerchster2010a} 
for this galaxy indicates a tentative
cluster membership, which is to be spectroscopically confirmed. 
A different process is seen at work in the red spectroscopic cluster member (ID01 in
Table\,\ref{tab_specmembers}) to the East of the green cross, which features a trail extending about 40\,kpc to the SE (label t).
This galaxy trail could be tidally stripped stellar material resulting from the gravitational interaction with the core
of the main halo and could end up as part of the Intracluster Light (ICL) \citep[e.g.][]{Zibetti2005a}. More detailed
studies on galaxies transformation processes in the dynamically active core region of XMMU\,J1230.3+1339 will require
high-resolution space-based imaging and spectroscopic information for the individual objects.

%tidal stripped material from outer parts of galaxies (e.g. Zibetti+2005)

%\clearpage
% D) Central Merger Component
% column width plot

%................................................
\subsubsection{Dark Matter halo properties}

The elongation in the total projected mass distribution seen in the weak lensing map and the high galaxy density
associated with the group bullet suggest that the initial mass of the merging group corresponds to a significant
fraction of the main cluster. With the available ground-based WL data the group mass cannot be further constrained at
the moment, but a reasonable assumption would be a bullet mass of the order of $10^{14}\,\mathrm{M_{\sun}}$ corresponding
approximately to a 1:4 merger scenario.

Hydrodynamical simulations to reproduce the observed properties of the merging system 1E\,0657-56 by
\citet{Springel2007a} have shown that the time dependent spatial separation of the mass peak of the bullet group
relative to the ICM and the shape of the contact discontinuity in the X-ray surface brightness depend sensitively on
the relative concentrations of the merging halos and their gas fractions. A first visual comparison of the merger
configuration in XMMU\,J1230.3+1339 to these simulations seem to be consistent with a low concentration parameter  of
$c\!\simeq\!2$-$3$ for the main halo and a time snapshot within about 100\,Myr of the closest core passage point.

%................................................
\subsubsection{Observed extended radio emission}

An additional point of particular interest is the observation of extended radio emission originating from a location
trailing the front of the infalling group very close to the approximate symmetry axis of the merger configuration. In
Fig.\,\ref{fig_CoreRegion}, the location of the detected 1.4\,GHz radio source (ID:\,J123016.1+133917) from the VLA
FIRST survey \citep{White1997a,Becker2003a} is indicated by the green cross in the center of the image. This source
with a signal-to-noise ratio of 7.2 is the only catalog entry within the field of the larger cluster environment %as
shown in Fig.\,\ref{fig_DynamicalState}. The radio source is clearly extended with good significance along the
North-South axis with a deconvolved FWHM of ($5.7 \pm 1.5$)\arcsec , but below the extent detection threshold along the orthogonal
East-West axis. The reconstructed source shape is indicated by the green ellipse in Fig.\,\ref{fig_CoreRegion}, where
the upper 1\,$\sigma$ limit of the extent along the minor axis of 1.5\arcsec \ was used. The white ellipse displays the
positional error region of the source center at 90\% confidence,  which is just small enough to exclude any of the
centers of the neighboring galaxies as origin for the radio emission at this confidence level\footnote{The nearest previously known foreground galaxy (label f in Fig.\,\ref{fig_CoreRegion}) SDSS\,J123015.88+133920.4 has a spatial separation of 4.3\arcsec \, from the radio source, corresponding to a distance of about a factor of 4 outside the  90\%  positional error ellipse of the FIRST source.}. The 1.4\,GHz radio
source has an integrated flux of ($1.69 \pm 0.19$)\,mJy, which translates into a total radio power of
$P_{1440\,\mathrm{MHz}}\!\simeq\!8.3\!\times\!10^{24}$ W\,Hz$^{-1}$ at the cluster redshift  under the assumption of a  typical
spectral index of -1 for the extended emission \citep[e.g.][]{Laing1980a}.

%................................................
\subsubsection{What is the origin of the radio emission?}

The simplest explanation for the observed radio emission close to the core region of XMMU\,J1230.3+1339 would be an
extended radio lobe associated with central AGN activity in one of the cores of the two red early-type galaxies
partially encompassed by the white positional error ellipse in Fig.\,\ref{fig_CoreRegion}.
% one is an tentative spec member, the other is not targeted; both are on RS
In fact, \citet{Dunn2010a} have found extended radio emission in more than half of the early-type galaxies of a
complete local sample of X-ray luminous E and S0 galaxies. However, only the two most radio luminous objects in this
sample (M87 \& NGC\,4696) would have comparable radio powers to the source in XMMU\,J1230.3+1339, but these are
associated with BCGs of local clusters. %, which is not the case here. 
With the currently available radio data on the field, the association of the
detected radio source with a nearby galaxy cannot be ruled out and is certainly a valid hypothesis.

%properties of ellipticals: Dunn et al. 2009, MNRAS submitted

However, the fact that we are most likely witnessing an ongoing merger event close to cluster core passage suggests %for
the consideration and discussion of an alternative scenario on the basis of turbulence-induced electron acceleration in
the wake of infalling substructure. This mechanism of turbulence (re)acceleration
\citep[e.g.][]{Brunetti2001a,Fujita2003a,Brunetti2004a} is now believed to be the source of relativistic electrons
associated with the giant radio halos observed in a growing number of massive clusters with recent merger activity such
as Coma \citep[e.g.][for reviews]{Feretti2008,Ferrari2008a}. The generation and time evolution of fluid turbulence
behind a moving subcluster can be probed and visualized with numerical hydrodynamic simulations
\citep[e.g.][]{Takizawa2005a,Dolag2008a,Vazza2009a}. The overall geometry and involved length scales seen in such
numerical studies are comparable to the observed configuration shown in Fig.\,\ref{fig_CoreRegion}, implying that
merger-generated  turbulence at the location of the observed radio emission in XMMU\,J1230.3+1339 is plausible and even
expected.

%% ----------------------------------------- Start Figs ----------------------------------------------
%% E) 3D Reconstruction
%% full width plot
%\begin{figure}[t]
%   \centering
%    \includegraphics[width=9cm, clip=true]{v3_3DGrafik.eps}
%      \caption{Sketch of the reconstructed 3D accretion geometry for the cluster environment. The horizontal direction shows a
%      cut through the main SE-NW cluster axis, whereas the tentative radial (i.e. redshift space) distances  are plotted along the vertical axis for
%      components I-V. The nominal cluster center is indicated by the central cross, the positions of the BCG and BCG2 by dots, and
%      the reference radii $R_{200}$ and $R_{500}$ by the solid and dashed circles.
%      The physical scale in the cluster reference frame is given in the upper left corner and the stated 3.24\,Gpc in
%      the lower part refers to the comoving distance to the cluster in the assumed cosmology. Green arrows indicate the approximate traversed
%      distance within 200\,Myr for linearly extrapolated trajectories under the assumption of matter infall velocities of 1000\,km/s
%      for the outer components (III,IV,V) and 2500\,km/s for the central merging group II.    }
%         \label{fig_3Dreconstr}
%\end{figure}
%
%
%% ----------------------------------------- End Figs ----------------------------------------------

An intriguing scenario would be that we are actually witnessing the injection point of re-accelerated
ultra-relativistic electrons into the ICM as the source for the later development of a giant radio halo filling a good
fraction of the cluster volume with low surface brightness radio emission. With the current radio data, we can only
perform some consistency checks to test the plausibility of this alternative hypothesis.

\citet{Cassano2007a} derived the expected scaling relations of the re-acceleration model for the total emitted radio
power with total cluster mass and other properties, which is reflected in the observed strong empirical correlations.
The latest radio power correlation measurements are provided by \citet{Brunetti2009a} for the
$P_{1440\,\mathrm{MHz}}$-$L^{\mathrm{bol}}_{\mathrm{X}}$ relation\footnote{Using the BCES Bisector parameters.}
yielding $\log(P_{1440\,\mathrm{MHz}}/\mathrm{W\,Hz}^{-1}) -24.5 = (0.077 \pm 0.057) + (1.76\pm
0.16)\,[\log(L^{\mathrm{bol}}_{\mathrm{X}}/\mathrm{erg\,s}^{-1})-45.4]$. The combination of this correlation with the
non-evolving $L^{\mathrm{bol}}_{\mathrm{X,500}}$-$M_{500}$ relation (Equ.\,\ref{e3_LX_M}) results in

\begin{equation}\label{e4_P1.4_M500_relation}
    M_{200} \! \simeq \! f_{\small{500 \rightarrow 200}} \!\left(\frac{f^{L_X}_{\small{\mathrm{tot} \rightarrow 500}}}{E^{\frac{4}{3}}(z)}
    \right)^{0.48}\! \left( \frac{P_{1440\,\mathrm{MHz}}}{8.6\!\times\! 10^{24}\,\mathrm{\small{W/Hz}}} \right)^{0.27} \!\times
    10^{15}\! \mathrm{M_{\sun}} \, .
\end{equation}
% ^{\small{1.4\,\mathrm{GHz}}}

\noindent Here $f^{L_X}_{\small{\mathrm{tot} \rightarrow 500}}\!\simeq\!0.95$ is a conversion factor from total lu\-minosity to
$L^{\mathrm{bol}}_{\mathrm{X,500}}$ and  $f_{\small{500 \rightarrow 200}}\!\simeq\!1.4$ as defined  in
Sect.\-\,\ref{s3_Tx_R200}. Applying this relation to the measured radio power in XMMU\,J1230.3+1339, we obtain a mass
estimate of $M^{1.4\,\mathrm{GHz}}_{200}\!\simeq\!(9.5 \pm 3.8)\!\times\!10^{14}\,\mathrm{M_{\sun}}$ (ME\,13). The error is dominated by the systematic
uncertainty in the redshift evolution of $L^{\mathrm{bol}}_{\mathrm{X,500}}$-$M_{500}$ (Sect.\,\ref{s3_Lx_lum}). Other
error contributions such as  the uncertainties for the total radio flux ($\pm$12\%), the K-correction from the assumed
spectral index ($\pm$20\%), and the used scaling relations are added in quadrature.

This plausibility check for the hypothesis that the observed marginally extended radio emission is related to the
origin of the radio halos in lower redshift clusters yields a mass in fairly good agreement with the best total mass
estimate for XMMU\,J1230.3+1339. Turning the argument around implies that for the inferred total cluster mass the
observed total radio power is just a factor  of a few higher than expected for the total radio halo emission, which might be
expected close to the core passage of a subcluster. Hence, the scenario of the turbulence-induced electron
re-acceleration model connected to the ongoing merging activity seems also to be a consistent alternative and should be
further pursued. Considering that current detailed measurements of diffuse radio emission in clusters are limited to
$z<0.6$ \citep{Brunetti2009a}, future low-frequency radio studies of the dynamically active core region of
XMMU\,J1230.3+1339 have the potential to gain new insights into the origin and evolution of radio halos.

\section{Discussion}
\label{s4_disc}

In the following section we discuss the dynamical state of the system and evaluate the different mass estimates and
scaling relations used to derive them. We then compare XMMU\,J1230.3+1339 to other high-z galaxy clusters and focus on
the connection to low redshift counterparts.
% and discuss future astrophysical and cosmological applications.

%................................................
\subsection{Dynamical state of XMMU\,J1230.3+1339}
\label{s4_dynstate}

The system XMMU\,J1230.3+1339 is a massive bona fide cluster with ongoing dynamical activity on all relevant scales. We
will argue in the following that this cluster, observed at a lookback time of 7.6\,Gyr, is likely a precursor of the
most massive non-cool-core clusters in the local Universe, %very 
similar to the Coma system.

In the optical, XMMU\,J1230.3+1339 (see Table\,\ref{table_results}) exhibits a galaxy population which is among the
richest of all known $z>0.9$ clusters. %(possibly even the richest).
On the X-ray side, the ICM is found to be hot, with a high luminosity and a rather extended surface brightness profile,
devoid of any detectable X-ray point source within
$R_{500}$. 
Moreover, the high total mass of the system enables %the clear detection and
complementary system diagnostics using %SZE and 
weak lensing measurements. 
The cluster center of XMMU\,J1230.3+1339 is
well defined, which indicates a virialized core region prior to the onset of the observed central merging event. The
projected density maps reveal the galaxy and total light peaks, the center of the total mass, and the X-ray surface brightness peak
%and center-of-mass
all within a few arcseconds of each other (Fig.\,\ref{fig_MultiLambda_DensityMaps}).

The location of the %BCG 
Brightest Cluster Galaxy at a projected distance of 18\arcsec \, or 140\,kpc from this nominal center
shows %, however,
a significant spatial offset. Considering the dynamical friction timescale of a few Gigayears,
spatial offsets of the BCG from the center of the main cluster halo are not surprising when observing systems in the
first half of cosmic time. However, in the local Universe the BCG location with respect to the X-ray center is known to
correlate strongly with the dynamical state of the cluster. \citet{Sanderson2009a} have shown that a large projected
BCG offset ($>0.02\,R_{500}$) implies on average a non-cool core host cluster with a shallow central ICM gas density
profile, a larger than average gas mass fraction, and very low star formation and AGN activity associated with the BCG
itself. These local ($z\!\sim\!0.2$) correlations and expectations for clusters with significant BCG offsets are
tentatively consistent with the properties of XMMU\,J1230.3+1339,  with the exception of the gas mass fraction, which is found to be close to the average local value (see Sect.\,\ref{s3_GasMass}).

The identified central merging event  gives rise to a dynamically highly active central region of the cluster with
various characteristic signatures as discussed in Sect.\,\ref{s3_GroupBullet}. Due to the  quadratic dependence of
thermal bremsstrahlung emissivity on the gas density, global ICM X-ray diagnostic measurements are expected to be
influenced most by central group mergers. In the specific case of XMMU\,J1230.3+1339, the resulting flattening of the
X-ray surface brightness within the wedge-like region in the wake of the infalling group is well visible in
Fig.\,\ref{fig_MultiLambda_DensityMaps} (top right panel). This feature is clearly a deviation from the underlying
assumption of spherical symmetry and results effectively in a less reliable radial profile fit
(Sect.\,\ref{s2_Chandra}) with an increased core radius $r_c$ and slope parameter $\beta$ with respect to the
pre-merger configuration. Additionally, significant deviations from an assumed hydrostatic equilibrium state
(Sect.\,\ref{s3_XrayResults}) can be expected in the core region. The local ICM temperature could be lowered due to the
cooler gas of the infalling group core, or temporarily boosted in the bow shock regions in conjunction with the X-ray
luminosity. However, in the light of total cluster mass proxies of Sect.\,\ref{s4_masses}  this merging event does not seem to affect the results significantly. The interaction region is rather to be regarded as a localized perturbation with little influence on the still well defined cluster center and only mild effects on global quantities. 

Standard cool core diagnostics classify XMMU\-J1230.3+1339 in its current state as a Non-Cool-Core cluster
(Sects.\,\ref{s2_Chandra}\,\&\,\ref{s3_XCoreProper}), albeit the relatively short central cooling time scale. 
Any possible previous onset for the development of  %of a  cool core, observable e.g as 
a centrally peaked surface brightness profile was already flattened out
by the group `bullet' (component II). The foreseeable core impact of the BCG2 group (component III) will likely further 
disrupt the dense
cluster center and hence erase the  central conditions required for the progress of significant cooling. This scenario
is consistent with the emerging picture from simulations on the different formation histories of NCC and CC clusters.
The simulations of \citet{Burns2008a} showed that Non-Cool-Core clusters are characterized by an increased mass
accretion rate ($\ga$50\% per Gyr) in early epochs ($z\!\ga\!0.8$), during which nascent cool cores are destroyed by major
merger events which also set the conditions to prevent cooling at later epochs. 
%Furthermore, these simulations suggest
%that the hydrostatic mass estimates for NCC at $z\!\sim\!1$ could be biased low by 15-20\%, albeit with large scatter.
%

%This apparent contradiction \new{?revise} can be explained by assuming that the observed ongoing merging event
%discussed in Sect.\,\ref{s3_GroupBullet} flattens out the central surface brightness profile (see
%Fig.\ref{fig_MultiLambda_DensityMaps}), and hence destroys the onset of a cool core. This scenario is predicted in
%\cite{Burns2008a} and can be observed in situ in XMMU\,J1230.3+1339.

%................................................
\subsection{Comparison of mass estimates and scaling relations}
\label{s4_masses}

% F) Mass Comparison
% column width plot
\begin{figure}[t]
   \centering
   \includegraphics[width=9cm, clip=true]{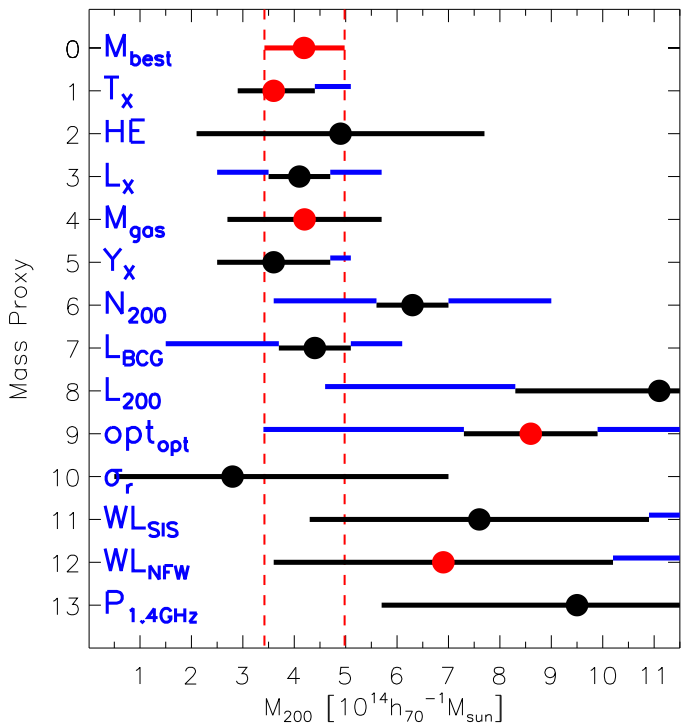}
      \caption{Comparison of different mass proxies for cluster XMMU\,J1230.3+1339. The combined mass estimate
      $M^{\mathrm{best}}_{200}\!\simeq\!4.19^{+0.79}_{-0.77}\times 10^{14}\,\mathrm{M_{\sun}}$, shown on top with $1\,\sigma$ uncertainties
      (dashed lines), is based on the combination of four quasi independent mass proxies (red points). The numbering along the vertical axis
      is equivalent to the mass estimate labels (ME\,1-13) in the text. Statistical $1\,\sigma$ uncertainties are shown by the black
      horizontal lines. The magnitude and direction of potential systematic offsets are displayed by the slightly shifted blue bars
      extending the statistical error for illustration purposes.}
         \label{fig_mass_comparison}
\end{figure}

In Sects.\,\ref{s3_results} \,\&\,\ref{s3_GroupBullet}, we have derived a total of 13 different mass proxies for
XMMU\,J1230.3+1339, which will be compared and discussed in the following.

%-----------------------------------------------------
\subsubsection{Combined mass estimate}

We have applied %four 
different and independent observational techniques to study and characterize the cluster 
%as summarized in the multi-wavelength view of Fig.\,\ref{fig_MultiLambda_DensityMaps}: 
based on (i) optical imaging data,
(ii) X-ray observations, %, (3) Sunyaev-Zeldovich effect measurements, 
(iii) weak gravitational lensing, and (iv) spectroscopic information for a first glimpse on the system dynamics. The objective in this
section is to combine several suitable measurements in order to obtain a (preliminary) best total mass estimate
$M^{\mathrm{best}}_{200}$ as a benchmark value and starting point for a comparison of the different high-z mass proxies.

We start by selecting four reliable, quasi-independent mass estimates from %each of the four 
different observational techniques as input for a combined mass estimate.
On the X-ray side, the concordant {\it Chandra} and XMM-{\it Newton} ICM temperature estimate  provides 
tight mass constraints based on the $T_X$-$M$ (ME\,1) scaling relation, which also encompasses a reliable redshift evolution prediction. 
%relation is favorable (ME\,1), since we have a well
%constrained, concordant Chandra and XMM-Newton measurement as input and a reliable redshift evolution prediction of the
%scaling relation. 
The {\it Chandra} determined $\beta$-model SB structure parameters were the basis for the gas mass measurements of the system 
%entered in the $Y_{\mathrm{SZ}}$ determination and hence in the $Y_{\mathrm{SZ}}$-$M$ mass proxy (ME\,11). 
and the corresponding $M_{200}^{\mathrm{gas}}$ estimate (ME\,4).
This constitutes a second reliable and quasi-independent  mass measurement,
since the cooling function for the soft %(0.5--2\,keV) 
X-ray band is practically insensitive to the ICM temperature (few per cent changes over the relevant $T_X$ range).
%since the ICM temperature entered only through a square root dependence in the cooling function and in the outer boundary $R_{500}$.
%, with a small influence on the total $Y_{\mathrm{SZ}}$ flux. 
Similarly, the temperature based $R_{200}$ was chosen as cluster boundary for the weak lensing
measurements, which was regarded as more robust %reliable 
compared to the self-consistent approach based on the current model
constraints from the ground-based weak lensing signal with larger uncertainties. We choose the NFW model estimate
(ME\,12) as WL input mass proxy for the $M^{\mathrm{best}}_{200}$ determination. As fourth reference mass, the
`optimal optical' mass tracer is considered (ME\,9).

Besides the statistical uncertainties arising from measurement errors and the scatter in the used scaling relations, we
also attempted to approximately quantify the influence of potential systematic offsets on the mass estimators. These
latter uncertainties were due to e.g. smaller analysis apertures compared to the used scaling relations ($T_{X}$), the
resulting influence on the derived cluster radii (WL), significant uncertainties in the redshift evolution
($L_{X}$, $L_{\mathrm{BCG}}$, optical apertures), and technical %uncertainties
ambiguities  arising from the adaptation of local
measurement recipes to high redshift ($N_{200}$, $L_{200}$). These technical uncertainties are in principle independent
from the statistical errors, however, their (unknown) statistical occurrence may not follow a Gaussian probability
distribution and are hence difficult to properly account for. As a first order approximation, we assume Gaussian
properties of the systematic technical uncertainties and add them in quadrature to the statistical error intervals to
obtain mass probability distributions for the individual proxies. The four\footnote{The velocity dispersion-based dynamical mass estimate (ME\,10) with the current uncertainty range has practically no statistical weight, its inclusion for the combined mass would only have a per cent effect.} quasi-independent mass reference proxies are
then combined with a maximum-likelihood approach, resulting in a final cluster mass estimate of
$M^{\mathrm{best}}_{200}\!=\!4.19^{+0.79}_{-0.77} \times 10^{14}\,\mathrm{M_{\sun}}$ with specified $1\,\sigma$ uncertainties.
This currently most likely total mass interval for XMMU\,J1230.3+1339 is shown in the mass proxy comparison plot in
Fig.\,\ref{fig_mass_comparison} (vertical dashed lines), together with the input reference proxies displayed by the red
filled circles, and all other mass estimates (black circles) with their corresponding uncertainties.

%-----------------------------------------------------
\subsubsection{Multi-wavelength mass proxies and challenges at high-z}

The combined cluster mass proxy $M^{\mathrm{best}}_{200}\!\simeq\!4.2 \times 10^{14}\,\mathrm{M_{\sun}}$ with $1\,\sigma$
uncertainties of about 19\% is currently the best reference value for XMMU\,J1230.3+1339  and is among the most accurate cluster
masses measured to date at $z>0.9$.
%one of the most
%constrained and least biased mass estimates for any known 
%$z>0.9$ galaxy cluster.

The X-ray derived mass estimates (ME\,1-5) provide a consistent picture of the cluster with mostly well constrained
mass intervals based on the medium deep {\it Chandra} and XMM-{\it Newton} data. The systematic uncertainties on $T_X$ (ME\,1)
arise from the unconstrained temperature structure of the ICM and the possible influence of the central merging event.
The constraints from the hydrostatic estimate (ME\,2) are mostly limited by the uncertainties in the
$\beta$-parameter, which is certainly influenced by the fly-through group bullet in addition to possible deviations
from the hydrostatic assumption in the central regions. The uncertainties of the gas mass (ME\,4) and $Y_X$ (ME\,5)
estimates are mostly limited by the depth of the available X-ray data. The cluster property with the smallest relative
uncertainty (see Table\,\ref{table_results}) is the X-ray luminosity $L_X$, which translates into small statistical
mass estimate errors based on the latest (local) $L_X$-$M$ scaling relation. However, a significant uncertainty is
added because of  %by 
the limited knowledge on the high-z evolution of the related $L$-$T$ scaling relation (see
Sect.\,\ref{s3_Lx_lum}). For the very good concordance of the $L_X$ based mass estimate with $M^{\mathrm{best}}_{200}$,
the $L_X$-$T$ relation was assumed to be non-evolving, which is different from the self-similar evolution expectations.
 However, such a breaking of self-similarity for the $L_X$-$M$ scaling in terms of a slower redshift evolution and a steeper slope of the relation is in  qualitatively good agreement with the latest results from the Millennium Gas Simulations after the inclusion of non-gravitational (pre-)heating mechanisms  \citep{Stanek2010a}.
Once an improved  and self-consistent calibration of the $L_X$ %-$T$ 
scaling relations at $z>0.9$ is available, the X-ray luminosity appears to
have the potential to be a good high-z mass proxy at relatively low observational cost.

%The mass estimate based on the Sunyaev-Zeldovich effect measurements of $Y_{\mathrm{SZ}}$ profitted from the very
%recent improvements of the empirical calibration of the $Y_{\mathrm{SZ}}$-$M$ relation and the availability of
%Chandra-measured $\beta$-model profile parameters. As expected from numerical simulations, this mass proxy has a %very
%high potential, which is reflected in our case in the very good match with $M^{\mathrm{best}}_{200}$. However,
%self-consistent accurate SZE mass estimates, i.e. with all parameters extracted from the SZE signal, will require much
%deeper exposure times. Further empirical support for the potential of the $Y_{\mathrm{SZ}}$-$M$ (and X-ray) scaling
%relation is provided by \citet{Marrone2009a}, who find no evidence for a hydrostatic mass bias in cluster cores based
%on their cross-calibration with weak lensing data.

Weak lensing mass estimates are potentially a %very 
powerful tool to provide independent low-bias measurements to make progress in the calibration of mass-observable
relations. However, the weak lensing analysis of XMMU\,J1230.3+1339  in Paper\,II %by \citet{Lerchster2010a} 
is at the current
feasibility limit for ground-based imaging observations, implying fairly large statistical uncertainties due to the
limited lensing signal strength. $M^{\mathrm{best}}_{200}$ is on the lower mass end of the uncertainty range but still
%The uncertainty range of $M^{\mathrm{best}}_{200}$ is on the lower mass end but still
%within the $1\,\sigma$ 
consistent within the errors with the WL mass proxies (ME\,11\,\& \,12). Significant progress and a self-consistent
parameter treatment in the WL analysis will require the availability of deep spaced-based imaging data.

The dynamical cluster mass estimate based on the measured radial velocity dispersion $\sigma_r$ (ME\,10) provides the
weakest statistical constraint for XMMU\,J1230.3+1339, due to the small dynamic range of $\sigma_r$ and the relatively
large uncertainties related to the limited number of available redshifts. % based on the limited number of available redshifts. 
The situation of having about a dozen spectroscopic cluster members %redshifts 
for a first order evaluation of the
velocity dispersion is typical under survey conditions with single  mask spectroscopic follow-up of newly discovered
high-z systems. More stringent mass constraints require large, dedicated follow-up programs to ideally increase the
number of cluster redshifts by an order of magnitude. Although %currently %not
hardly  feasible in the general case for $z\!>\!0.9$
clusters due to the magnitude limit of the available spectrographs, the high richness of XMMU\,J1230.3+1339 might
realistically enable the identification of about 100 spectroscopic cluster members. However, biases arising from
infalling structures, which could lead to positive and negative offsets depending on the line-of-sight angle, are
likely to %further 
limit the achievable accuracy of dynamical mass estimates at high redshifts.

Mass proxies based on optical/NIR imaging data have the advantage of being observationally cheap and usually readily
available for distant cluster searches or studies. Although significant progress has been made in the calibration of
optical scaling relations at lower redshifts, unbiased  mass proxies at $z>0.9$ to better than 40-50\% seem to require
significant further work to overcome  technical and  evolutionary challenges. We have obtained a consistent picture
that the optical galaxy population of XMMU\,J1230.3+1339 is rather %extreme 
remarkable for a high-z cluster, both in terms of
total galaxy numbers ($N_{200}$, R$^{\mathrm{rich}}$) and total luminosity ($L_{200}$, $L^{\mathrm{tot}}_{500}$). 
%This conclusion is based on galaxy counting and total luminosity determination along the
%richly populated red ridgeline in the  VLT/FORS\,2 imaging data set, and independently from measurements of the
%overdensity and the total cluster luminosity with statistical background subtraction in the wide-field LBT/LBC data.
Interestingly, the optical BCG luminosity (ME\,7) yields a mass estimate which is fully consistent with
$M^{\mathrm{best}}_{200}$, followed by the suggested higher system masses of the $N_{200}$ %estimate 
proxy (ME\,6), the proposed optimal optical mass tracer
(ME\,9), and the $L_{200}$ luminosity (ME\,8). In the case of XMMU\,J1230.3+1339, the optimal optical mass tracer of 
\citet{Reyes2008a} does not seem to improve the individual proxies based on $N_{200}$ and $L_{\mathrm{BCG}}$.

The use of luminosity-based mass proxies (e.g. $L_{200}$) at high-z requires as input
rather large evolution and K-correction factors to relate them to the scaling relations calibrated at lower
redshifts.
Counting galaxies along the typically well-defined red-sequence of early-type galaxies (e.g. $N_{200}$) seems to be a more
straightforward approach. % for applicability at high-z. 
However, even this apparently simple task %of counting %red cluster galaxies
gets increasingly complicated at high-z, where various subtleties and potential biases are to be taken into account if a
somewhat reliable mass estimate is to be achieved.  Even under the assumption that a distant cluster is sufficiently rich to exhibit an easily identifiable red-sequence,  three main critical technical aspects remain and may impose significant mass biases: (i) finding  the correct RS slope, (ii) the definition of appropriate color cuts, and (iii) the determination of a self-consistent measurement aperture. All three items are in principle functions of cluster redshift, and the first two  depend moreover on the applied filter combination and the depth and quality of the data. As examples of  redshift dependent biases, we can consider galaxy evolution effects in clusters, which lead to an observed increase with lookback time of the fraction of objects bluer than the red-sequence and  also an increasing relative effective age (and hence color) difference  of the average stellar populations as a function of cluster-centric radius  \citep[e.g.][]{Rosati2009a}. For a fixed  color cut  definition, a redshift dependent positive or negative richness and mass bias could thus be imposed through an increasing contamination of bluer non-passive galaxies in the red ridgeline selection, or by  excluding a growing fraction of slightly bluer passive galaxies on the cluster outskirts. Similar redshift dependent mass biases may arise through a non-consistent  scaling of the measurement apertures.

An alternative  approach applicable  in the absence of a robust estimate for the cluster radius $R_{200}$ or for direct cluster-to-cluster comparisons is to measure relevant cluster quantities inside a fixed physical projected radius of e.g. $1\,h^{-1}_{70}\,$Mpc. The hereby imposed positive mass bias with redshift  might in many cases be an acceptable trade-off for otherwise large possible uncertainties in the cluster radius. In the case of XMMU\,J1230.3+1339, the ICM temperature determined $R_{200}$ is within 2\% of $1\,h^{-1}_{70}\,$Mpc, implying effectively only percent level differences for  most of the total mass estimates (ME\,1-5, 7, 10-13), when re-scaled to the fixed physical aperture. However, for the optical mass proxies based on richness and total luminosity (ME\, 6, 8, 9) a significant reduction of the total mass estimate of the order of 30\%  applies for the fixed radius, which would move these proxies in much better agreement with the current best estimate $M^{\mathrm{best}}_{200}$.

%velocity dispersion poor mass estimator

%velocity dispersion not good estimator with survey mode spectroscopy

%galaxies outside R200 with low potential energy -- biased low
%infalling groups with non virilized velocities -- biased high

%vel disp: what is observed as a mix of the squashed redshift ellipsoid due to coherent infall and the fingers-of-God
%due to virialized region (e.g. Guzzo+ 2008)

%merging event: not HSE and thermal equilibrium in center

%final best mass estimate

%Ettori+04: We detect hints of negative evolution in the L - T, Mgas - T and L - Mtot relations, thus suggesting that
%systems at higher redshift have lower X-ray luminosity and gas mass for fixed temperature.
% Such results point toward a scenario in which a relatively lower gas density is present in high-redshift
%objects, thus implying a suppressed X-ray emission, a smaller amount of gas mass and a higher entropy level.

%Lx: preheating drives relations away from self-similar in the observed direction (student of Gus Evrard)

%masses within cooling flow region can be underestimated by up to factor 2: Markev+Vikl 2007, p27 result should always
%be an underestimate, but center only contains a fraction of the total mass

%-----------------------------------------------------
%\subsubsection{Challenges of accurate high-z mass estimates}

%................................................
\subsection{Comparison to similar X-ray clusters at $z\!\sim\!1$ }
\label{s4_comparison}

The growing number of known $z\!\sim\!1$ galaxy clusters with available multi-wavelength data will soon allow
comprehensive population studies  from the group regime to the most massive systems at this important cosmic epoch. A
point of  particular interest will be the intrinsic cluster-to-cluster variations at a given epoch for systems with
comparable masses, e.g. with respect to the thermodynamic properties of the ICM, the formation history of the galaxy
population, or the structure of the Dark Matter halo. This intrinsic physical scatter sets on one hand the ultimate
limit of accuracy for the calibration of high-z scaling relations and reflects on the other hand the different
conditions during cluster formation, e.g. the collapse epoch, the influence of the LSS environment, or AGN feedback.

Within the XDCP survey, the cluster XMMU\,J1229.5+0151 at $z\!=\!0.975$ \citep{Santos2009a} is almost the `twin brother'
of XMMU\,J1230.3+1339. Besides the concordant redshift, XMMU\,J1229.5+0151 features a comparable X-ray luminosity, a
similar ICM temperature within the errors ($T_X\simeq6.4$\,keV), and a consistent velocity dispersion
($\sigma_{\mathrm{r}}\simeq 683$\,km/s). The cluster is also optically rich with a well populated red sequence and two
bright galaxies in the center. A third bright galaxy of comparable magnitude and mass  is located on the outskirts of
the cluster, at a projected distance similar to the BCG2 group in XMMU\,J1230.3+1339. Overall, the basic global
properties of the two clusters seem to be well consistent within the observational uncertainties.

The two most distant clusters from the Wide Angle ROSAT Pointed Survey (WARPS) \citep{Perlman2002a} are further
appropriate candidates for a direct comparison with  XMMU\,J1230.3+1339 as they are similar in mass and redshift.
%Besides a detailed X-ray study \citep{Maughan2006a}, these systems have also measured SZE parameters
%\citep{Muchovej2007a}. 
The most distant WARPS cluster, Cl\,1415.1+3612 at $z\!=\!1.03$, has an ICM temperature of
$T_{X}\!\simeq\!5.7$\,keV, a luminosity of L$^{\mathrm{bol}}_{\mathrm{X,200}}\!\simeq\!10.4\times 10^{44}$\,erg/s,  a
hydrostatic total mass estimate of $M_{200}\!\simeq\!3.8 \times 10^{14}\,\mathrm{M_{\sun}}$ \citep{Maughan2006a}, 
%, SZE parameters of $\Delta T_{\mathrm{SZE}}\simeq -420\,\mu$K, and y$_{\mathrm{0}} \simeq 0.80 \times 10^{-4}$, 
and a velocity dispersion of
$\sigma_{\mathrm{r}}\!\simeq\!810$\,km/s \citep{Huang2009a}. The second system Cl\,1429.0+4241 at $z\!=\!0.92$ is
characterized by the parameters $T_{X}\!\simeq\!6.2$\,keV, L$^{\mathrm{bol}}_{\mathrm{X,200}}\!\simeq\!9.6\times
10^{44}$\,erg/s, and $M_{200}\!\simeq\!4.5 \times 10^{14}\,\mathrm{M_{\sun}}$. 
%$\Delta T_{\mathrm{SZE}}\simeq -750\,\mu$K, and y$_{\mathrm{0}} \simeq 1.44 \times 10^{-4}$. 
With core radii of $r_c\!\sim\!100$\,kpc (for $\beta\!=\!2/3$) both systems are
more compact in their ICM structure compared to XMMU\,J1230.3+1339 and their bolometric X-ray luminosities are 40-50\%
higher,  while the total mass and temperature estimates are %consistent 
similar within the uncertainties. 
\subsection{Appearance at $z=0$}
\label{s4_z0look}

We now address the question of the expected system properties of XMMU\,J1230.3+1339 after an additional 7.6\,Gyr of
cluster evolution (i.e. $z\!=\!0$) and compare these to the characteristics of the local Coma cluster ($z\!\simeq\!0.023$).

A lower limit for the total system mass at $z\!=\!0$, without considering further accretion, can be obtained from the
self-similar evolution model. Due to the decreasing reference density $\rho_{\mathrm{cr}}(z)$ with time, the system
radius is expected to evolve as $R_{200}(z\!=\!0)\!\simeq\!E^{2/3}(z)\cdot R_{200}(z)$, %(see Sect.\,\ref{s3_opt_richness}),
which translates approximately linearly into the total cluster mass $M_{200}(z\!=\!0)\!\simeq\!E^{2/3}(z)\cdot M_{200}(z)\!\simeq\!1.4 \cdot M_{200}(0.975)$ (see e.g. Equ.\,\ref{e2_HSmass_profile}). However, once including the continuing mass
accretion the cluster growth rate since $z\!\simeq\!1$ is much faster. The simulations of \citet{Burns2008a} predict
average mass growth rates of $\simeq$5-10\% per Gyr at late times and $\simeq$30-50\% per Gyr during active mass
assembly periods around $z\!\sim\!1$. As a benchmark, the time evolution of the largest local $\sim$$10^{15}\,\mathrm{M_{\sun}}$
halo in the Millennium-II simulation \citep{Boylan2009a} exhibits a total mass increase since $z\!\sim\!1$ by more than
a factor of 5. A realistic prediction for the evolved total cluster mass of XMMU\,J1230.3+1339 at the current epoch is
hence $M_{200}(z\!=\!0)\!\sim\!(1.5$-$2)\times 10^{15}\,\mathrm{M_{\sun}}$, i.e. ranked among the most massive local clusters and
likely even higher in mass compared to the Coma system with $M\!\simeq\!10^{15}\,\mathrm{M_{\sun}}$.

From the X-ray perspective, Coma is a prototype Non-Cool-Core cluster featuring a shallow surface brightness profile
with a large core radius of $r_c\!\sim\!300$\,kpc \citep[e.g.][]{Briel1992a}. A %very 
similar X-ray appearance can be
expected for XMMU\,J1230.3+1339, which shows already in the current configuration a very extended NCC SB distribution.
The core impact of the BCG2 group and subsequently accreted matter (Sect.\,\ref{s4_dynstate}) will likely further
disperse the remaining central ICM density peak within the next Gigayear
(Figs.\,\ref{fig_SBprof}, \ref{fig_MultiLambda_DensityMaps}) resulting in an even more flattened  X-ray SB profile for
the expected new equilibrium configuration a few Gyr later.

The upcoming merging with the BCG2 group will lead to another close resemblance to Coma in the optical, namely the
existence of two central, almost equally massive brightest galaxies. The core impact velocity of the BCG2 group with
the main halo is too high (see Sect.\,\ref{s3_GroupBullet}) to allow efficient  merging with other central galaxies,
i.e. BCG2 is expected to retain its entity as a massive galaxy. After relaxation within the main cluster halo and
several Gyr of dynamical friction at work,  BCG2 can be expected to settle close to the cluster center at low
redshifts and form a pair of two dominant cluster galaxies together with the nominal BCG. In terms of the general
optical richness, XMMU\,J1230.3+1339 is already comparable to Coma (richness class R=2) when considering the accessible
bright end of the galaxy population. With the continuing accretion of galaxies from the surrounding filaments, the new
system is likely to surpass Coma in optical richness by $z\!=\!0$.

Coma is also the prototype for  clusters featuring a diffuse radio halo. In Sect.\,\ref{s3_GroupBullet}, we discussed
the possibility that such a radio halo originates from merging events such as the one observed in XMMU\,J1230.3+1339.
If this scenario prevails, the evolved cluster  might additionally show similarity to Coma in the radio regime at certain epochs.

\begin{table}[t]    % h = here ; positioning
\caption{Multi-wavelength properties of the cluster XMMUJ\,1230+1339.} \label{table_results}

\centering
\begin{tabular}{ l l l l}
\hline \hline

Property & Value & Unit &  M$^{\mathrm{est}}_{200}$ [$10^{14}\,$M$_{\sun}$]  \\
%& $z$  & $f_{\mathrm{X}}$(0.5--2.0\,keV) & $r_{\mathrm{c}}$  &  $L_{\mathrm{X}}$(0.5--2.0\,keV)  \\ %& $M_{500}$ \\
% & &  Eq. 2000 &  Eq. 2000  &    & erg\,s$^{-1}$cm$^{-2}$ & \arcsec &  erg\,s$^{-1}$  \\ %& $M_{\sun}$ \\

\hline

% General
RA      &  12:30:16.9 &   &    \\
DEC     &  13:39:04  &   &   \\
N$_{\mathrm{H}}$ & $2.62 \times 10^{20}$   &  cm$^{-2}$   &    \\
z       &  $0.975\pm 0.002$    &   &    \\
M$^{\mathrm{best}}_{200}$ &   $4.19^{+0.79}_{-0.77} \times 10^{14}$  & M$_{\sun}$   &    \\

\hline

R$_{\mathrm{X,200}}$ & $1017^{+68}_{-60}$ & kpc &  \\
R$_{\mathrm{X,500}}$ & $670^{+45}_{-39}$ & kpc  &  \\
R$_{\mathrm{X,2500}}$ & $297^{+20}_{-17}$ & kpc  &  \\

\hline
% X-ray
T$_{\mathrm{X,2500}}$ & $5.30^{+0.71}_{-0.62}$ & keV & $3.61^{+0.77 \, (+0.73)}_{-0.68}$ \\
T$_{\mathrm{X,71\arcsec}}$ & $6.0^{+1.6}_{-1.2}$ & keV & \\
Z$_{\mathrm{X,2500}}$ & $0.43^{+0.24}_{-0.19}$ & Z$_{\sun}$  &   \\

f$^{0.5-2\,\mathrm{keV}}_{\mathrm{X,500}}$ & $ (5.14 \pm 0.54)\times 10^{-14}$ & erg\,s$^{-1}$\,cm$^{-2}$  & \\
L$^{0.5-2\,\mathrm{keV}}_{\mathrm{X,500}}$  & $(1.92 \pm 0.20) \times 10^{44}$ & erg\,s$^{-1}$ &  \\
L$^{0.5-2\,\mathrm{keV}}_{X,[0.15-1]500}$  & $(1.53 \pm 0.21) \times 10^{44}$ & erg\,s$^{-1}$ &  \\
L$^{\mathrm{bol}}_{\mathrm{X,500}}$  & $(6.50 \pm 0.68) \times 10^{44}$ & erg\,s$^{-1}$ & $4.1^{\pm 0.6 \, (\pm 1.0)}$  \\

Z$_{\mathrm{2500}}$  & $0.43^{+0.24}_{-0.19}$  & Z$_\odot$ &   \\
$\beta$$_{\mathrm{X}}$ & $0.84 \pm 0.47$ &  & $4.9^{\pm 2.8}$  \\
r$_{\mathrm{c}}$ & $215 \pm 110$ & kpc &  \\
c$_{\mathrm{SB}}$ & 0.07 &  &  \\
$n_{\mathrm{e,0}}$ & $(2.3 \pm 1.4) \times 10^{-2}$ & cm$^{-3}$ &  \\
t$_{\mathrm{cool,20kpc}}$ & $4.0 \pm 2.4$ & Gyr &  \\
$c_{\mathrm{sound}}$ & 1160 & km/s &   \\

M$_{\mathrm{gas,500}}$  & $(3.0 \pm 0.9)\times 10^{13}$  & $\mathrm{M_{\sun}}$ & $4.2^{\pm 1.5}$ \\
Y$_{\mathrm{X,500}}$ & $1.6^{\pm 0.5 \, (+0.2)}\times 10^{14}$  & $\mathrm{M_{\sun}}$\,keV & $3.6^{\pm 1.1  \, (+0.4)}$  \\
f$_{\mathrm{gas,500}}$  & $0.100 \pm 0.035$ &  &  \\

\hline
% Optical
R$^{\mathrm{rich}}$ & 2 &  &  \\
N$_{200}$ & 57 &  & $6.3^{\pm 0.7 \, (\pm 2.0)}$  \\
$L^{\mathrm{z}}_{200}$ & $(9.2 \pm 2.3)\times 10^{12}$ &  $\mathrm{L_{\sun}}$ & $11.1^{\pm 2.8 \, (\pm 3.7)}$  \\

$L^{\mathrm{tot,z}\arcmin}_{500}$  & $(1.06 \pm 0.16)\times 10^{13}$ & $\mathrm{L_{\sun}}$ & \\

R$-$z & $2.05\pm 0.10$ & mag &  \\
$m^{\mathrm{z}}_{\mathrm{BCG}}$  & $19.90\pm 0.02$ & mag &  \\
$M^{\mathrm{z}}_{\mathrm{BCG}}$ & $-24.95\pm 0.03$ & mag  & $4.4^{+0.7\,(+1.0)}_{-0.7\,(-2.2)})$ \\
d$^{\mathrm{center}}_{\mathrm{BCG}}$ & 140 & kpc &  \\
$\sigma_{\mathrm{r}}$ & $658 \pm 277$  & km/s &  $2.8^{+5.2}_{-2.3}$  \\
$M/L^{\mathrm{z}\arcmin}$  & $46.7 \pm 11.3$ & M$_{\sun}$/L$_{\sun}$ & \\

opt$_{\mathrm{opt}}$  &  & & $8.6^{+1.3 \, (+2.8)}_{-1.3 \, (-3.9)}$   \\

%\hline
% SZE
%$\Delta T_{\mathrm{SZE}}$  & $-217 \pm 53$ & $\mu$K$_{\mathrm{CMB}}$ & \\
%$y_{\mathrm{0}}$ & $(0.88\pm 0.22) \times 10^{-4}$ &  &  \\
%Y$_{\mathrm{SZE}}$  & \new{$1 \times 10^{-3}$} &  & \new{revise} \\
%$Y_{\mathrm{SZ},500}$ & $1.44^{\pm 0.46\, (+0.09)} \times 10^{-11}$ & & $4.1^{\pm 1.0 \, (+0.3)}$ \\

\hline

WL  $\sigma^{\mathrm{SIS}}_{\mathrm{r}}$ & $1\,271 \pm 275$  & km/s & $7.6^{+3.3 \, (+3.0)}_{-3.3}$   \\
WL  NFW &   &  & $6.9^{+3.3 \, (+2.2)}_{-3.3}$   \\

\hline

P$_{\mathrm{1440\,MHz}}$  & $8.3 \times 10^{24}$ & W\,Hz$^{-1}$ & $9.5^{\pm 3.8}$ \\

%\hline

%$D_{\mathrm{A}}$  & $994^{+ 620 \, (+250)}_{-570 \, (-250)}$ & Mpc &  \\
%$D_{\mathrm{A}}$  & $1876^{+ 1160 \, (+420)}_{-1060 \, (-420)}$ & Mpc &  \\
%1876^{+ 1160 \, (+420)}_{-1060 \, (-420)

\hline
\end{tabular}
\end{table}
% ----------------------------------------- End Table -----------------------------------------------
% Copy&Paste:  XMMU\,J1230.3+1339

%============== SECT 6 ===============================================================================================
%______________________________________________________________

\section{Summary and conclusions}
\label{s5_concl}

We have presented a multi-wavelength analysis and characterization of the galaxy cluster XMMU\,J1230.3+1339 at
$z\!=\!0.975$, which is summarized in the following. An overview of the system parameters and the derived mass
estimates is provided in Table\,\ref{table_results}.
%has numerous features motivates a more detailed investigation for addressing a number of fundamental questions

   \begin{enumerate}
    \item The galaxy cluster was discovered within the XMM-{\it Newton} Distant Cluster Project (XDCP) as a bright, extended X-ray source
    in an archival XMM-{\it Newton}  observation targeting the Virgo galaxy NGC\,4477.

   \item We have characterized the X-ray properties of the system with a joint analysis of the XMM-{\it Newton} data and %an 
   additional archival
   {\it Chandra} observations finding a bolometric luminosity of $L^{\mathrm{bol}}_{\mathrm{X,500}}\!\simeq\!6.50 \times
   10^{44}$\,erg\,s$^{-1}$ ($\pm\,10\%$), an ICM temperature of $T_{\mathrm{X,2500}}\!\simeq\!5.30$\,keV ($\pm\,13\%$),
   a gas mass of $M_{\mathrm{gas,500}}\!\simeq\!3.0 \times 10^{13}\,\mathrm{M_{\sun}}$ ($\pm\,30\%$), a gas mass fraction of
   f$_{\mathrm{gas,500}}\!\simeq\!0.10$ ($\pm\,35\%$) consistent with local values, and a $Y_X$ parameter of
   $Y_{\mathrm{X,500}}\!\simeq\!1.6 \times 10^{14}\,\mathrm{M_{\sun}}$\,keV ($\pm\,38\%$).

   \item The {\it Chandra} analysis of the X-ray surface brightness distribution yielded an approximate core radius of
   215\,kpc, a $\beta$ profile parameter of 0.84, and an overall classification as a Non-Cool-Core cluster. The very central
   part of the SB profile is better represented by a double $\beta$-model with an additional more compact radial
   component, from which we derive a central electron density of $n_{\mathrm{e,0}}\simeq 2.3 \times 10^{-2}$\,cm$^{-3}$
   ($\pm\,60\%$) and a corresponding cooling time of about 4\,Gyr.

    \item The cluster features a very rich galaxy population (Abell richness R=2, $N_{200}\!\simeq\!57$) with a prominent
    red-sequence in the color-magnitude-diagram
    and a total z\arcmin-band luminosity
    of $L^{\mathrm{tot,z}\arcmin}_{500,\mathrm{ap}}\!\simeq\!1.06\times 10^{13}\,\mathrm{L_{\sun}}$ ($\pm\,15\%$).

    \item The Brightest Cluster Galaxy is located at a projected distance of 140\,kpc from the nominal cluster center
    and has a luminosity of $L^{\mathrm{z}}_{BCG}\simeq 6.1 \times 10^{11}\,\mathrm{L_{\sun}}$, two magnitudes brighter than the
    characteristic $L*$ luminosity at this redshift.

    \item Based on 13 secure spectroscopic cluster members, we find a radial velocity dispersion
    estimate of $\sigma_{\mathrm{r}}\!\simeq\!658$\,km/s ($\pm\,42\%$).

%    \item We detect the Sunyaev-Zeldovich effect for the cluster on the $4\,\sigma$ significance level and find a
%    central temperature decrement at 150\,GHz of $\Delta T_{\mathrm{SZE}}\simeq -217\,\mu$K$_{\mathrm{CMB}}$
%    ($\pm\,24\%$). Based on the X-ray determined $\beta$ model parameters we obtain an integrated Compton-$y$
%    parameter of $Y_{\mathrm{SZ},500}\simeq 1.44 \times 10^{-11}$ ($\pm\,46\%$).

    \item 
    As presented in detail in the accompanying  Paper\,II of \citet{Lerchster2010a},
    we %also 
    detect a weak gravitational lensing signal for the cluster on the $3.5\,\sigma$ significance level,
    which currently sets the redshift limit for the extraction of a weak lensing signal based on ground imaging.
    %Details of the weak lensing  analysis are presented in an accompanying  Paper\,II. %paper by \citet{Lerchster2010a}.

    \item We have performed a multi-wavelength analysis of the cluster structure based on projected density maps of (i) 
     red galaxies, (ii) the X-ray surface brightness, (iii) the total z\arcmin-band light, %the SZE temperature decrement, 
    and (iv) the weak lensing mass distribution. % map. 
    We find a well defined concordant cluster center and a main elongation axis along the SE-NW direction present
    in all maps.

    \item Several cluster-associated sub-components were identified on different cluster-centric distance scales ranging from  a
     merging event in the central region to two infalling groups on the cluster outskirts and two galaxy filaments
     extending beyond the nominal cluster radius.

    \item One of the groups has recently entered the $R_{500}$ radius of the cluster along a radial infall orbit.
    The central dominant galaxy of this group has also BCG properties and is only marginally fainter than the nominal
    Brightest Cluster Galaxy.

    \item We find evidence for a central fly-through group `bullet' close to cluster core passage from a wedge-like
    X-ray surface brightness feature, an associated off-center galaxy density and luminosity peak with the front end aligned to the
    X-ray surface brightness jump, indications for tidally disrupted galaxies, blue-shifted redshifts in the
    cluster-centric reference frame, and a matching elongation in the weak lensing mass map.

    \item Marginally extended 1.4\,GHz radio emission  located in the wake region of the
    central merging group along the axis of the apparent `bullet' trajectory is detected by the VLA FIRST survey. 
    The origin of this radio emission could be associated to
    an extended radio jet connected to the center of one of the nearby cluster galaxies, or possibly the direct observable result of
    ultra-relativistic particle acceleration in the turbulent wake region of the merger, as predicted by simulations.
    A cross-check with the scaling relation for the total power in lower-z radio halos with the observed radio source in the cluster
    is reasonably consistent with the possibility that they share a related physical origin.

    \item We have attempted a tentative reconstruction of the  3D-accretion geometry of the cluster environment that
    connects the identified associated components with available redshift information to the main halo.
    %based on  was obtained of the different cluster components with

   \item We have derived 13 proxies for the total cluster mass from the different wavelength regimes. The combination
   of four reliable quasi-independent mass estimates yielded a joint best estimate for the total cluster mass
   of $M^{\mathrm{best}}_{200}\!\simeq\!4.19\times 10^{14}\,\mathrm{M_{\sun}}$ ($\pm\,19\%$). All X-ray derived mass estimates 
   %and the SZE mass proxy 
   are in good agreement %fully consistent 
   with this preferred mass interval,  the weak lensing results are consistent %s are
   within the currently larger statistical %$\sigma$ 
   uncertainties, and the tested optical mass proxies tend to suggest
   significantly higher total system masses, albeit with large systematic uncertainties.

%    \item Based on the combination of the X-ray and SZE data, we have attempted the first absolute distance measurement
%    at this redshift, finding an estimate for the angular diameter distance to the cluster of $D_{\mathrm{A}}\simeq 1876$\,Mpc,
%    in good agreement with the concordance cosmological model.

    \item We argued that the cluster will likely evolve into  a Coma-like system with expected similarities in the ICM structure, the optical
    appearance with two dominant central galaxies, and possibly even surpass Coma in terms of galaxy richness and total
    mass.

    %\item The listed features of the newly identified cluster make it a prime target for future in depth follow-up
    %observations in all wavebands

%\item The cluster XMMU\,J1230.3+1339 is an ideally suited astrophysical laboratory for detailed follow-up studies
%   to observationally address various fundamental questions in cluster physics and cosmology spanning more than six orders of
%   magnitude in physical scale. Such studies could shed new light on the origin of ultra-relativistic electrons and
%  diffuse radio emission in clusters, various galaxy interaction processes in different environments, hydrodynamic merger
%  shocks in the intracluster medium, the origin of the Cool Core/Non-Cool-Core Cluster dichotomy, the accretion modes
% of (dark) matter, the connection to large-scale structure filaments, and on the absolute geometric distance to the cluster.
%

   \end{enumerate}

% Final Statement and outlook:

\noindent
The newly identified galaxy cluster XMMU\,J1230.3+1339 is an excellent test laboratory at $z\!\sim\!1$ for detailed %follow-up 
observational studies to %observationally 
address various fundamental questions related to the assembly phase of massive systems and their Dark Matter, ICM, and galaxy components.  Recent follow-up campaigns of the cluster environment include additional optical spectroscopy at VLT/FORS\,2, deeper X-ray coverage with  {\it Chandra}, and mm  observations with APEX-SZ. Upcoming studies will e.g. aim to derive a more detailed picture on the 
interaction processes of the different components related to the central merging event or to characterize the galaxy populations as a function  
of cluster-centric distance in different sub-components and the large-scale structure filaments.

\begin{acknowledgements}
We thank the referee for useful comments and suggestions that helped to improve this paper.
RF would like to thank Andrea Merloni, Robert Dunn, Angela Bongiorno, Miguel Verdugo, and Klaus Dolag 
for fruitful discussions and helpful comments.
This work was supported by the DFG under grants Schw536/24-1,
Schw 536/24-2, BO 702/16-3, and the German DLR under grants 50 QR 0802 and 50 OR 0405. 
HQ thanks the FONDAP Centro de Astrofisica for partial support.
%GL was supported by DLR under contract  500OX201, JK is supported by DFG under contracts Schw536/24-1 and BO702/16-2.
We acknowledge additional support from the excellence cluster {\em Universe EXC153}. 
The XMM-{\it Newton} project is an ESA Science Mission with instruments and contributions directly funded by ESA Member
States and the USA (NASA). The XMM-{\it Newton} project is supported by the Bundesministerium f\"ur Wirtschaft und
Technologie/Deutsches Zentrum f\"ur Luft- und Raumfahrt (BMWI/DLR, FKZ 50 OX 0001), the Max-Planck Society and the
Heidenhain-Stiftung. 
%\new{APEX-SZ instrument acknowlegdement} 
This research has made use of the NASA/IPAC Extragalactic
Database (NED) which is operated by the Jet Propulsion Laboratory, California Institute of Technology, under contract
with the National Aeronautics and Space Administration. 
\end{acknowledgements}

%:Bibliography
% Bibliography
\bibliographystyle{aa} % style aa.bst with Journal Definitions
\bibliography{../../BIB/RF_BIB_10}
%\bibliography{RF_BIB_09}
% .bib file has to be in same directory, linking does not work

% Styles:
% Single Citation
% \citep{jon90}                 --> (Jones et al. 1990)
% \citep[see][]{jon90}          --> (see Jones et al. 1990)
% \citep[see][chap.~2]{jon90}   --> (see Jones et al. 1990, chap. 2)
% \citet{jon90}                 --> Jones et al. (1990)
% \citep[e.g. the South Pole Telescope and APEX-SZ][]{Carlstrom2006,Dobbs2006})
% Multiple Citations
%\citep{jon90,jam91}            --> (Jones et al., 1990; James et al. 1991)
%\citep{jon90,jon91}            --> (Jones et al. 1990, 1991)
%\citep{jon90a,jon90b}          --> (Jones et al. 1990a,b)
%\citet{jon90,jam91}            --> Jones et al. (1990); James et al. (1991)
% \citeyear{key} ==>>             1990

% Symbols:
% \sun
% \degr
% \arcsec
% \arcmin
% \farcs --> fractional arcseconds ".
% \farcm --> fractional arcmins    .
% \la --> <~
% \ga --> >~

%\begin{thebibliography}{}
%\end{thebibliography}

\end{document}